\documentclass[%
 reprint,
superscriptaddress,
nofootinbib,
 amsmath,amssymb,
 aps, prl
floatfix,
]{revtex4-2}

\usepackage{graphicx}
\usepackage[caption=false]{subfig}
\usepackage{dcolumn}
\usepackage{bm}

\usepackage{siunitx}
\usepackage{physics}
\usepackage{placeins}
\usepackage{gensymb}
\usepackage{tikz}
\usetikzlibrary{external}
\usepackage{adjustbox}
\usepackage{todonotes}
\usepackage{multirow}
\usepackage{placeins}

\newcommand{\snr}{\text{SNR}}
\newcommand{\tem}{TEM$_{00-18}$ }
\newcommand{\temns}{TEM$_{00-18}$}

\newcommand{\Ap}{A^{\prime}}
\newcommand{\betaterm}{\frac{\beta}{\beta+1}}
\newcommand{\gagg}{g_{a\gamma \gamma}}
\newcommand{\veff}{V_{eff}}
\newcommand{\cost}{\langle \cos^2\theta \rangle_T}
\newcommand{\ldb}{\lambda_{\text{dB}}}

\newcommand{\ff}{\mathcal{F}(f)}

\begin{document}

\preprint{APS/123-QED}

\title{ADMX-Orpheus First Search for \SI{70}{\mu eV} Dark Photon Dark Matter: Detailed Design, Operations, and Analysis}

\author{R. Cervantes}%
  \email[Correspondence to: ]{raphaelc@fnal.gov}
  \affiliation{University of Washington, Seattle, WA 98195, USA}
  \affiliation{currently Fermi National Accelerator Laboratory, Batavia IL 60510}
\author{G. Carosi}
\affiliation{Lawrence Livermore National Laboratory, Livermore, CA 94550, USA}
 
\author{C. Hanretty}
  \affiliation{University of Washington, Seattle, WA 98195, USA}
\author{S. Kimes}
  \affiliation{University of Washington, Seattle, WA 98195, USA}
  \affiliation{currently Microsoft Quantum, Microsoft, Redmond, WA 98052, USA}
   \author{B. H. LaRoque}
  \affiliation{Pacific Northwest National Laboratory, Richland, WA 99354, USA}
\author{G. Leum}
  \affiliation{University of Washington, Seattle, WA 98195, USA}
\author{P. Mohapatra}
  \affiliation{University of Washington, Seattle, WA 98195, USA}
  \affiliation{currently Joby Aviation, San Carlos, CA 94063, USA}
  \author{N. S. Oblath}
  \affiliation{Pacific Northwest National Laboratory, Richland, WA 99354, USA}
\author{R. Ottens}
  \affiliation{University of Washington, Seattle, WA 98195, USA}
  \affiliation{currently NASA Goddard Space Flight Center Greenbelt, MD, United States}
\author{Y. Park}
  \affiliation{University of Washington, Seattle, WA 98195, USA}
  \affiliation{currently University of California, Berkeley, CA 94720}
  \author{G. Rybka}%
  \affiliation{University of Washington, Seattle, WA 98195, USA}
\author{J. Sinnis}
  \affiliation{University of Washington, Seattle, WA 98195, USA}
    \author{J. Yang}%
  \affiliation{University of Washington, Seattle, WA 98195, USA}
  \affiliation{currently Pacific Northwest National Laboratory, Richland, WA 99354, USA}

  \date{\today}

\begin{abstract}
  Dark matter makes up 85\% of the matter in the universe and 27\% of its energy density, but we do not know what comprises dark matter. It is possible that dark matter is composed of either axions or dark photons, both of which can be detected using an ultra-sensitive microwave cavity known as a haloscope. The haloscope employed by ADMX consists of a cylindrical cavity operating at the TM$_{010}$ mode and is sensitive to the QCD axion with masses of few \si{\mu eV}. However, this haloscope design becomes challenging to implement for higher masses. This is because higher masses require smaller-diameter cavities, consequently reducing the detection volume which diminishes the detected signal power. ADMX-Orpheus mitigates this issue by operating a tunable, dielectrically-loaded cavity at a higher-order mode, allowing the detection volume to remain large. This paper describes the design, operation, analysis, and results of the inaugural ADMX-Orpheus dark photon search between \SI{65.5}{\mu eV} (\SI{15.8}{GHz}) and \SI{69.3}{\mu eV} (\SI{16.8}{GHz}), as well as future directions for axion searches and for exploring more parameter space.
\end{abstract}
\maketitle


\section{Introduction}
Dark matter is the non-luminous, non-absorbing matter that makes up about 84.4\% of the matter of the universe and 26.4\% of its energy density~\cite{Zyla:2020zbs}. The Lambda cold dark matter ($\Lambda$CDM) model describes dark matter as feebly interacting, non-relativistic, and stable on cosmological timescales. The evidence for dark matter is abundant and includes the galactic rotation curves~\cite{Rubin:1982kyu,10.1093/mnras/249.3.523}, gravitational lensing~\cite{1998gravitational_lensing,10.1093/mnras/stw3385}, the bullet cluster~\cite{Markevitch_2004}, and the cosmic microwave background~\cite{2020Planck}. However, little is known about the nature of dark matter or what constitutes it. Many hypothetical candidates include WIMPS, sterile neutrinos, axions, axion-like particles, and dark photons~\cite{Zyla:2020zbs}. Dark matter may be made up of a combination of many of these particles (a dark sector).

Of particular interest are dark matter candidates that have wavelike properties. For a particle to be wavelike, the de Broglie wavelength $\ldb$ is much greater than the interparticle spacing. In other words, if the number of particles $N_{dB}$ inside a de Broglie volume $\ldb^3$ is large, the set of particles is best described as a classical wave~\cite{doi:10.1146/annurev-astro-120920-010024}. The dark matter density in our halo is fixed to $\rho\sim \SI{0.45}{GeV/cm^3}$~\cite{Read_2014}, so a smaller dark matter mass results in a larger $N_{dB}$. Because fermions cannot occupy the same phase space, wavelike dark matter must be bosonic.

A subset of dark matter candidates can be wavelike. Examples of wavelike dark matter include axions, axion-like particles (ALPs), and dark photons~\cite{Arias_2012, hui2021wave, essig2013dark, Zyla:2020zbs}. Axions are a compelling candidate because they solve an outstanding problem in particle physics known as the Strong CP problem~\cite{PhysRevLett.38.1440, PhysRevLett.40.279, PhysRevLett.40.223}. ALPs arise naturally from string theory, and dark photons arise from the simplest extension to the Standard Model (SM). 

The dark photon (DP) is a vector boson associated with an added Abelian U(1) symmetry to the Standard Model~\cite{essig2013dark, PhysRevD.104.092016, PhysRevD.104.095029}. The dark photon is analogous to the Standard Model (SM) photon in that the SM photon is also a vector boson associated with an Abelian U(1) gauge symmetry. The dark photon interacts with the SM photon through kinetic mixing~\cite{HOLDOM198665, HOLDOM1986196} via the Lagrangian

\begin{align}
  \mathcal{L} = -\frac{1}{4}F_1^{\mu \nu}F_{1\mu \nu} + -\frac{1}{4}F_2^{\mu \nu}F_{2\mu \nu} + \frac{1}{2}\chi F_1^{\mu \nu}F_{2\mu \nu} + \frac{1}{2} m_{\Ap}^{2} A^{\prime 2}
  \label{eqn:dp_mixing_lagrangian}
\end{align}
where $F_1^{\mu \nu}$ is the electromagnetic field tensor, $F_2^{\mu \nu}$ is the dark photon field tensor, $\chi$ is the kinetic mixing strength, $m_{\Ap}$ is the DP mass, and $\Ap$ is the DP gauge field. The consequence is that the dark photon and SM photon can oscillate into each other (reminiscent of neutrino oscillations)\footnote{More rigorously, the dark photon generates an effective current which can generate electromagnetic fields depending on the boundary condition.}. The photon frequency $f$ is related to the dark photon energy $E_{\Ap}$ by the relationship $f= E_{\Ap}$ (using natural units). For non-relativistic dark photons, $f\approx m_{\Ap}$.

The rate that dark photons decay into three photons is $\Gamma_{\Ap \rightarrow 3\gamma} = (\num{4.70e-8})\times \alpha^3 \alpha^{\prime} m_{\Ap}^9/m_e^9$,
where $\alpha$ is the fine structure constant, $m_e$ is the mass of the electron, and $\alpha^{\prime}\equiv\frac{(e\chi)^2}{4\pi}$ is the dark photon counterpart to the fine structure constant~\cite{PhysRevD.78.115012}. If the dark photon has a sufficiently small kinetic mixing, then it is stable on cosmological timescales and makes for a compelling dark matter candidate.  The dark photon lifetime is about the same as the age of the universe if $m_{A^{\prime}} (\chi^2 \alpha)^{1/9} < \SI{1}{keV}$. This condition is met for dark photons with $m_{\Ap} \approx \SI{1e-4}{eV}$ and $\chi < \num{e-12}$.

Several mechanisms would produce cosmic dark photons. The simplest mechanism is through quantum fluctuations during inflation~\cite{PhysRevD.93.103520}. These quantum fluctuations seed excitations in the dark photon field, resulting in the cold dark matter observed today in the form of coherent oscillation of this field. The predicted mass from this mechanism is $m_{\Ap}\approx\SI{e-5}{eV}\left (\SI{e14}{GeV}/H_I\right )^4$, where $H_I$ is the Hubble constant during inflation. Measurements of the cosmic microwave background tensor-to scalar-ratio constrain $H_I < \SI{e14}{GeV}$~\cite{2016Planck}, making the search for $m_{\Ap}>\SI{e-5}{eV}$ well-motivated. 

Another mechanism is the misalignment mechanism that is similar to that for axions~\cite{PRESKILL1983127}. However, the misalignment mechanism would inefficiently produce dark photon CDM. Due to its vector nature, the dark photon would have its energy density redshifted as the universe expands (in the same way SM photons redshift during cosmic expansion) and would not contribute to the energy density of matter $\Omega_m$~\cite{PhysRevD.104.095029, Arias_2012}. A nonminimal coupling to gravity needs to be invoked for the misalignment mechanism to generate the correct relic abundance. This leads to instabilities in the longitudinal DP mode, resulting in dark photons with a fixed polarization within a cosmological horizon. Dark photons can also be produced from topological defects like cosmic strings~\cite{PhysRevD.99.063529}. They may also be produced thermally through processes like $e^- + \gamma \rightarrow e^- + \Ap$~\cite{PhysRevD.101.063030}. However, due to the small coupling to SM particles, this process would be inefficient and not produce the observed abundance of dark matter.

This paper serves as a detailed companion to Ref.~\cite{PhysRevLett.129.201301}. Section~\ref{sec:dielectric_haloscope} motivates the use of dielectric cavities for $\mathcal{O}\left(\SI{100}{\mu eV}\right)$ wavelike dark matter searches and describes the Orpheus experiment on a conceptual level. Section~\ref{sec:cavity} describes the Orpheus electromagnetic design.  Section~\ref{sec:mechanics} describes the mechanical design. Section~\ref{sec:operations} describes the electronics, software controls, and experimental procedures. Section~\ref{sec:analysis} describes the analysis of the data collected from the inaugural dark photon search used to exclude dark photon dark matter with kinetic mixing strength $\chi > \num{1e-13}$ between \SI{65.5}{\mu eV} and \SI{69.3}{\mu eV}. This section also describes the electrodynamic simulation and characterization of the Orpheus cavity. Finally, Section~\ref{sec:future} discusses the possibility of using the Orpheus experiment to search for the QCD axion.

\section{Dielectric Haloscopes for Detecting Axions and Dark Photons: Orpheus Conceptual Design}\label{sec:dielectric_haloscope}

Dark photon dark matter (DPDM) can be detected through their mixing with the SM photon. The electromagnetic field produced by the dark photon is~\cite{PhysRevD.104.095029}

\begin{align}
  |\vb{E}_0| = \left | \frac{\chi m_{\Ap}}{\epsilon} \vb{\Ap}_0 \right |
\end{align}
where $\epsilon$ is the dielectric constant of the medium. The SM photon polarization is determined by the dark photon polarization. 

If dark photons oscillate into SM photons inside a microwave cavity with a large quality factor, then a feeble EM signal accumulates inside the cavity, which can be read by ultra-low noise electronics. This method is often deployed to search for axions~\cite{PhysRevLett.51.1415}, but the method works the same for dark photons without the need for an external magnetic field. The dark photon signal power is, in natural units~\cite{PhysRevD.104.092016}, 

\begin{align}
  & P_{S} = \eta \chi^2 m_{\Ap} \rho_{\Ap} V_{eff} Q_L \betaterm L(f, f_0, Q_L)\label{eqn:dp_power}  \\
  & L(f, f_0, Q_L) = \frac{1}{1+4\Delta^2}; \quad \Delta \equiv Q_L \frac{f-f_0}{f_0}\\
  & V_{eff} = \frac{\left (\int dV \vb{E}(\vec{x}) \vdot \vb{\Ap}(\vec{x})\right )^2}{\int dV \epsilon_r |\vb{E}(\vec{x})|^2|\vb{\Ap}(\vec{x})|^2}\label{eqn:veff}
\end{align}
 where $\eta$ is a signal attenuation factor (described in Section~\ref{sec:analysis}), $\rho_{\Ap}$ is the dark matter local density, $\veff$ is the cavity's effective volume, $Q_L$ is the loaded quality factor, and $\beta$ is the cavity coupling coefficient. $L(f, f_0, Q_L)$ is the Lorentzian term and depends on the SM photon frequency $f$, cavity resonant frequency $f_0$, and $Q_L$. $\veff$ is the overlap between the dark photon field $\vb{\Ap}(\vec{x})$ and the dark photon-induced electric field $\vb{E}({\vec{x}})$\footnote{In more traditional axion haloscope terms, it is the physical volume of the cavity times the form factor.}. Equation~\ref{eqn:dp_power} assumes the cavity size is much smaller than the dark photon de Broglie wavelength and the cavity bandwidth is much larger than the dark matter velocity dispersion, $Q_L << Q_{DM}$~\cite{PhysRevLett.55.1797, Kim_2020}. This type of detector that searches for a direct signal from the dark matter halo is known as a haloscope~\cite{PhysRevLett.51.1415}.

Haloscope experiments search for dark matter as a spectrally-narrow power excess over a thermal noise floor. The noise power $P_n$ is composed of the cavity's blackbody radiation and the added Johnson noise from the receiver that extracts power from the cavity. The noise power is written as $P_n=G k_b b T_n$, where $h$ is Planck's constant, $k_b$ is the Boltzmann constant, $G$ is the system gain, $b$ is the frequency bin width, and $T_n$ is the system noise temperature referenced to the cavity. The measured $P_n$ is often the average of thousands of power spectra and consequently follows the Central Limit Theorem, so $\sigma_{P_n} = P_n/\sqrt{N}$, where N is the number of averaged spectra. The number of spectra is $N = b \Delta t$, where $\Delta t$ is the integration time. The SNR for a haloscope signal is $\snr = P_S/\sigma_{P_n} = (P_S/P_n)\sqrt{b \Delta t}$.  If $Q_L < Q_{DM}$, a haloscope is sensitive to dark matter within its cavity bandwidth ${\Delta f = f_0/Q_L}$. The instantaneous scan rate is then

\begin{align}
  \dv{f}{t} = \frac{\Delta f}{\Delta t} = \frac{f_0 Q_L}{b}\left (\frac{\eta \chi^2 m_{\Ap} \rho_{\Ap} \veff \beta}{\snr T_n(\beta+1)}\right )^2. 
  \label{eqn:scan_rate}
\end{align}
Often, the bin width is taken to be of the same order of magnitude as the dark matter signal linewidth ${b \sim f_0/Q_{\Ap}}$, where $Q_{\Ap}\sim \mathcal{O}(\SI{e6}{})$~\cite{PhysRevD.42.3572}. 

The same haloscope immersed in a strong DC magnetic field can search for axions. Axions, in the presence of a magnetic field, oscillate into photons. In a microwave cavity, the axion signal power $P_a$, in natural units, is

\begin{align}
  & P_a = \eta \frac{\gagg^2}{m_a} m_a \rho_a B_0^2 V_{eff} Q_L \betaterm  L(f, f_0, Q_L)\label{eqn:axion_power} \\
  & V_{eff} = \frac{\left |\int dV \vb{B}_0\vdot \vb{E}_a \right |^2}{B_0^2 \int dV \epsilon_r |E_a|^2}\label{eqn:axion_veff}
\end{align}
where $\gagg$ is the axion-photon coupling constant, $m_a$ is the axion mass, $\rho_a$ is the axion local density, $\vb{B}_0$ is the external DC magnetic field, $\veff$ is the effective volume, and $\vb{E}_a$ is the axion-induced electric field. $\veff$ in Equation~\ref{eqn:axion_veff} is different from that in Equation~\ref{eqn:veff} because the SM photon polarization is aligned with the $\vb{B}_0$. $\vb{B}_0$ is known, fixed, and controlled by the lab, so the haloscope receiver can be optimized to detect the full power of the axion-induced electric field.

Established programs like ADMX use haloscopes consisting of a cylindrical cavity inside a solenoid field and operate at the TM$_{010}$ mode to search for QCD axions with masses around a few \si{\mu eV}~\cite{PhysRevLett.120.151301, PhysRevLett.124.101303, PhysRevLett.127.261803}. Unfortunately, this haloscope design becomes increasingly difficult to implement at higher frequencies. Increasing mass corresponds to higher frequency photons. Operating at the TM$_{010}$ mode would require smaller-diameter cavities, and a smaller cavity volume reduces the detector signal power. The volume would scale by $V_{eff} \propto f^{-3}$ if one wanted to keep the same aspect ratio. Furthermore, the anomalous skin effect causes the unloaded quality factor $Q_0$ to decrease with frequency, further decreasing the signal power. The surface resistivity is $R_s \propto f^{2/3}$~\cite{chambers1952anomalous, pippard1947surface, chou1995anomalous}, so $Q_0 \propto f^{-2/3}$. The quantum noise limit also increases linearly with $f$. The combination of these effects causes the axion signal power to scale as $P_s \propto f^{-3.66}$ and the scan rate to scale as $\dv{f}{t} \propto f^{-8.66}$. To put this in perspective, for Run 1B, ADMX employed a \SI{136}{L} cylindrical cavity that resulted in an axion signal power of about ${P_a = \SI{2.2e-23}{W}}$ and a scan rate $df/dt = \SI{543}{MHz/yr}$ operating at around \SI{740}{MHz}~\cite{PhysRevD.103.032002}. At \SI{15}{GHz}, $\veff$ would scale to \SI{6.53}{mL}, the $P_a$ would scale to \SI{3.62e-28}{W}, and $df/dt$ would scale to \SI{2.6}{mHz/year}. Thus it would be challenging to implement a cylindrical cavity haloscope operating at the TM$_{010}$ mode to search for DFSZ axions beyond \SI{10}{GHz}. 

The detection volume can be increased by combining many cylindrical cavities. This is the plan for future ADMX runs~\cite{10.1007/978-3-030-43761-9_7}. However, once the frequency approaches \SI{15}{GHz}, the freespace wavelength is about \SI{1}{cm}. To have $\veff = \SI{1}{L}$, one would need to coherently power combine more than \num{1000} cylindrical cavities. Instrumenting and operating these many cavities in a coordinated way is challenging with current technology.

One can keep the cavity volume large and operate at a higher-order mode. But without further measures, the cavity modes would couple poorly to dark photons or axions.  Let $\vu{d}$ be the polarization of either $\vb{B}_0$ in an axion search or $\vb{\Ap}$ in a dark photon search. For higher-order modes, the spatial oscillations in $\va{E}$ result in $\veff \propto \qty|\int dV {\vb{E}\vdot \vu{d} }| \approx 0$, even though the physical volume is large. Thus, there is little benefit to operating an empty cylindrical cavity at a higher-order mode.

As a concrete example, the ORGAN experiment~\cite{MCALLISTER201767} operates at the TM$_{020}$ mode at a fixed frequency ${f_0 = \SI{26.5}{GHz}}$. The form factor is defined as $C \equiv \veff/V$, where V is the physical volume of the cavity. This iteration of ORGAN has $C=0.13, V = \SI{7.78}{mL}$, resulting in $\veff \approx \SI{1}{mL}$. Table 1 of reference~\cite{MCALLISTER201767} demonstrates that for a fixed frequency $f_0$, $\veff$ remains fixed at $\sim \SI{1}{mL}$ for the TM$_{010}$, TM$_{020}$, and TM$_{030}$ modes even though the physical volume has increased. However, $Q_L$ becomes worse with higher order modes. 

\begin{figure}[htp]
  \centering
  \includegraphics[width = \linewidth]{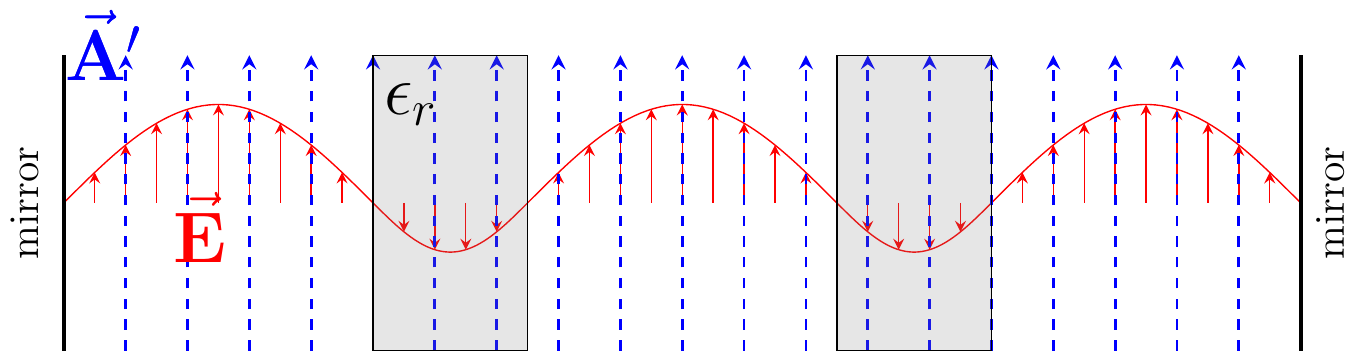}
  \caption{A multiwavelength dielectrically-loaded cavity. Dielectrics suppress electric fields. So dielectrics are placed strategically such that $V_{eff} \propto \int dV \vb{E} \vdot \vb{A} > 0$. The dashed blue lines represent the dark photon field. The red lines represent the dark photon-induced electric field. The grey box represents the dielectric material.}
  \label{fig:dielectric_haloscope}
\end{figure}

Fortunately, the coupling of higher-order modes to dark matter can increase when dielectrics are placed inside the cavity. Dielectrics suppress electric fields. If dielectrics are placed strategically, then the overlap between $\vb{E}$ and $\vu{d}$ is greater than zero, as shown in Fig.~\ref{fig:dielectric_haloscope}. Because the cavity size is no longer constrained by the photon wavelength, the effective volume can become arbitrarily large, and the dark matter signal power is greater than what it would have been for a cylindrical cavity operating at the TM$_{010}$ mode. This makes higher-order mode dielectric cavities suitable for higher-frequency dark matter searches.

Orpheus is the implementation of the dielectric haloscope concept to search for axions and dark photons around \SI{70}{\mu eV}. Orpheus\footnote{Orpheus was initially designed to have a spatially alternating magnetic field rather than a periodic dielectric structure~\cite{PhysRevD.91.011701}. However, this alternating magnetic field design is difficult to scale to many Tesla.} is a dielectrically loaded Fabry-Perot open cavity. Dielectric plates are placed every fourth of a half-wavelength to shape the electric field and increase the \tem mode's coupling to the dark photon. Orpheus is designed to search for axions and dark photons around \SI{16}{GHz} with over \SI{1}{GHz} of tuning range. The cavity tunes by changing its length, and the dielectric plate positions are adjusted appropriately.

There are several benefits to the open resonator design. Fewer metallic walls lead to smaller ohmic losses and fewer resonating modes. This leads to a sparse spectrum and fewer mode crossings, making it easier to follow the mode of interest while the cavity tunes. 	
However, this experiment has many challenges. First, the optics should be designed so that the \tem mode has a  \SI{1}{GHz} tuning range. This includes choosing the right radius of curvature for the Fabry-Perot mirrors and appropriate dielectric thicknesses. In addition, the mechanical design for such a cavity is complicated because there are many moving parts that have to work in a cryogenic environment. Finally, in addition to ohmic losses, both diffraction and dielectric losses will decrease the quality factor\footnote{This iteration of Orpheus achieved $Q_L\approx 10000$. However, this cavity electrodynamic design has not yet been optimized. The literature suggests that quality factors of \num{5e4}~\cite{Dunseith_2015} and even \num{2e5}~\cite{Clarke_1982} are achievable for GHz range Fabry-Perot cavities.}.

Because of their potential to search for higher-mass dark matter than current experiments, dielectric haloscopes are being developed by other collaborations. Examples include MADMAX~\cite{Brun2019, PhysRevLett.118.091801}, LAMPOST~\cite{PhysRevD.98.035006, chiles2021constraints}, MuDhi~\cite{PhysRevD.105.052010}, and DBAS~\cite{PhysRevApplied.9.014028, PhysRevApplied.14.044051}. 

\section{Orpheus Cavity Design}\label{sec:cavity}
The Fabry-Perot cavity consists of a flat aluminum mirror and a curved aluminum mirror with a radius of curvature, $r_0 = \SI{33}{cm}$. $r_0$ is chosen to be about twice the cavity optical length near \SI{18}{GHz} so that the flat mirror is at the focus of the curved mirror. Both mirrors are \SI{15.2}{cm} in diameter. The cavity tunes by changing the distance between mirrors, and the dielectric plates are adjusted appropriately. The curved mirror, bottom dielectric plate, and top dielectric plate are each controlled by a pair of threaded rods driven by a room-temperature stepper motor. Thus the cavity has three degrees of freedom. A pair of scissor jacks constrain the inner two dielectric plates so that they are evenly spaced between the top and bottom dielectric plate. The mechanical design is described in further detail in Section~\ref{sec:mechanics}.

The dielectric plates consist of 99.5\% alumina sheets purchased from Superior Technical Ceramic. The alumina plates are about $\SI{151.6}{mm}\times\SI{151.6}{mm}\times\SI{3}{mm}$. The plates are octagonal to approximate a circular shape, but the straight lines are easier to machine. The dielectric constant is $\epsilon_r = 9.8$ and the loss tangent is $\tan\delta < 0.0001$~\cite{stc_alumina}. The dielectric plates are \SI{3}{mm} thick because that is approximately half a wavelength inside a $\epsilon_r = 9.8$ dielectric at \SI{16.5}{GHz}.

The Orpheus cavity modes resemble the Gaussian TEM modes of the empty Fabry-Perot cavity~\cite{1447049, Clarke_1982, Dunseith_2015}. A theoretical summary, simulations, and measurements of the empty Fabry-Perot cavity pertinent to Orpheus are described in~\cite{cervantes2021search}. However, it is infeasible to describe the fields of the dielectrically-loaded cavity analytically, and simulations are required to understand the electric field. 

\begin{figure}[htpb]
  \centering
  \includegraphics[height=50 mm]{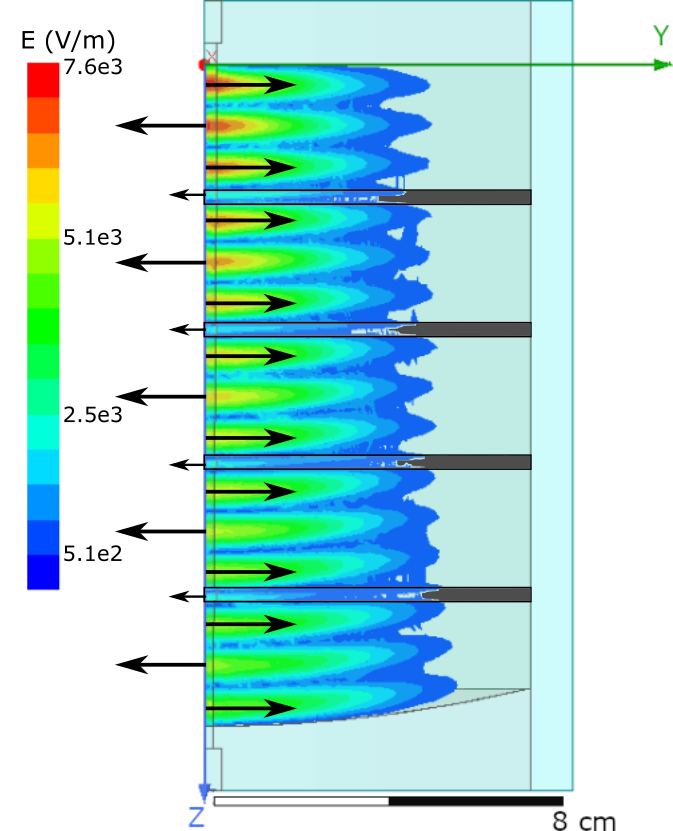}
  \caption{The simulated electric field magnitude of the \tem mode at $f_0 = \SI{15.97}{GHz}$, $L=\SI{15.4919}{cm}$. The electric field magnitude scales linearly from blue to red in accordance with the color bar.} 
  \label{fig:orpheus_simulations1}
\end{figure}

The \tem mode is designed to be the mode of interest. The mode has 19 antinodes (or half-wavelengths) across the cavity, and a dielectric plate is placed every fourth antinode, as shown in Fig.~\ref{fig:orpheus_simulations1}\footnote{Orpheus would have a larger $\veff$ if dielectric plates were placed every other antinode, but the mechanical design would be more challenging.} The amplitude of the field along the transverse plane follows a Gaussian function.

Power is both injected into and extracted from the cavity via aperture coupling connected to WR-62 rectangular waveguides. There are two apertures that make up the strongly-coupled port and the weakly-coupled port. The strongly coupled port consists of an aperture \SI{5.4}{mm} in diameter and \SI{3.8}{mm} thick located on the flat mirror. This was empirically determined to have an acceptable $\beta$ without too much detriment to the mechanical stability or $Q_0$. The weakly-coupled port consists of an aperture \SI{4.0}{mm} in diameter and about \SI{3.8}{mm} thick.

The cavity can be modeled as a two-port network. Networks can be fully described by the scattering matrix~\cite{pozar}, each element defined as $S_{ij} = \eval{V_i^{-}/V_j^{+}}_{V_k^+=0\text{ for } k\neq j}$
where $V_j^+$ is the amplitude of the voltage wave incident on port $j$ and $V_i^-$ is the amplitude of the voltage wave reflected from port $i$. $S_{11}$ is the reflection coefficient $\Gamma$ of port 1, and $S_{21}$ is the transmission coefficient $T$ from port 1 to port 2. The transmitted and reflected power near a cavity resonance is Lorentzian

\begin{align}
  &\abs{T}^2 = \frac{\delta_y}{1+4\Delta^2} + C_1\label{eqn:lorentzian_transmission}\\
  &\abs{\Gamma}^2 = -\frac{\delta_y}{1+4\Delta^2} + C_2\label{eqn:lorentzian_reflection}
\end{align}
where $\delta_y$ is the depth of the Lorentzian, and $C_1$ and $C_2$ are constant offsets.

To extract $f_0$, $Q_L$, and $\beta$, ``narrow scans'' are taken, where the vector network analyzer (VNA) measures $S_{21}$ and $S_{11}$ within a few Q-widths of the \tem mode. These narrow scans are fitted to the Lorentzian functions in Equations~\ref{eqn:lorentzian_transmission} and~\ref{eqn:lorentzian_reflection}. From the Lorentzian fits, one can extract $f_0$ and $Q_L$. 

The cavity coupling coefficient $\beta$ can extracted from the value of the reflection coefficient on resonance and the phase change on resonance. First, whether the cavity is overcoupled (${\beta <1}$) or undercoupled (${\beta >1}$) can be determined from $\angle \Gamma$; an undercoupled (overcoupled) cavity will have an increase (decrease) in phase on resonance. $\beta$ can then determined from the depth of the Lorentzian on resonance,

\begin{align}
 \beta = 
  \begin{cases} 
   \frac{1-{\abs{\Gamma_c(f_0)}}}{1+\abs{\Gamma_c(f_0)}} & \text{if undercoupled} \\
   \frac{1+\abs{\Gamma_c(f_0)}}{1-\abs{\Gamma_c(f_0)}} & \text{if overcoupled} 
  \end{cases}
\end{align}
where $\Gamma_c$ is the reflection coefficient of the strongly-coupled port. Note that the measured $S_{11}$ is affected by the transfer function of the network connected to the cavity, and this effect must be removed to obtain $\Gamma_c$. $\Gamma_c(f_0)$ can be obtained from the fitting parameters in Equation~\ref{eqn:lorentzian_reflection}, $\Gamma_c(f_0) = \sqrt{1-\delta_y/C_2}$.

Once $Q_L$ and $\beta$ are extracted, the intrinsic quality factor $Q_0$ can be calculated with $Q_0 = Q_L(1+\beta)$.

\section{Orpheus Cryogenic and Mechanical Design}\label{sec:mechanics}

This section describes the mechanical structure that moves the mirrors and dielectric plates. This structure is designed to work in a vacuum and cryogenic environment. The machining tolerances need to be tight enough to maintain good alignment and achieve good $Q$, but the machining tolerances should be large enough to allow the tuning mechanism to work even with small amounts of misalignment.

\subsection{Orpheus Cavity Mechanical Design}\label{sec:cavity_mechanics}
\begin{figure}[htp]
  \centering
  \includegraphics[width=0.95\linewidth]{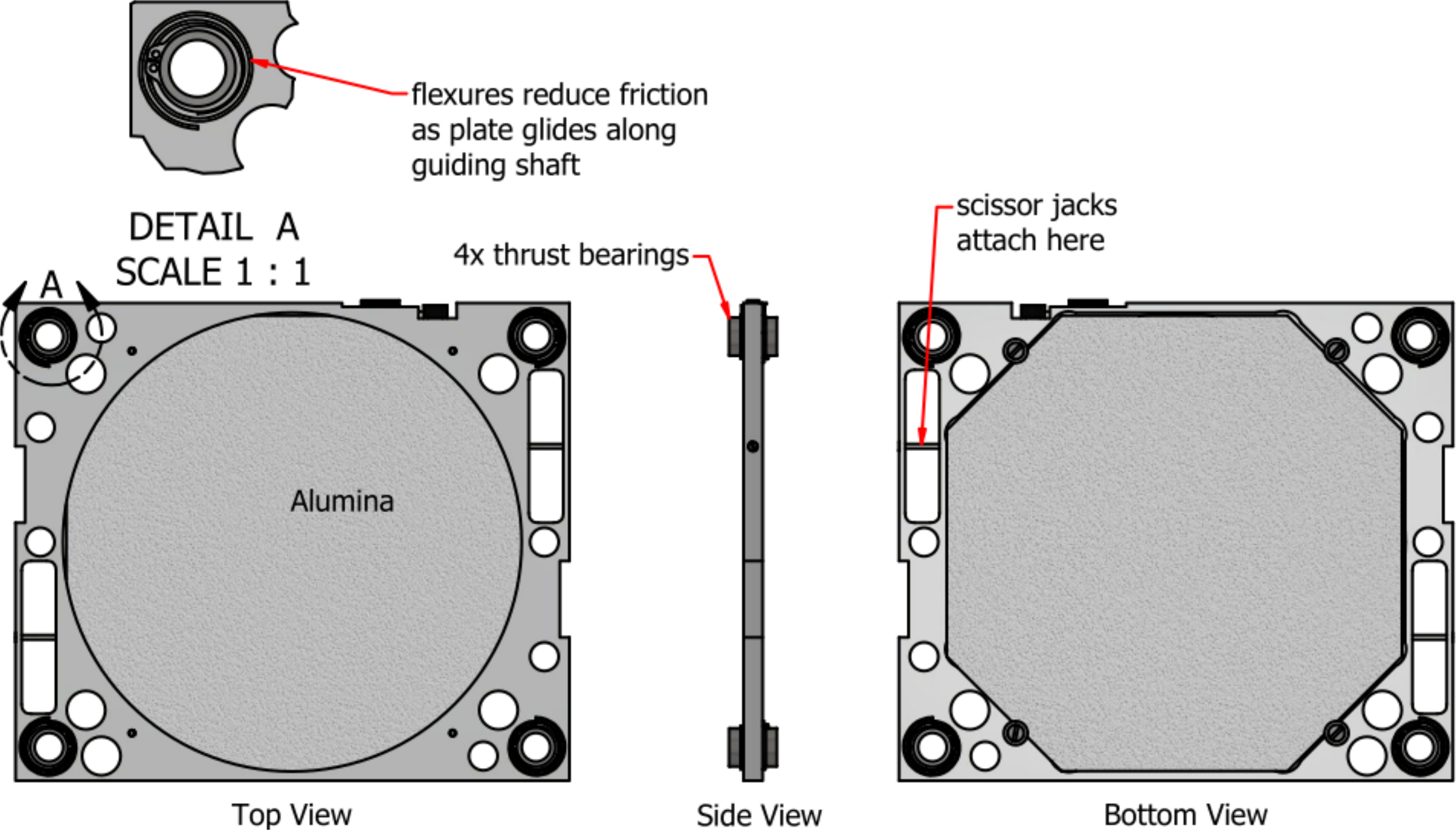}
  \caption{CAD model for one of the dielectric plates.}
  \label{fig:orpheus_plate_cad}
\end{figure}

The mirror and alumina plates are held in place by aluminum holders (Fig.~\ref{fig:orpheus_plate_cad}). They rest on a lip and rely on gravity for mechanical stability. Each dielectric plate sits inside an aluminum holder with lots of slack to accommodate the differential thermal expansion coefficients during cooldown. If the tolerances were tight at room temperature, the aluminum would contract faster and crush the alumina. Nothing is clamping down the alumina because the clamping mechanism would contract more quickly than the alumina and crush it. The mirrors are made out of aluminum, so there is no differential expansion between the mirror holders. Tighter tolerances are viable and desired to keep the coupling apertures in the cavity axis.

\begin{figure}[htp]
  \centering
  \includegraphics[width=0.98\linewidth]{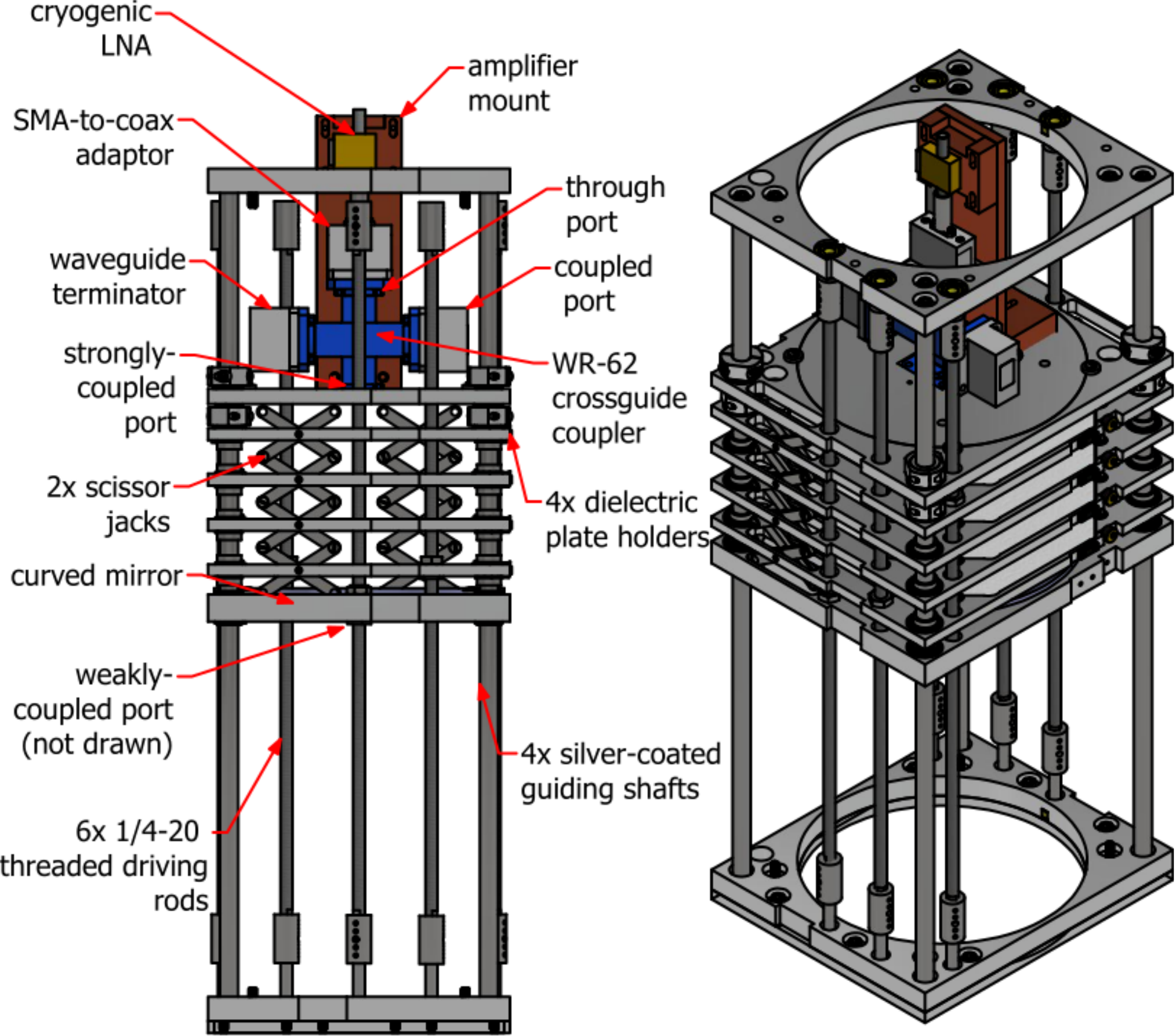}
  \caption{CAD model for the Orpheus cavity.}
  \label{fig:orpheus_cavity_cad}
\end{figure}

The aluminum holders containing the curved mirror, top dielectric plate, and bottom dielectric plate are moved vertically by a pair of 1/4''-20 stainless steel threaded rods. This results in a tuning mechanism with three degrees of freedom. The middle two dielectric plates are constrained to be spaced evenly between the top and bottom dielectric plate because they are attached to a pair of scissor jacks. The scissor jacks are connected to all dielectric plates but not the mirrors. All the plates slide along four guide rails. To compensate for any possible misalignment that would cause the plates to jam, each aluminum holder has spiral cut flexures around the bearing surface.

Cavity materials were chosen to accommodate thermal contractions. Alignment is maintained after cooldown because all vertical structures are made from stainless steel, and all horizontal structures are made from aluminum. Bearings are made out of stainless steel so that the fitting tolerance between the shaft and bearing remains the same after cooldown. However, having stainless steel thrust bearings rub against the stainless steel guide rails would lead to galling. All bearing surfaces are coated with silver to prevent galling and reduce friction.

Without further measures, the friction between the bearing and the guide rail would cause the plates to tilt as the cavity tunes. Two measures are taken to mitigate this tilting. First, the thrust bearings (\SI{0.615}{in.}) are much longer than the alumina holder thickness (\SI{0.25}{in}). The bearing length gives the plates less leeway to tilt. Second, weak compression springs are placed around the guide rails between the dielectric plates and mirror plates. For this experiment,  LP 024L 03 S316 from Lee Spring was used. It has a spring constant of \SI{0.37}{lb/in}. If the springs are too stiff, they deform the spiral flexures and cause significant position errors.

The cavity is shown in Fig.~\ref{fig:orpheus_cavity_cad}. The cavity tunes accurately and maintains good alignment while tuning. Measurements with a circular bubble level demonstrated that plates did not tilt more than 0.5\degree. Measurements demonstrated that the position error was often around \SI{0.2}{mm} and rarely exceeded \SI{0.4}{mm}. These possible misalignments increase the uncertainty in $\veff$ (See Section~\ref{sec:position_error} for more details). 

\subsection{Insert}
\begin{figure}[htp]
  \centering
  \includegraphics[width=0.50\linewidth]{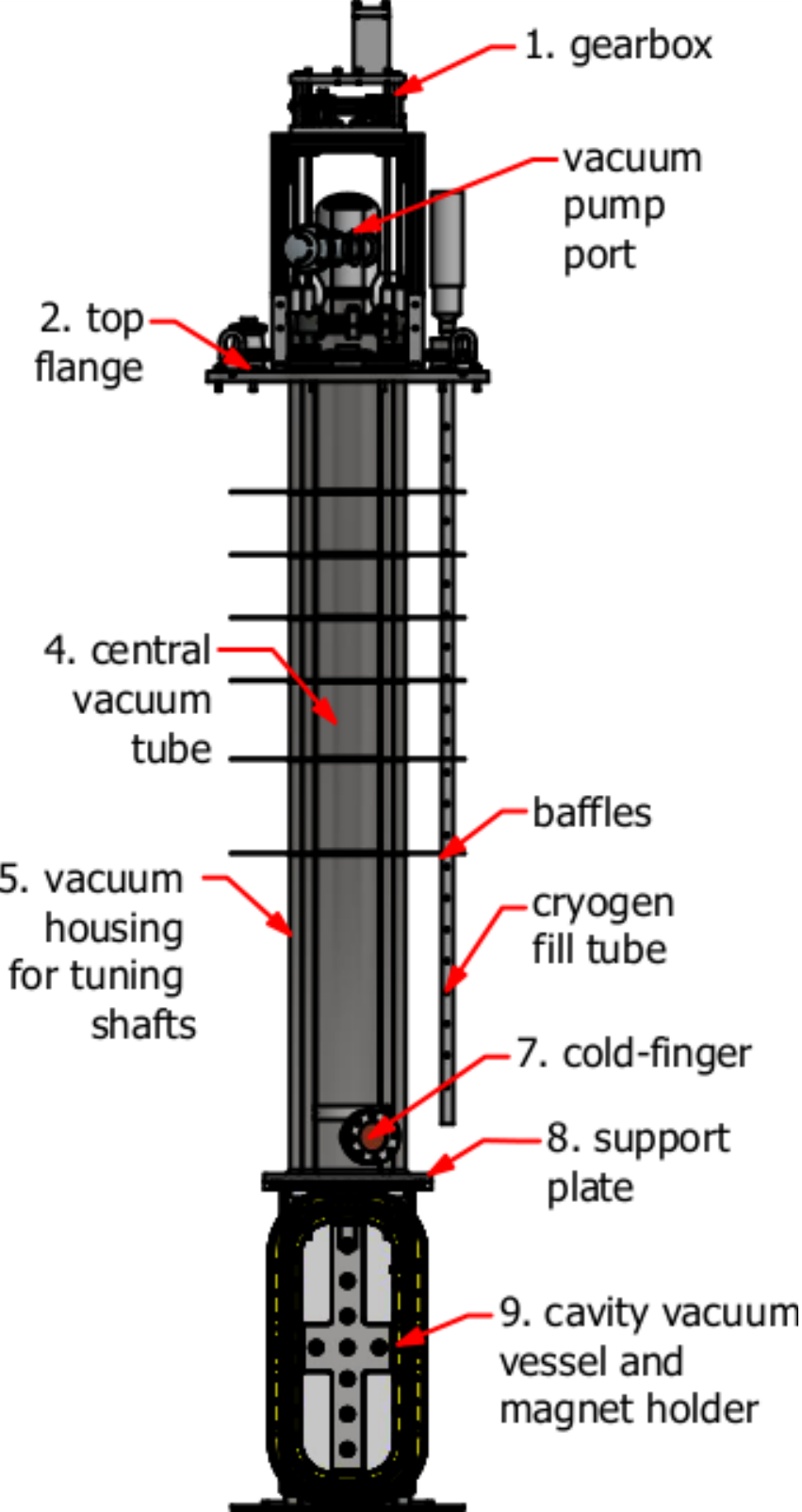}
  \caption{Orpheus cryogenic insert.}
  \label{fig:insert_cad}
\end{figure}

The cavity is the core of the experiment, but the cavity needs to be cooled down to \SI{4}{K} under vacuum. The cavity is kept in a vacuum instead of directly under cryogenic fluids because the cavity is not designed to remain mechanically stable under a boiling fluid. The cavity is to stay cold for about a week to allow enough time to scan the tuning range. Room-temperature stepper motors tune the cavity, and the stepper motors are located away from the superconducting dipole magnet that will be commissioned in the next run. The insert in Fig.~\ref{fig:insert_cad} was designed to meet these requirements. The main components are described below.

\begin{enumerate}
  \item Motor stage shown in Fig.~\ref{fig:top_flange}. Each stepper motor (Applied Motion Products STM23S-2EE~\cite{stm}) drives a pair of miter gears that transmit power to the vertical shafts connected to the vacuum rotary feedthroughs on the top flange.

  \item Top flange that sits on top of the Dewar (Fig.~\ref{fig:top_flange}). Has many of the vacuum and cryogenic ports listed below.
    \begin{itemize}
      \item Six MDC Precision 670000 rotary motion feedthroughs.
      \item AMI vapor-cooled magnet leads with superconducting busbars. They have a \SI{8}{L/day} boil off rate.
      \item Kurt-Lesker IFDRG197018B 19 pin electrical feedthrough. 
      \item Isocross for vacuum port, 51 pin micro type-D port (MDC 9163004), and RF coaxial port.
      \item cryogen fill port
      \item cryogen vent port
    \end{itemize}
    \begin{figure}[htp]
      \centering
      \includegraphics[width=0.85\linewidth]{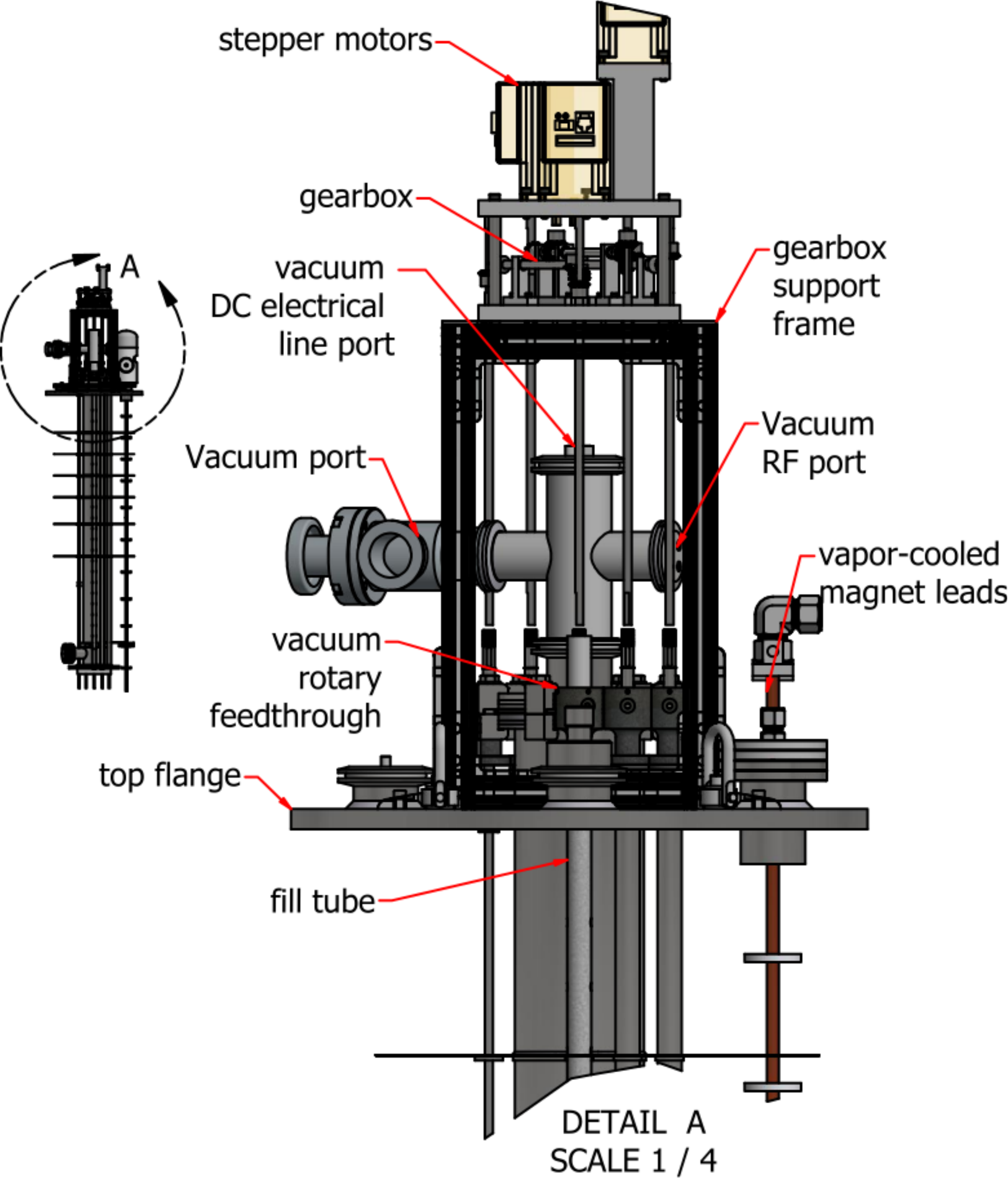}
      \caption{Insert top flange with all the vacuum ports, cryogen ports, and motor stage.}
      \label{fig:top_flange}
    \end{figure}

  \item \SI{300}{L} Dewar with a \SI{18}{L/day} boil off rate.
  \item 50'' long, 4'' OD stainless steel tube. 
  \item Six 50'' long, 0.75'' ID stainless steel tubes that house motor shafts in vacuum space.
  \item Six motor shafts. Each motor shaft consists of a stainless steel tube coupled to a G10 fiberglass tube. The fiberglass tube reduces thermal contact with the top flange. The stainless steel tube is not as thermally isolating as the fiberglass but has less elasticity and would lead to less mechanical backlash.
  \item Copper plug to act as cold finger. Used to thermally sink cryogenic fluid to the cavity top.
  \item Rectangular experimental support plate welded to the bottom of the central vacuum tube. The cavity attaches to the experimental support plate.
  \item Cavity vacuum vessel. In the future, the dipole magnet will attach to this vessel. 
\end{enumerate}

Unfortunately, several preventable mechanical alignment issues caused motor stalls and thermal gradient issues (see Fig.~\ref{fig:temp_v_time}) that increased the noise power and uncertainty of the noise power (see Section~\ref{sec:parameter_extraction}) and limited the total data-taking time to two days. These issues are talked about in detail in~\cite{cervantes2021search} and will be addressed before the next run.

\section{Electronics, Data Acquisiton System, and Operations}\label{sec:operations}
This section focuses on how measurements in the Orpheus experiment are taken. The section explains the radio frequency (RF) and intermediate frequency (IF) electronics, the steps taken for each data-taking cycle, and the software that controls and monitors the experiment. 

\subsection{Cryogenic Electronics}
\begin{figure}[htp]
  \centering
  \includegraphics[width=0.95\linewidth]{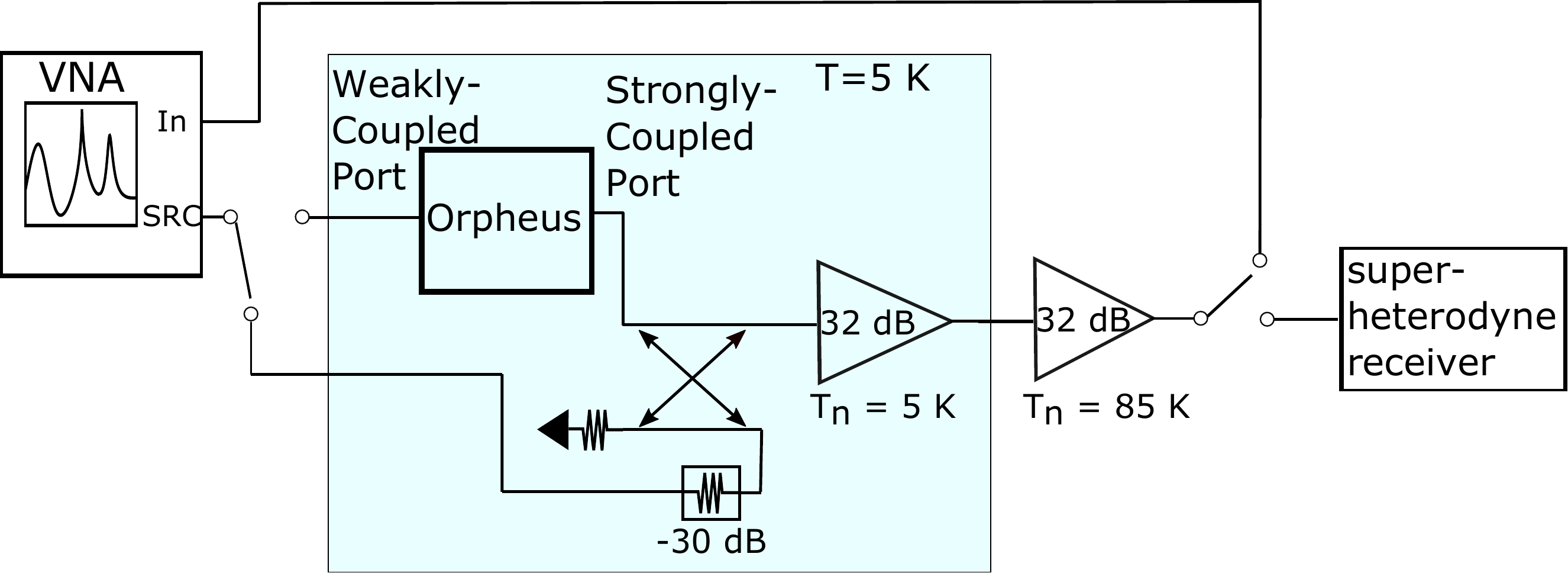}
  \caption{A diagram of how transmission, reflection, and power measurements are taken. Room temperature Teledyne switches are used to switch between transmission and reflection measurements, and from VNA measurements to power measurements.}
  \label{fig:cold_electronics}
\end{figure}

The diagram for the cold electronics is shown in Fig.~\ref{fig:cold_electronics}. The cavity has a weakly-coupled port and a strongly-coupled port. The strongly-coupled port is connected to a WR-62 \SI{20}{dB} crossguide coupler (PEWCP1047). The crossguide coupler is attached to a waveguide-to-coax adapter (PE9803). The coax adaptor connects directly to the cryogenic low noise amplifier (LNF-LNC6\_20C). The cryogenic amplifier output is connected to an RG405 coaxial cable that connects directly to the room-temperature SMA bulkhead\footnote{This direct connection causes a large heat leak but was easy to implement. Future runs will take steps to mitigate this heat leak.}. 

During a science run, the strongly-coupled port transmission and reflection coefficients are measured every time the cavity is tuned to extract $f_0$, $Q$, and $\beta$. Transmission and reflection measurements are performed using a Keysight E5063A VNA. For a transmission measurement, VNA port 1 is connected to the weakly-coupled port. For a reflection measurement, the VNA port 1 is connected to the crossguide coupler coupled port, as shown in Fig.~\ref{fig:cold_electronics}. The signal then travels to the strongly-coupled port, gets reflected, and reaches the input of the cryogenic amplifier. The crossguide coupler is needed because the VNA source signal needs to bypass the cryogenic amplifier to reach the strongly-coupled port. The remaining port in the crossguide coupler is terminated with a waveguide terminator (PE6804). A Teledyne SMA switch is used to switch between reflection and transmission measurements.

The VNA output power is set to \SI{-5}{dBm}. During a reflection measurement, this power is reflected off the cavity and into the first stage amplifier. To prevent saturation of this amplifier, a 30 dB cryogenic attenuator is connected to the coupled port of the crossguide coupler (Fig.~\ref{fig:cold_electronics}). The amplifier was confirmed to not be saturated by varying the VNA output power and observing that the transmission and reflection measurements were not changed.

The cryogenic attenuator also attenuates the room-temperature thermal noise that is injected by the VNA into the directional coupler.

The cavity temperature $T_{cav}$ must be measured to calculate the system noise temperature $T_n$. The temperatures of the flat mirror and curved mirror are monitored with calibrated Lakeshore cernox resistors, model: CX-1010-AA-0.1L. The temperature sensors are bolted on the flat mirror holder and curved mirror holder with a layer of Apiezon N grease between the mating surfaces. The resistance is measured by the Agilent 34970A multiplexer using a  4-wire measurement. The uncertainty of the measurements is thought to be \SI{5}{mK} from Lakeshore specifications~\cite{lakeshore_2022}. It is possible that the temperature sensors have poor thermal contact with the mirrors. In this case, the measured temperature is an overestimate, and the extracted systen noise temperature is a conservative overestimate.

\subsection{Room-temperature Electronics}
\begin{figure}[htp]
  \centering
  \includegraphics[width=0.95\linewidth]{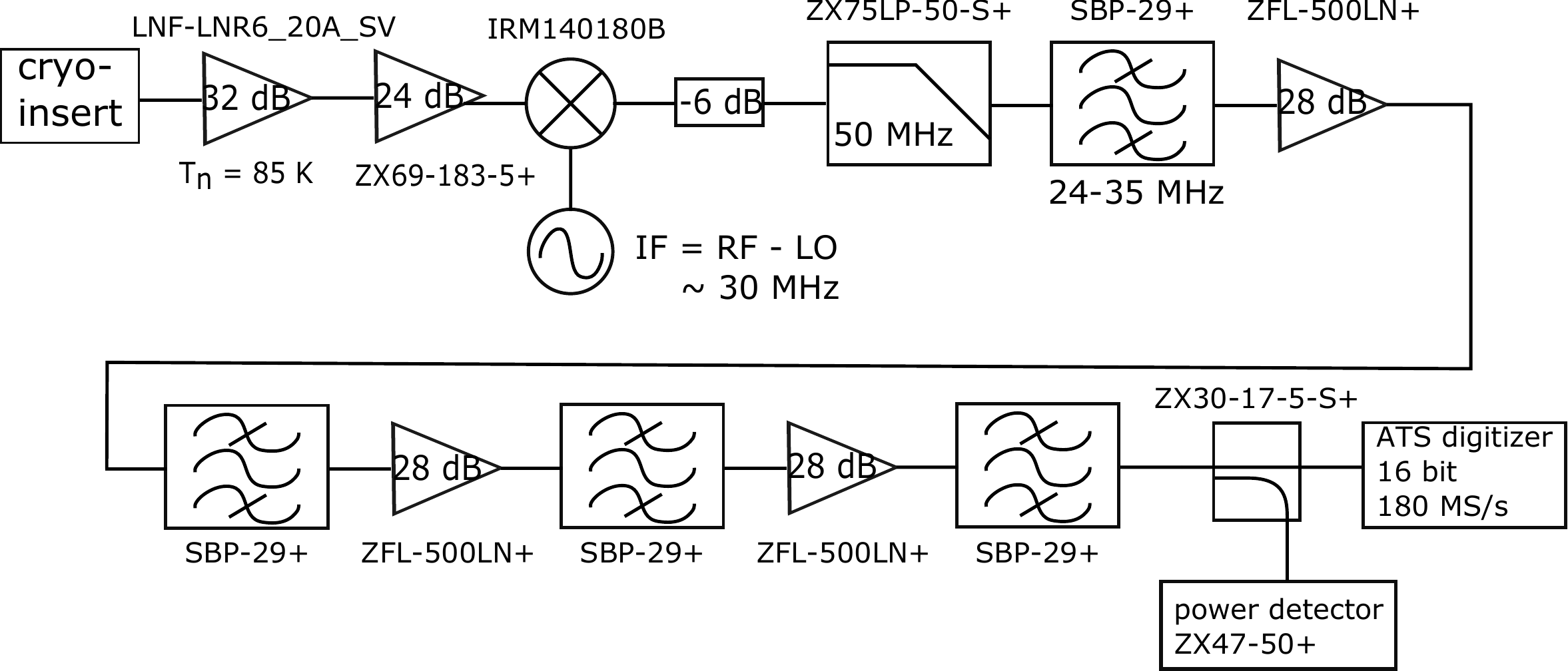}
  \caption{Diagram of the superheterodyne receiver used to measure power spectra.}
  \label{fig:warm_box_schematic}
\end{figure}

The schematic for the room-temperature electronics is shown in Fig.~\ref{fig:warm_box_schematic} and the list of components with the associated cascade analysis is shown in Table~\ref{tab:cascade}.

After the signal leaves the cryogenic insert, it is directed to a room-temperature low noise amplifier (LNF-LNR6\_20A SV~\cite{lnf}) to ensure sufficient gain so that the rest of the electronics do not degrade the SNR. The signal is amplified again by the ZX60-183-S amplifier. The \SI{16}{GHz} signal needs to be frequency mixed so that it can be digitized by a \SI{180}{MSPS} digitizer card. This frequency downconversion is achieved by sending the RF signal to an Image-Reject Mixer (IRM140180B). The mixer uses the nonlinear semiconductor properties to multiply the RF signal with angular frequency $\omega_{RF}$ by a local oscillator signal with angular frequency $\omega_{LO}$. The result is a signal with spectral components with both the sum and difference of the two frequencies (and higher-order harmonics), i.e. $\sin(\omega_{RF}t)\sin(\omega_{LO}t) \propto \sin((\omega_{RF}-\omega_{LO})t) + \sin((\omega_{RF}+\omega_{LO})t)$. The signal component with ${\omega = \omega_{RF}+\omega_{LO}}$ is removed using a low-pass filter, and the signal component with ${\omega = \omega_{RF}-\omega_{LO}}$ is digitized. The resulting frequency $(\omega_{RF}-\omega_{LO})/(2\pi)$ is known as the Intermediate Frequency (IF). The IF is chosen to be about \SI{30}{MHz} to be compatible with the mixer. 

After the RF signal is multiplied by the local oscillator signal, the resulting signal passes through a low-pass filter, and only the IF frequency survives. There is an attenuator between the mixer and low-pass filter to eliminate standing waves which may cause the filter to detune. After this, the IF signal passes through a series of IF amplifiers and bandpass filters. The IF amplifiers amplify the IF signal to the input range of the digitizer card, and the bandpass filters attenuate noise outside the frequency of interest. The amplified IF signal then goes to a directional coupler. The coupled port is connected to a power detector (ZX47-50+), allowing for the real-time measurement of the IF signal power. The directional coupler's through port is connected to the Alazartech digitizer card (ATS9462), a 16-bit digitizer with a input range of $\pm\SI{200}{mV}$ to $\pm\SI{16}{V}$

The Alazartech digitizer is set to a sampling rate of \SI{125}{MSPS}. This rate is chosen so that the Nyquist frequency is well above the IF band. Each subspectrum contained \num{50000} time samples, resulting in a \SI{2.5}{kHz} bin width. Depending on the run settings, each spectrum is either the average of \num{75000} (30 seconds) subspectra or \num{250000} (100 seconds) subspectra. No window function is applied to each subspectrum (equivalently, a rectangular window is applied).  

The cascade analysis in Table~\ref{tab:cascade} shows that the gain from the RF and IF electronics is about \SI{140}{dB}. For $T_n \approx \SI{10}{K}$, the noise power without the system gain integrated over the $\SI{10}{MHz}$ IF bandwidth is $P_n \approx \SI{-119}{dBm}$. Thus the noise power measured by the digitizer is $P_n = \SI{21}{dB}$. The digitizer has a $\SI{50}{\ohm}$ impedance, so the measured voltage is $\SI{7.4}{V_{p-p}}$, which is well within the maximum input range of the digitizer. 

The digitizer was also confirmed to be configured properly for the measurement by injecting a large SNR signal into the weakly-coupled port of the cavity and observing that the digitized signal was not clipped or compressed. The digitized power was also linear with small perturbations in the cavity temperature, further demonstrating that the digitizer is appropriately configured to detect small power excesses deposited from the dark matter halo\footnote{The digitized power is not linear with large perturbations in cavity temperature because the amplifier gain changes with temperature.}.

\begin{table*}[htp]
\centering
\resizebox{\textwidth}{!}{%
  \begin{tabular}{lllllll}
\textbf{Part Description}                                                          & \textbf{Vendor}     & \textbf{Part Number} & \textbf{Relative Gain (dB)} & \textbf{Absolute Gain (dB)} & \textbf{Device Noise Temp. (K)} & \textbf{Cascaded Noise Temp. (K)} \\
LNF-LNC6\_20C s/n 1556Z                                                            & Low Noise Factory   &                      & 33                           & 33                     & 4.7                                 & 9.4                  \\
SMA Male to SMA Male Cable Using RG405 Coax, RoHS                                  & Pasternack          & PE3818LF-72          & -7.6                         & 25.4                   & 0                                 & 9.4                    \\
C3146 SMA Hermetic Bulkhead Adapter 18Ghz                    			   & Centric RF          & C3146                & -0.2                         & 25.2                   & 0                                 & 9.4                   \\
Low Loss Test Cable 12 Inch Length, PE-P142LL Coax 				   & Pasternack          & PE341-12             & -0.8                        & 24.4                  & 0                                 & 9.4                   \\
Hand-Flex Interconnect, 0.086" center diameter, 18.0 GHz                           & Minicircuits        & 086-SBSMR+           & -0.3                         & 24.1                  & 0                                 & 9.4                    \\
LNF-LNR6\_20A\_SV s/n 1257Z                                                        & Low Noise Factory   &                      & 32                           & 56.1                  & 100                               & 9.7                    \\
Right Angle Semi-Flexible Cable                                     		   & Pasternack          & PE39417-6            & -2                           & 54.1                  &                                   & 9.7                       \\
Wideband Microwave Amplifier 6 to 18 GHz                                           & Minicircuits        & ZX60-183-S+          & 24                           & 78.1                  & 1400                            & 9.7                    \\
Semi-Flexible Cable                               				   & Pasternack          & PE39417-9            & -3                           & 75.1                  &                                   & 9.7	\\
Hand-Flex Interconnect, 0.086" center diameter, 18.0 GHz                           & Minicircuits        & 086-4SM+             & -0.4                        & 74.7                   & 0                                 & 9.7                    \\
Hand-Flex Interconnect, 0.086" center diameter, 18.0 GHz                           & Minicircuits        & 086-4SM+             & -0.4                        & 74.3                  & 0                                 & 9.7                    \\
DC–18 GHz/DC-22 GHz SPDT Coaxial Switch                          		   & Teledyne            & CCR-33S/CR-33S       & -0.4                         & 73.9                  & 0                                 & 9.7  \\
Hand-Flex Interconnect, 0.086" center diameter, 18.0 GHz                           & Minicircuits        & 086-2SM+             & -0.3                        & 73.6                   & 0                                 & 9.7                    \\
IMAGE-REJECT MIXER 14.0 – 18.0 GHz                                                 & Polyphase Microwave & IRM140180B           & -8.5                         & 65.1                   & 3400                            & 9.7                    \\
3 dB Fixed Attenuator                                                              & Pasternack          & PE7005-3             & -3                           & 62.1                   & 0                                 & 9.7                    \\
Low Pass Filter                                                                    & Minicircuits        & ZX75LP-50-S+         & -1.4                        & 60.7                  & 0                                 & 9.7                    \\
Hand-Flex Interconnect, 0.086" center diameter, 18.0 GHz                           & Minicircuits        & 086-2SM+             & -0.3                        & 60.4                   & 0                                 & 9.7                    \\
Lumped LC Band Pass Filter, 24 - 35 MHz, 50$\Omega$                                & Minicircuits        & SBP-29+              & -0.9                        & 59.5                  & 0                                 & 9.7                    \\
Low Noise Amplifier, 0.1 - 500 MHz, 50$\Omega$                                     & Minicircuits        & ZFL-500LN+           & 28                           & 87.7                  & 275                               & 9.7                    \\
Lumped LC Band Pass Filter, 24 - 35 MHz, 50$\Omega$                                & Minicircuits        & SBP-29+              & -0.9                        & 86.9                  & 0                                 & 9.7                    \\
Low Noise Amplifier, 0.1 - 500 MHz, 50$\Omega$                                     & Minicircuits        & ZFL-500LN+           & 28                           & 115.9                 & 275                               & 9.7                    \\
Lumped LC Band Pass Filter, 24 - 35 MHz, 50$\Omega$                                & Minicircuits        & SBP-29+              & -0.9                        & 114.2                 & 0                                 & 9.7                    \\
Low Noise Amplifier, 0.1 - 500 MHz, 50$\Omega$                                     & Minicircuits        & ZFL-500LN+           & 28                           & 142.4                 & 275                               & 9.7                    \\
Lumped LC Band Pass Filter, 24 - 35 MHz, 50$\Omega$                                & Minicircuits        & SBP-29+              & -0.9                        & 141.5                 & 0                                 & 9.7                    \\
17.5 dB Directional Coupler, 5 - 2000 MHz, 50$\Omega$                              & Minicircuits        & ZX30-17-5-S+         & -0.7                         & 140.8                 & 0                                 & 9.7                   
\end{tabular}%
}
\caption{List of electronics and the cascaded system gain and noise temperature estimated from typical values from vendor data sheets and assuming a cavity temperature of \SI{4.7}{K}.}
\label{tab:cascade}
\end{table*}

\subsection{System Noise Temperature}\label{sec:thermal_model}

For modeling the system noise temperature, the cryogenic electronics in Fig.~\ref{fig:cold_electronics} can be approximated as a cavity connected to the first-stage amplifier by a transmission line. The system noise temperature is then 

\begin{align}
  T_n = T_{cav}(1-|\Gamma_c|^2) + T_{amp, input}|\Gamma_c|^2 + T_{rec}
  \label{eqn:orpheus_tsys}
\end{align}
where $T_{cav}$ is the physical temperature of the cavity, $T_{amp, input}$ is the noise temperature coming from the input of the cryogenic amplifier, $T_{rec}$ is the noise temperature of the receiver chain from the output of the cryogenic amplifier outward, $\Gamma_c$ is the reflection coefficient of the strongly-coupled port, and $|\Gamma|^2$ is the fraction of power reflected. From Equation~\ref{eqn:lorentzian_reflection}, $|\Gamma|^2$ depends on both the cavity coupling coefficient $\beta$ and the detuning factor $\Delta$. Boson statistics do not need to be taken into account in the Raleigh Jeans limit ($k_b T_n >> hf$).

To understand Equation~\ref{eqn:orpheus_tsys}, consider several limiting cases. When the receiver is critically coupled to the cavity, the cavity on resonance looks like a blackbody. Thermal photons are in thermal equilibrium with the cavity, and $T_n = T_{cav} + T_{rec}$. If the receiver is poorly coupled or if the RF frequency is far off resonance, the cavity looks more like a mirror. In this scenario, thermal photons are emitted from the input of the amplifier. These thermal photons reach the flat mirror and are reflected back into the amplifier. Thus, $T_n = T_{amp, input} + T_{rec}$. Generally, the cavity is only partially reflecting, and the noise temperature is described by Equation~\ref{eqn:orpheus_tsys}.

\begin{figure}[htp]
  \centering
  \includegraphics[width=\linewidth]{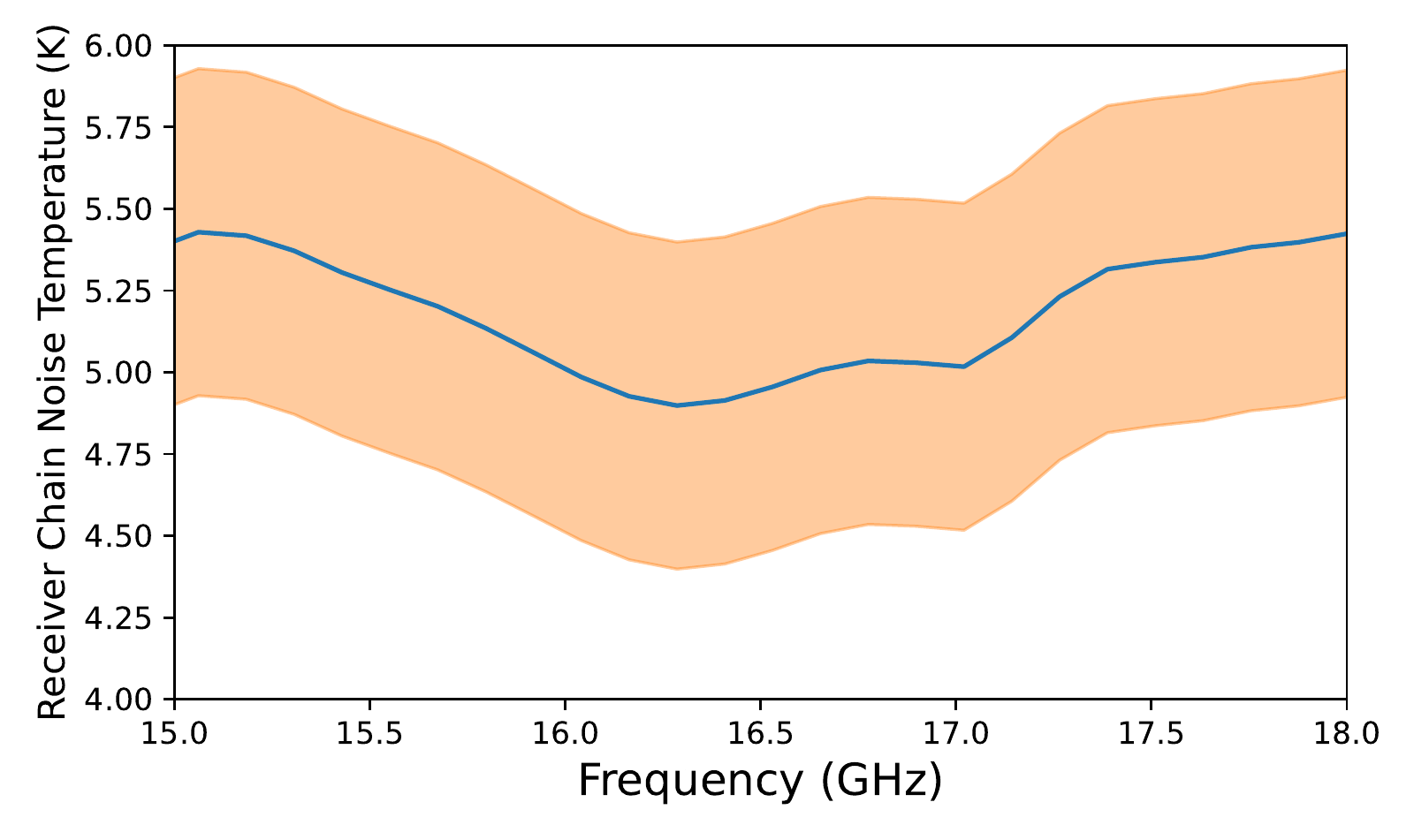}
  \caption{The receiver noise temperature $T_{rec}$ derived from the frequency-dependent amplifier noise temperature measured by Low Noise Factory and the Friis cascade equation (Equation~\ref{eqn:friis}). The orange band represents the uncertainty in $T_{rec}$. The receiver noise temperature can be applied to Equation~\ref{eqn:orpheus_tsys} to obtain the system noise temperature as referenced to the cavity.}
  \label{fig:receiver_noise_temperature}
\end{figure}

The noise temperature of the receiver is derived from calibrations performed by Low Noise Factory~\cite{lnf} and the Friis cascade equation~\cite{friis} 
\begin{align}
  T_{rec} = T_1 + \frac{T_2}{G_1} + \frac{T_3}{G_1G_2} + ...
  \label{eqn:friis}
\end{align}
where $T_1$ and $G_1$ are the noise temperature and gain of the first-stage cryogenic amplifier. $T_2$ and $G_2$ are the noise temperature and gain of the second-stage room-temperature Low Noise Amplifier (LNA). $T_3$ is the noise temperature of the third stage amplifier. The third term is negligible compared to the first two terms. The uncertainty of each amplifier noise temperatures is thought to be about \SI{0.5}{K}. This uncertainty comes from private correspondence with Low Noise Factory, and additional references include~\cite{1603866_tamp_uncertainty, 5540248_tamp_uncertainty}. The derived receiver noise temperature is shown in Fig.~\ref{fig:receiver_noise_temperature}.

\subsection{Software Stack}
The control software consists of modular, self-healing, loosely-coupled services with a standardized messaging protocol for all serial communication between devices. The control software stack consists of Python, Postgresql, RabbitMQ, Dripline, Docker, Kubernetes, Helm, and Grafana. Except for the digitizer driver, all software is open source. Dripline~\cite{dripline} is the standardized messaging protocol that communicates with different hardware such as the digitizer, VNA, and stepper motors; and software services such as the database and message broker. Each dripline service is run inside a docker container, and Kubernetes manages the lifecycle of these containers. Grafana provides real-time data visualization. Python is the scripting language of choice.

\subsection{Data-taking procedure}\label{sec:cadence}
The cavity is tuned continuously to scan for dark photons with different masses. For each cavity length, a series of ancillary measurements are taken to extract a noise power calibration and expected dark photon signal power. The power spectrum is then measured out of the cavity to search for a spectrally-narrow power excess that may correspond to a dark matter signal. The intended data-taking procedure for each tuning step is as follows:

\begin{enumerate}
  \item The state of the system is recorded. The temperature sensors and motor encoders are logged into the database. These measurements allow the extraction of the noise power, cavity length, and dielectric plate positions.
  \item The transmission and reflection coefficients are measured \SI{20}{MHz} around the \tem mode. Parameters $f_0$, $Q_L$ and $\beta$ are extracted from these measurements. The extracted parameters are logged into the database. This data is re-analyzed offline for more sophisticated uncertainty analysis.

  \item The VNA output is disabled, and the power spectrum is measured. The power spectrum is integrated for either \SI{30}{s} or \SI{100}{s}.
  \item The motors are tuned in a coordinated way so that the curved mirror and dielectric plates are moved by a specified amount.
\end{enumerate}
This procedure is repeated for each tuning step. Every 20 data-taking cycles, the transmission coefficient is measured with a frequency range of \SI{15}{GHz} to \SI{18}{GHz}, i.e., a wide scan measurement is taken.

The dielectrics were intended to maintain an even spacing between the mirrors while the cavity was tuning. However, issues during operations resulted in unevenly-spaced dielectric plates. The top dielectric plate motor stalled after cooldown, preventing automated tuning of the top dielectric plate. In response, this motor was turned off and the top dielectric plate was tuned by hand every few hours to correct position errors. Also, programming errors caused the bottom dielectric plate to always undershoot the intended position. This effect was discovered after the run. Fortunately, the motor encoder values allow us to extract the actual plate positions, and $\veff$ can be simulated using the measured dielectric plate positions.
 
\section{Dark Photon Search Analysis}\label{sec:analysis}
The data collected between 9/3/2021 and 9/7/2021 was used to search for dark photons between \SI{65.5}{\mu eV} (\SI{15.8}{GHz}) and \SI{69.5}{\mu eV} (\SI{16.8}{GHz}). All measured power was consistent with thermal noise, so a 90\% confidence level exclusion limit was placed on the kinetic mixing strength $\chi$ in this mass range. The procedure for deriving the exclusion limits follows the procedure developed by ADMX and HAYSTAC~\cite{PhysRevD.64.092003, PhysRevD.96.123008, PhysRevD.103.032002}, and is adapted for dark photon searches~\cite{PhysRevD.104.095029, PhysRevD.104.092016}. 

The dark photon search strategy is to look for a spectrally narrow power excess (Equation~\ref{eqn:dp_power}) over the noise floor. In broad strokes, the strategy is to first remove the low-frequency structure from each power spectrum, such that the population mean of each bin is zero and deviation from zero is either from statistical fluctuation due to the noise temperature of the detector or from a coherent RF signal. This results in a unitless power excess normalized to the system noise power. In searching for potential dark photon candidates, the SNR is the figure of merit. Thus the power excess is rescaled so that it is in units of single-bin dark photon power. In other words, a bin's population mean is one in the presence of a single-bin dark photon signal. However, the dark photon power is spread across many bins, reducing the SNR. The SNR of a potential signal is increased by applying a matched filter with dark photon kinetic energy distribution as the template. The different, partially overlapping spectra are then combined using a maximum likelihood weighting procedure to form a combined spectrum. In the absence of any dark photon signal candidates, the sample mean and sample standard deviation of each bin in the combined spectrum can be used to place a 90\% confidence exclusion limit on the scanned dark photon mass ranges.

\subsection{Parameter extraction}\label{sec:parameter_extraction}
For each tuning step, ancillary measurements were taken to extract the parameters needed to determine the noise temperature and the dark photon power. The necessary parameters are the cavity temperature, cavity length, dielectric positions, resonant frequency, loaded Q, cavity coupling coefficient, and effective volume.

\begin{figure}[htp]
  \centering
  \includegraphics[width=\linewidth]{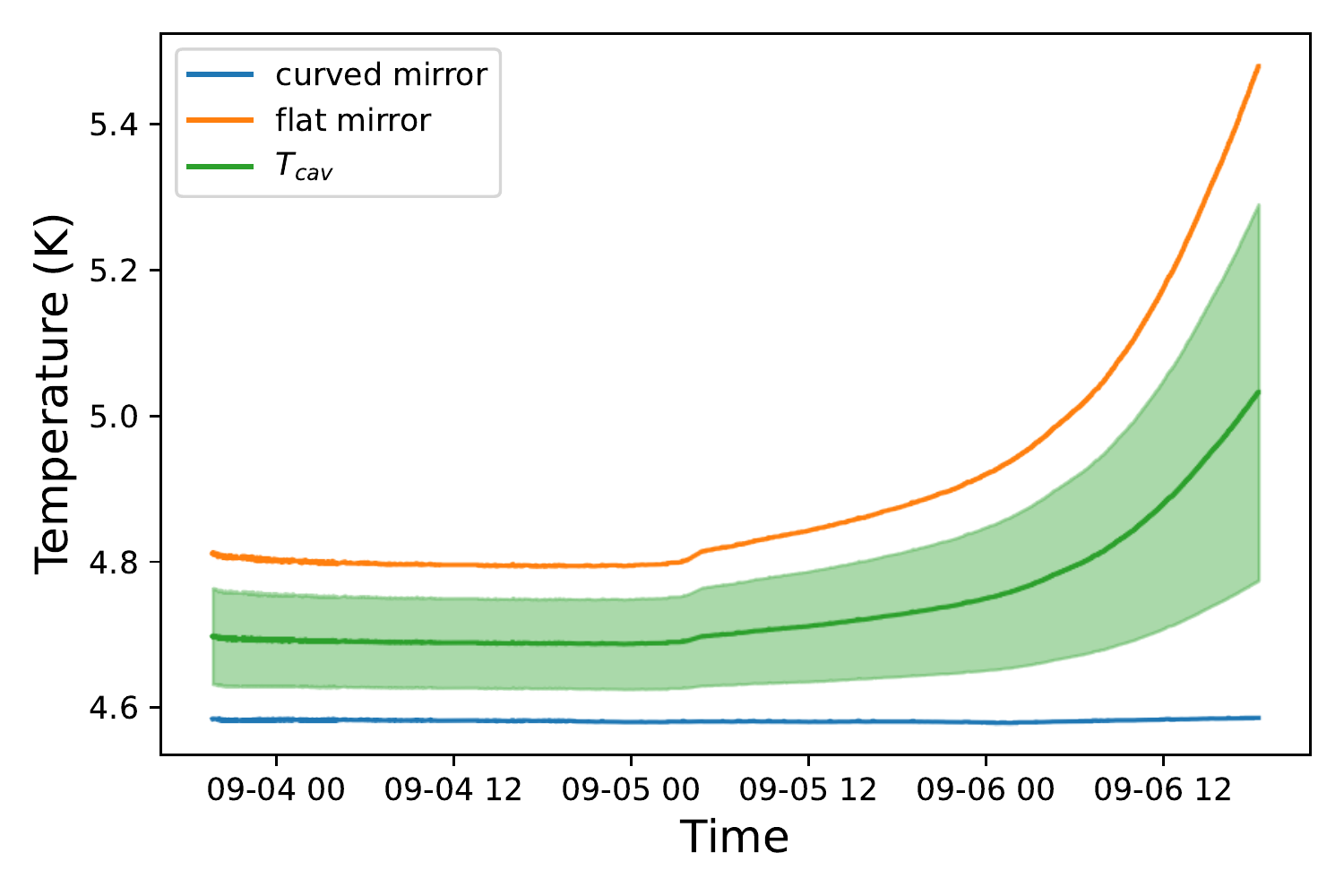}
  \caption{The cavity mirror temperatures throughout the dark photon search.}
  \label{fig:temp_v_time}
\end{figure}

The cavity temperature $T_{cav}$ is taken to be the mean of the flat mirror and curved mirror temperatures as measured by calibrated Cernox resistors. The measured values are shown in Fig.~\ref{fig:temp_v_time}, and the uncertainty is taken from a continuous uniform distribution, $\sigma_{T_{cav}} = \left(T_{flat} - T_{curved}\right)/\sqrt{12}$. The temperature $T_{cav}$ is thought to be an overestimate of the mean temperature of the thermal photons coming from the cavity and is, therefore, a conservative choice. This is because the curved mirror is more representative of the temperature of the cavity. The flat mirror was hotter than the rest of the cavity because it had poor thermal contact and was subject to greater heat leak from the two coaxial cables connected directly to a room-temperature port. 

\begin{figure}[htp]
  \centering
  \includegraphics[width=\linewidth]{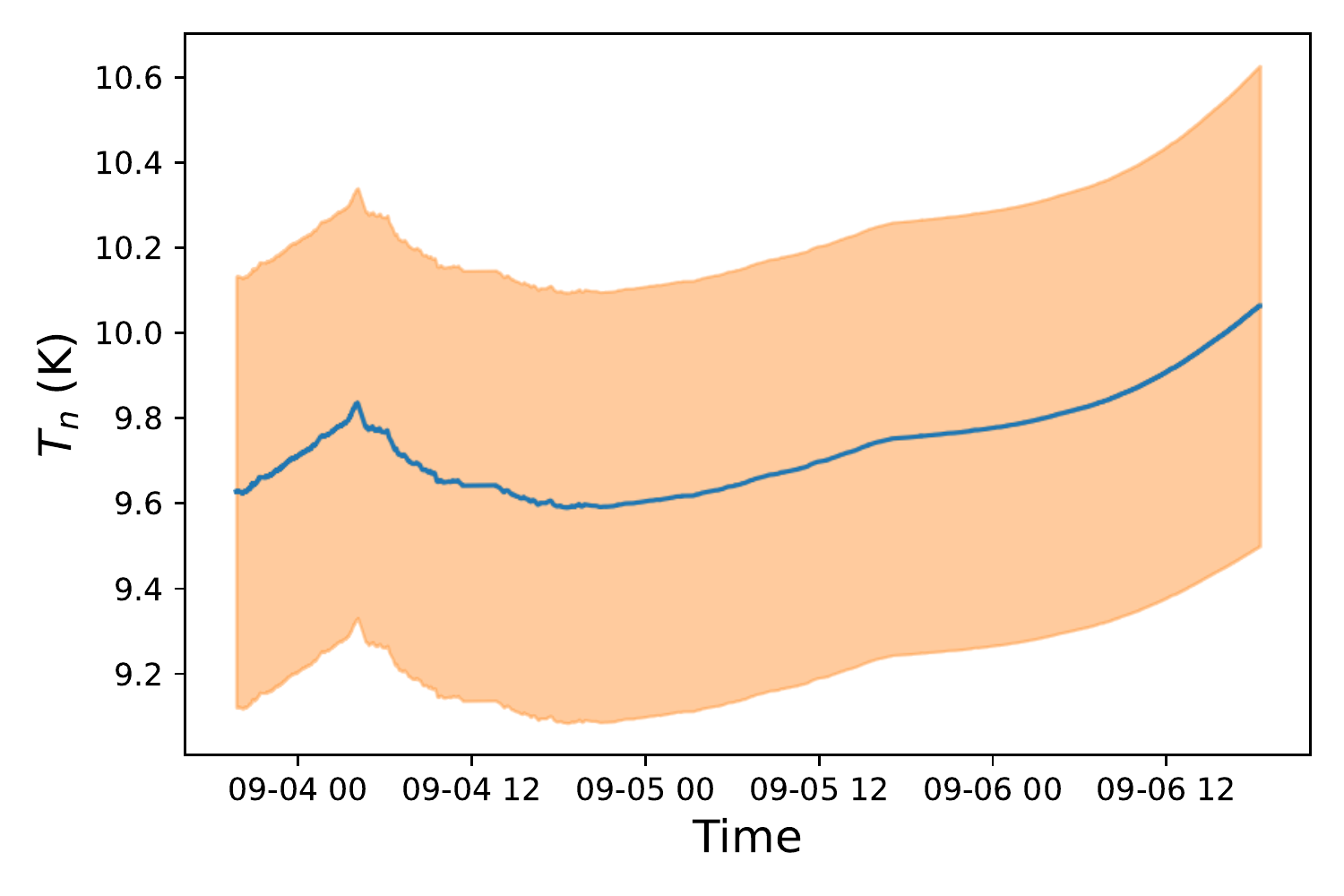}
  \caption{The system noise temperature $T_n$ the dark photon search. This is derived from Equation~\ref{eqn:orpheus_tsys}. The orange band is the uncertainty.}
  \label{fig:tn_v_time}
\end{figure}

Once the cavity temperature is extracted, the system noise temperature (Section~\ref{sec:thermal_model}) is calculated using Equation~\ref{eqn:orpheus_tsys}. $T_{rec}$ is shown in Fig.~\ref{fig:receiver_noise_temperature} and is derived from the manufacturer's datasheets. The uncertainty in the manufacturer's calibration is thought to be \SI{0.5}{K}. $T_n$, shown in Fig.~\ref{fig:tn_v_time}, is typically \SI{9.7}{K}. The uncertainty in the cavity temperature and amplifier noise temperature add in quadrature, and the relative uncertainty is about 5.2\%. This is determined by taking the mean of relative uncertainty $\sigma_{T_n}/T_n$ for each measured spectrum. 

The cavity length and dielectric positions are calculated using the motor encoder values. The motor encoders are first set to 0 when the cavity length is \SI{13.6}{cm} long. The steps counted by the motor encoders are then used to calculate relative changes in distance. One complete revolution corresponds to \num{20000} motor steps, and twenty revolutions correspond to a plate moving one inch along the 1/4''-20 threaded rod. However, backlash causes a systematic bias in the measured cavity length. This systematic bias can be corrected by comparing the measured $f_0$ to simulations, as will be seen in Fig.~\ref{fig:dpsearch_modefreq}.

The discussion of the extracted $f_0$, $\veff$, $Q_L$, and $\beta$ for the \tem mode is deferred to Section~\ref{sec:orpheus_simulations}. 

\subsection{Simulation and Characterization of the TEM$_{00-18}$ Mode}\label{sec:orpheus_simulations}

The crux of the Orpheus experiment is using the dielectric structure to increase $\veff$. Since $\vb{E}$ cannot be measured directly, it is simulated using Finite Element Analysis simulation software (specifically, ANSYS\textsuperscript{\textregistered} HFSS 2021 R1). Simulations are also used to identify which of the many measured cavity modes corresponds to \temns. The \tem fields are simulated and shown in Fig.~\ref{fig:orpheus_simulations1}. The fields resemble their free space Gaussian counterparts in the empty cavity case.

From the simulated field, $\veff$ was calculated using Equation~\ref{eqn:veff}. Because of the orientation of the WR-62 waveguide, the receiver is only sensitive to $\vb{E}_y$, so ${\veff = \left (\int dV \vb{E}_y(\vec{x})\right)^2/\left (\int dV \epsilon_r |\vb{E}_y(\vec{x})|^2 \right ) \langle \cos^2\theta \rangle_T}$, where $\theta$ is the angle between the electric field along $\hat{y}$ and the dark photon field. $\theta$ is unknown, but $\cost =1/3$ if the dark photon is randomly polarized~\cite{Arias_2012, PhysRevD.104.092016, PhysRevD.104.095029}.

\subsubsection{Simulations and Tabletop Measurements of Evenly-spaced Configuration} 
\begin{figure}[htp]
  \centering
  \includegraphics[width=\linewidth]{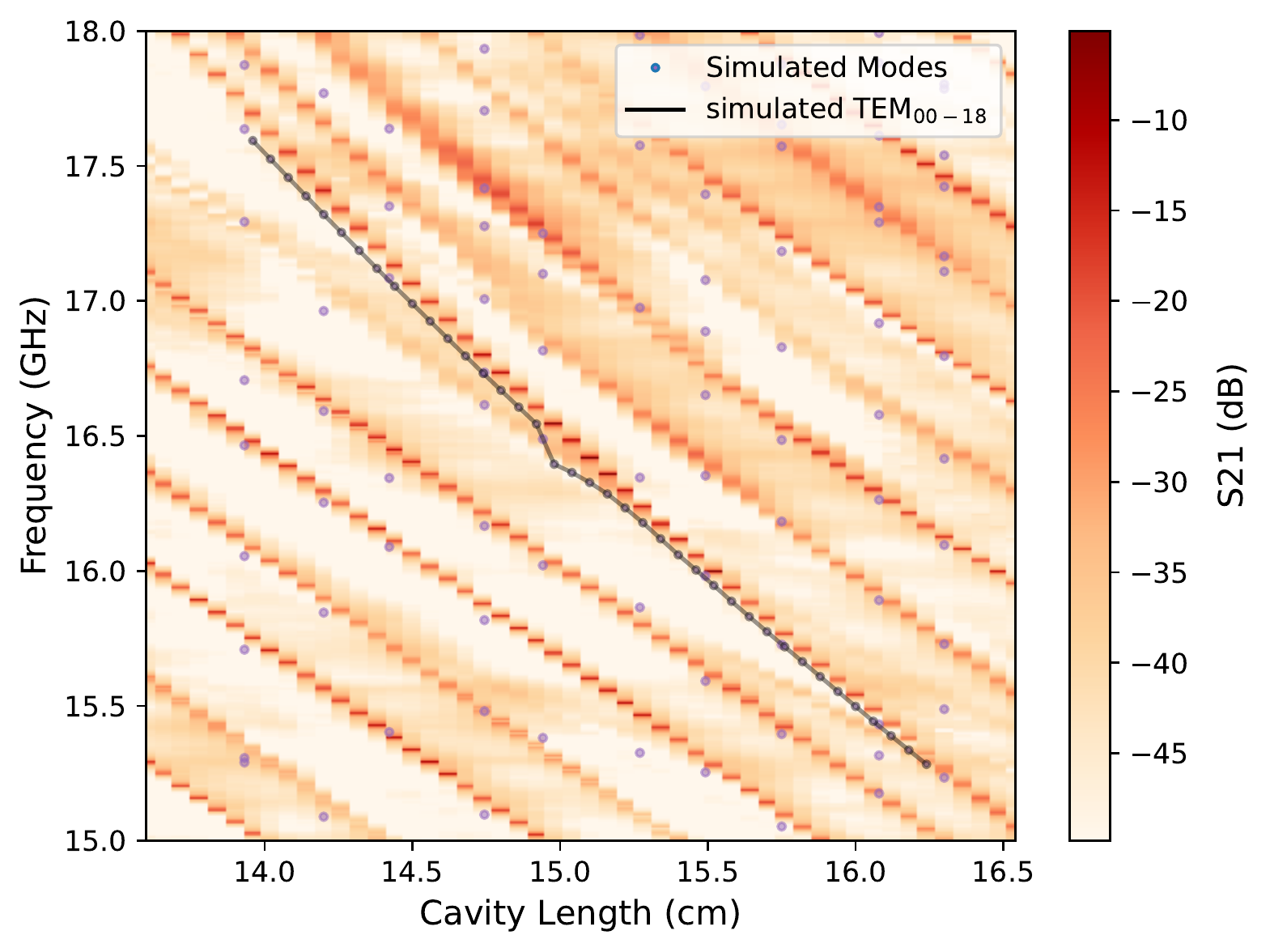}
  \caption{Measured mode map for dielectrically-loaded Fabry-Perot cavity implementing the evenly-spaced configuration. The dark lines correspond to the cavity modes where transmission is highest. The simulated modes, shown as circles, are overlaid on the measurement. The simulated \tem mode is shown as a black line along the diagonal of the mode map. This configuration suffers from a mode crossing at about \SI{16.4}{GHz}. This mode crossing was mitigated in the dark photon search by deviating from the evenly-spaced configuration.}
  \label{fig:tabletop_orpheus_modemap}
\end{figure}

The dielectric plates were originally intended to maintain an even spacing throughout the cavity as it tuned. Exploratory tabletop measurements implemented this evenly-spaced configuration. The mode spectrum of the cavity was measured and is visualized as a mode map (Fig.~\ref{fig:tabletop_orpheus_modemap}), which is a 2D plot of the transmitted power through the cavity as a function of frequency and cavity length. The dark lines correspond to the different cavity modes.

The \tem mode is simulated for different cavity lengths.  The modes of the evenly-spaced configuration were simulated for the entire tuning range and are found to agree with the measured mode spectrum of a tabletop setup seen in Fig.~\ref{fig:tabletop_orpheus_modemap}. However, the evenly-spaced configuration resulted in a wide mode crossing near \SI{16.4}{GHz} where two degenerate modes hybridize, significantly reducing $\veff$ and $Q$. The simulation also had trouble accurately simulating the mode crossing, as shown by the kink in the black curve. Perhaps more simulation precision is required for the mode crossing. The simulations for the evenly-spaced configuration are described in detail in Ref.~\cite{cervantes2021search}.

\subsubsection{Simulations and Characterization of Cryogenic Dark Photon Search with Measured Position Errors}
While the evenly-spaced measurements and simulations are illuminating, they do not apply to the dark photon search in this paper. The dark photon search did not follow this evenly-configured configuration because of mechanical tuning and software issues described in Section~\ref{sec:cadence}. The evenly-spaced configuration is discussed only because the motor issues prevented the same mode map measurement similar to Fig.~\ref{fig:tabletop_orpheus_modemap}. This accidental deviation mitigated the mode crossing and improved the detector's sensitivity throughout its entire tuning range. The simulation results in this paper pertain to the dielectric plate positions measured in the dark photon search rather than the ``idealized'' evenly-spaced configuration.  

\begin{figure}[htp]
  \centering
  \includegraphics[width=\linewidth]{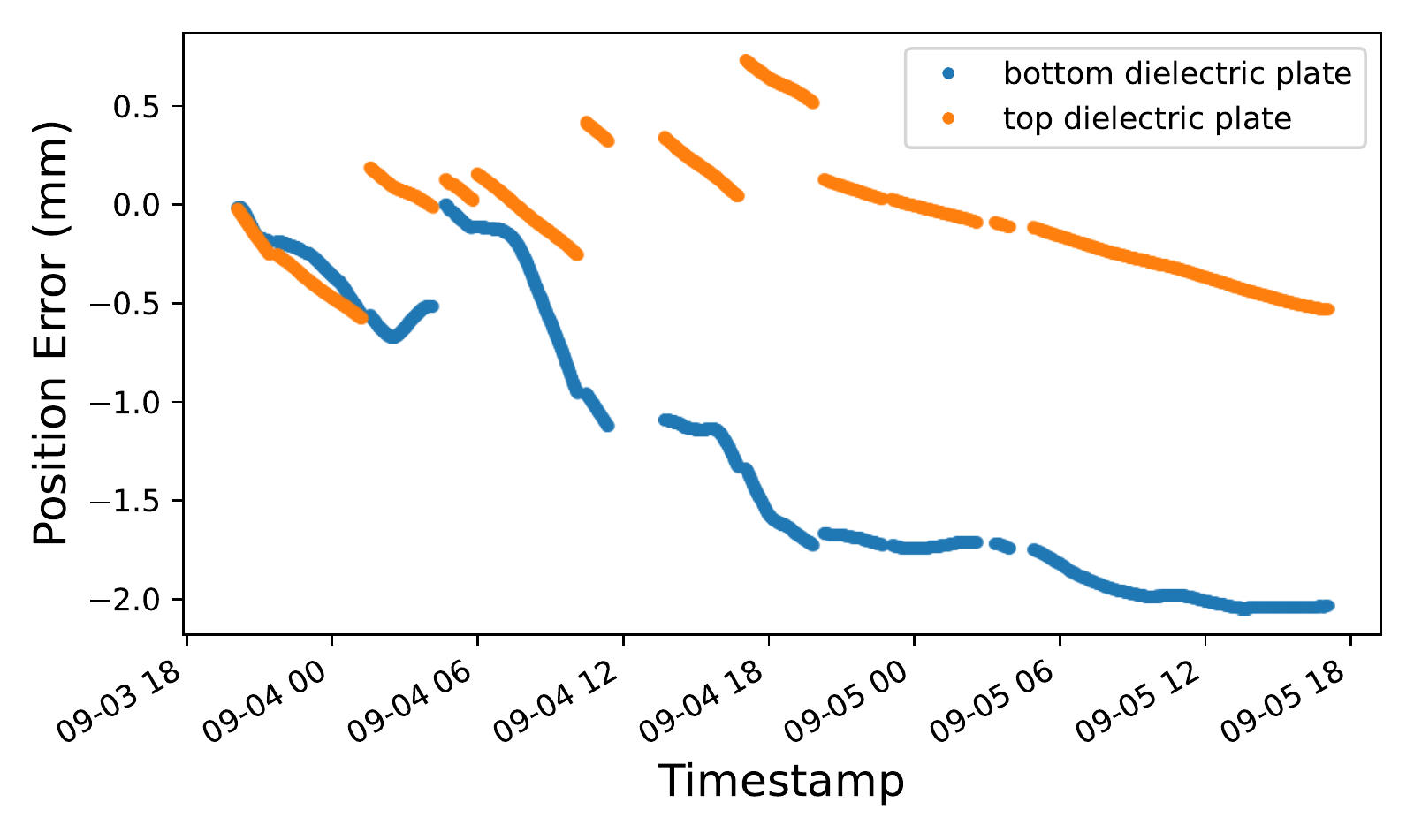}
  \caption{The deviation of the plate positions from the evenly-spaced configuration. This position error was used to estimate the uncertainty in $V_{eff}$.}
  \label{fig:motor_step_error}
\end{figure}

This deviation is parametrized by the position error $\delta$, defined as the deviation from the evenly spaced configuration (actual position - evenly-spaced position). Orpheus is oriented vertically such that the flat mirror is on the top and the curved mirror is on the bottom, as shown in Fig.~\ref{fig:orpheus_cavity_cad}. Naturally, the bottom dielectric plate is closest to the curved mirror, and the top dielectric plate is closest to the flat mirror. A positive position error means the dielectric plate is closer to the curved mirror than what would be intended by the evenly-spaced configuration. This sign convention is consistent with the direction of increasing or decreasing cavity length (the flat mirror is fixed, and the curved mirror moves to adjust the cavity length). The middle two dielectric plates are constrained by the scissors jacks and stay evenly spaced between the top and bottom dielectric plates. The position errors measured throughout the entire search are plotted in Fig.~\ref{fig:motor_step_error}.

\begin{figure}[htp]
  \centering
  \includegraphics[width=\linewidth]{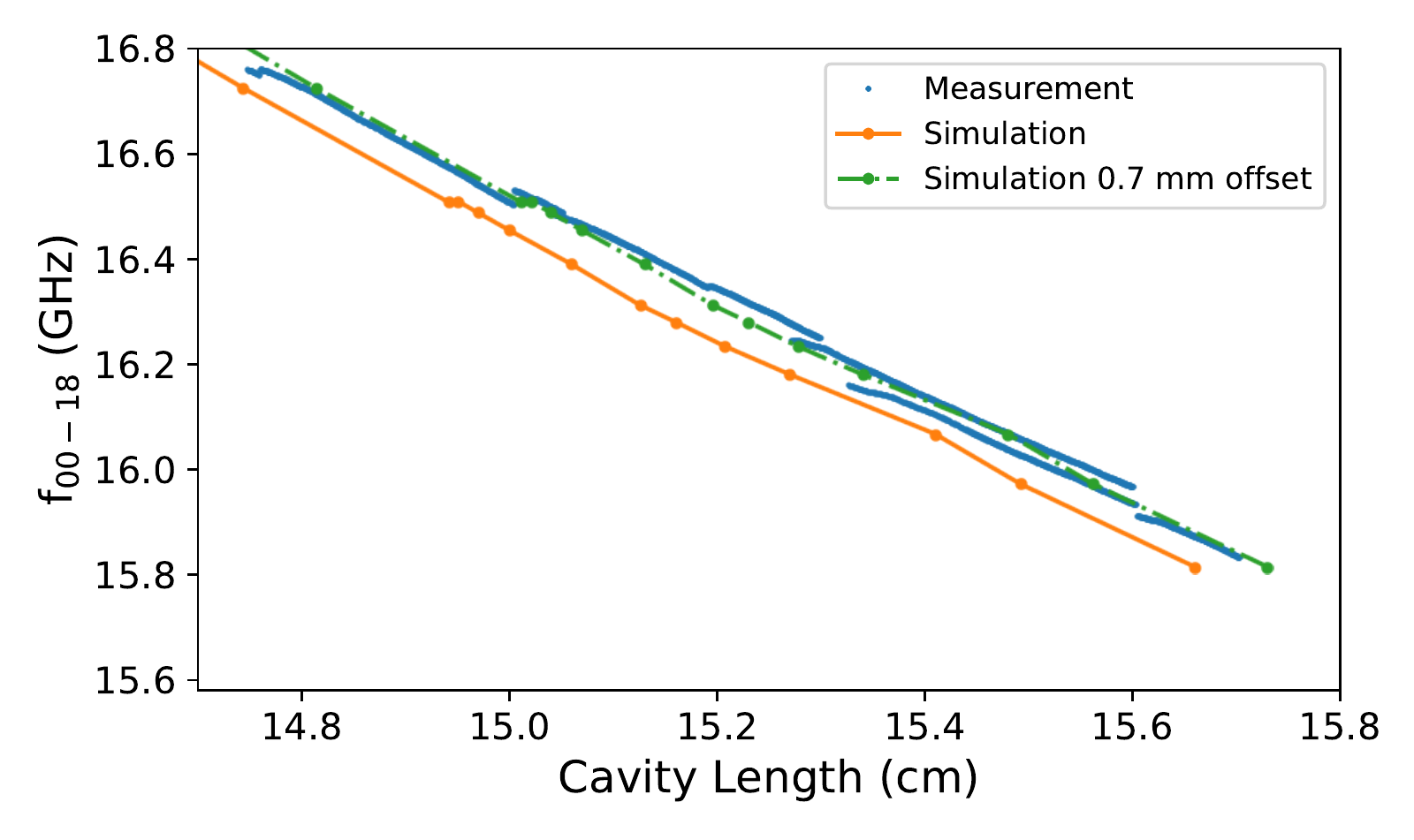}
  \caption{The \tem mode frequency vs. the cavity length (as determined by the motor encoder values).  The measured $f_{00-18}$ is double-valued at some measured cavity lengths because of the hysteresis in the tuning mechanism. The simulated mode (orange) is found to be offset from the measured mode (blue) by \SI{0.7}{mm}, suggesting a systematic error in the absolute cavity length. Thus, it is thought that the measured $f_{00-18}$ is better for determining cavity length than the motor encoder values.}
  \label{fig:dpsearch_modefreq}
\end{figure}

\begin{figure}[htp]
  \centering
  \includegraphics[width=\linewidth]{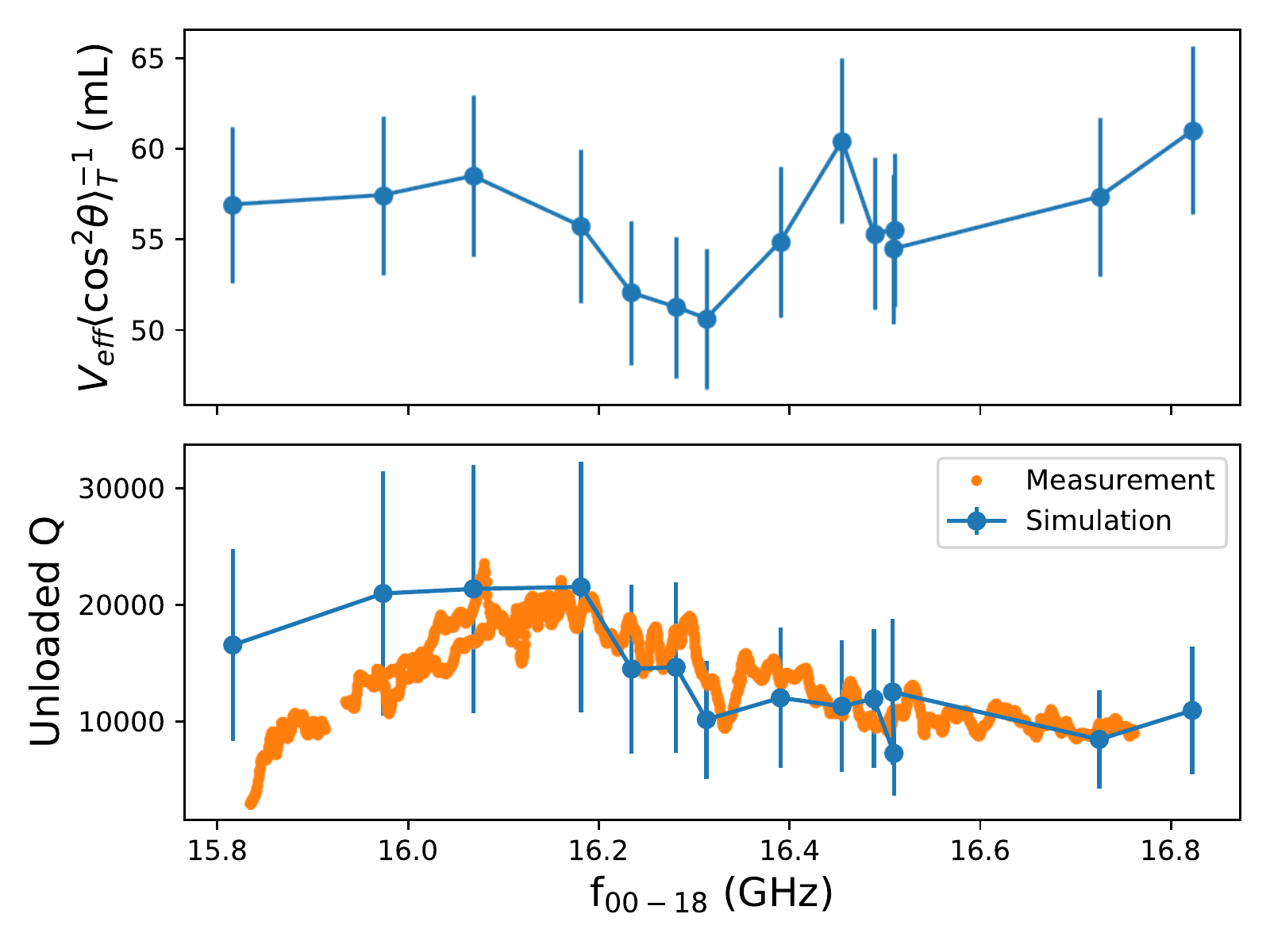}
  \caption{$\veff$ and $Q_0$ as a function of mode frequency. The simulated $Q_0$ is consistent with the measured $Q_0$ within the simulated uncertainties. The measured $Q$ is double-valued at some frequencies because of the hysteresis in the tuning mechanism.}
  \label{fig:dpsearch_q_v_freq}
\end{figure}

The simulated $f_0$, $Q_0$, and $\veff$ are plotted in Figs.~\ref{fig:dpsearch_modefreq} and~\ref{fig:dpsearch_q_v_freq}. The plot divides $\veff$ by $\cost$ to make the results more comparable to $\veff$ for axion experiments (where $\cost = 1$). $\veff \cost ^{-1} \sim\SI{55}{mL}$ for much of the tuning range, which is about ten times larger than the ADMX Run 1B haloscope resized so the TM$_{010}$ mode corresponds to \SI{16}{GHz}. The form factor  ($\veff/V_c$, where $V_c$ is the volume of the cavity) is about 2\% but can be improved upon by optimizing mirror curvatures, dielectric thicknesses, dielectric positioning, and by adding more dielectrics.

$f_0$, $Q_L$ and $\beta$ are extracted from the transmission and reflection measurements discussed in Section~\ref{sec:cavity}. $f_0$ and $Q_L$ are determined from the transmission measurement, and $\beta$ is determined from the reflection measurement. It is possible to determine $Q_L$ from the reflection measurement, but crosstalk-like effects in the waveguide coupler distort the Lorentzian shape of the reflection measurement, making the reflection $Q_L$ fit less reliable than that obtained from the transmission measurement. Adding extra parameters to the reflection fitting measurement improves the fit to the distorted Lorentzian, but the additional fitting parameters have a degeneracy, and extracting $\beta$ seems infeasible~\cite{cervantes2021search}. 

The fitted $f_0$ is shown in Fig.~\ref{fig:dpsearch_modefreq}. Some cavity lengths are double-valued because of the hysteresis in the tuning mechanism. The same VNA measurements taken with the same motor encoder values at different times do not result in the same $f_0$ because the motor backlash would lead to different cavity lengths. $f_0$ is discontinuous at cavity lengths of around \SI{15}{cm}, \SI{15.3}{cm}, and \SI{15.6}{cm} because the top dielectric plate was abruptly tuned by hand (motor stall issues prevented automated tuning of the top dielectric plate). The simulated mode frequency is also plotted (orange line) and shown to deviate from the measured mode frequency. However, the measured and simulated frequencies match if a \SI{0.7}{mm} offset is added to the simulated frequency. This suggests there is a systematic uncertainty in the measured absolute cavity length which may be caused by mechanical contractions during cooldown or by tuning hysteresis. After accounting for the systematic bias, the measured $f_0$ matches the simulated $f_0$ often by less than one part-per-thousand.

\begin{figure}[htp]
  \centering
  \includegraphics[width=0.8\linewidth]{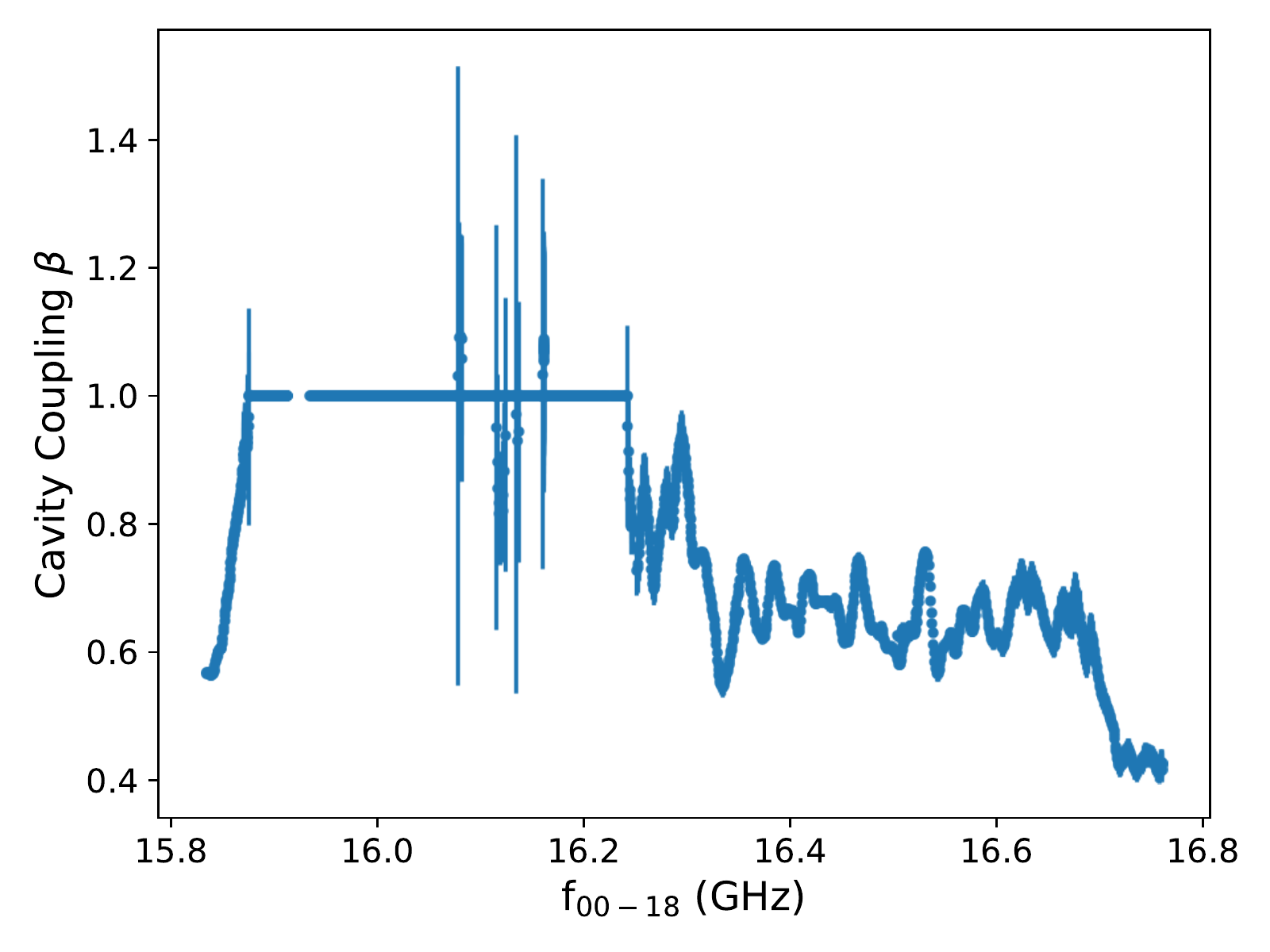}
  \caption{Cavity coupling as a function of mode frequency.}
  \label{fig:dpsearch_beta_v_freq}
\end{figure}

The fitted $Q_0$ and $\beta$ are shown in Fig.s~\ref{fig:dpsearch_q_v_freq} and~\ref{fig:dpsearch_beta_v_freq}. $Q_L$ and $\beta$ are measured directly from the transmission and reflection measurements described previously, and $Q_0 = Q_L(1+\beta)$. Because the Lorentzian fit does not capture the full physics of a reflection measurement (most noticeably the skewed Lorentzian exemplified in Fig.~\ref{fig:s11} of the appendix), there are often fits, particularly when the cavity is critically coupled, when the fitted reflected power is negative on resonance. When this occurs, the Lorentzian dip is often greater than \SI{15}{dB}, which is very close to critical coupling. In this case, the coupling coefficient is taken to be 1 with no uncertainty. This choice is safe because, near critical coupling, the dark photon power is insensitive to uncertainty in $\beta$.

The loaded Q drops off below \SI{16}{GHz} and above \SI{16.2}{GHz}, suggesting Orpheus has a natural bandwidth. This makes sense because of the fixed dielectric thickness. The more the dielectric thickness deviates from $\lambda/2$, the more destructively interfering the dielectrics become. If the dielectric thickness is $\lambda/4$, the wave destructively interferes in the dielectric and would not transmit through the cavity. This system is reminiscent of a cavity filter. 

\subsection{Uncertainty in $V_{eff}$}\label{sec:position_error}
$\veff$ is an important parameter that can only be obtained through simulation. Thus, it is essential to understand how $\veff$ is affected by mechanical misalignments and uncertainties in the dielectric properties. These perturbations are simulated to obtain an uncertainty in $\veff$. 

Different perturbations of the cavity parameters were simulated when $f_0\approx \SI{16}{GHz}$. This frequency is chosen because it corresponds to where the DPDM search had the most integration time and, therefore, the best sensitivity. These parameters include the flat mirror tilt angle, the third dielectric plate tilt angle, position errors $\delta_{top}$ and $\delta_{bottom}$, dielectric constant $\epsilon_r$ and dielectric loss $\tan \delta$. Uncertainties of these parameters are estimated from the literature and tabletop measurements (see the end of Section~\ref{sec:cavity_mechanics} for measurement). The uncertainties in the angles of the plates for the in situ cryogenic measurements are doubled from the tabletop measurement because of the motor tuning issues discussed at the end of Section~\ref{sec:cadence}. Uncertainties in the position errors $\delta_{top}$ and $\delta_{bottom}$ are statistical and derived from the tabletop measurements. They do not include systematic bias because of backlash. Because of motor backlash, it is determined that the $f_0$ is the best way to extract the cavity length rather than the motor encoders. It should also be noted that shifting the green curve in Fig.~\ref{fig:dpsearch_modefreq} by $\pm\SI{0.2}{mm}$ covers most of the measured resonant frequencies (blue dots).  Relative uncertainties $\delta f_0/f_0$, $\delta Q_0/Q_0$, and $\delta \veff/\veff$ are estimated from the parametric sweep results and the parameter uncertainties.

The simulations associated with Fig.~\ref{fig:dpsearch_q_v_freq} do not include the mechanical structure described in Section~\ref{sec:mechanics} because of a lack of computational resources. A simplified version of the mechanical structure was simulated for $f_0 \approx \SI{16}{GHz}$. This simplified structure consisted of perfectly conducting metal surrounding the mirrors and dielectric plates to approximate the mirror and dielectric plate holders. These simulated plate holders are connected by perfectly conducting rods\footnote{The rods are made of stainless steel, which is a poor electrical conductor. However, simulating these rods as stainless steel requires more computational resources than what we have available.} at the outer corners of the holders to approximate the guiding rails. This simplified structure caused the simulated $Q_0$ to drop by 30\% and the simulated $\veff$ to drop by 2\%.

The simulated uncertainties caused by parameter perturbations are added in quadrature to derive total relative uncertainty in $f_0$, $Q_0$, and $\veff$. The results are shown in Table~\ref{tab:veff_uncertainties}. It is found that $Q_0$ is much more sensitive to perturbations than $\veff$.

\begin{table}[htp]
\resizebox{\linewidth}{!}{%
\begin{tabular}{lllll}
perturbation source        & uncertainty & $\delta f_0/f_0 (ppm)$       & $\delta Q_0/Q_0$     & $\delta \veff/\veff$  \\ \hline
flat mirror tilt angle        & 1\degree                     & \num{214} & 0.054 & 0.035 \\
dielectric tilt angle    & 1\degree                     & \num{37.6} & 0.000 & 0.000 \\
$\delta_{top}$    & \SI{0.2}{mm}                     & \num{190.} & 0.061 & 0.028 \\
$\delta_{bottom}$    & \SI{0.2}{mm}                     & \num{190.} & 0.061 & 0.028 \\
$\epsilon_r$                 & 0.1                   & \num{645.} & 0.267 & 0.048 \\
$\tan \delta$                & \num{1e-4}                  & \num{50.1} & 0.280 & 0.014 \\
effects of structure         &                       & \num{90.0} & 0.299 & 0.022 \\
total   relative uncertainty &                       & \num{714.} & 0.499 & 0.076
\end{tabular}%
}
\caption{Different sources of perturbations were simulated at $f_0 \approx \SI{16}{GHz}$ to derive an relative uncertainties $\delta f_0$, $\delta Q_0$, $\delta \veff$.  }
\label{tab:veff_uncertainties}
\end{table}

\begin{figure}[htp]
  \centering
  \subfloat[]{\includegraphics[width=0.8\linewidth]{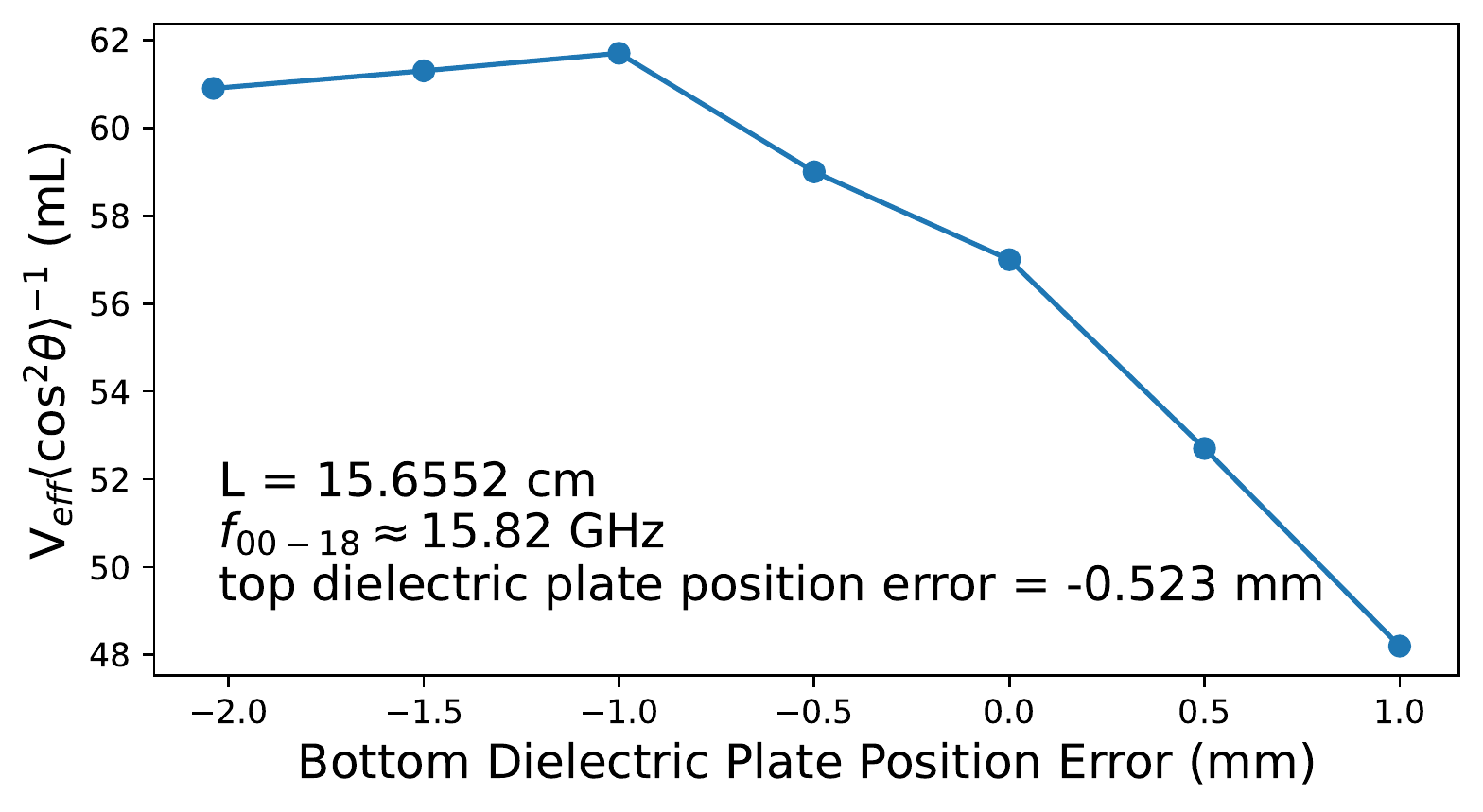}}\\
  \subfloat[]{\includegraphics[width=0.8\linewidth]{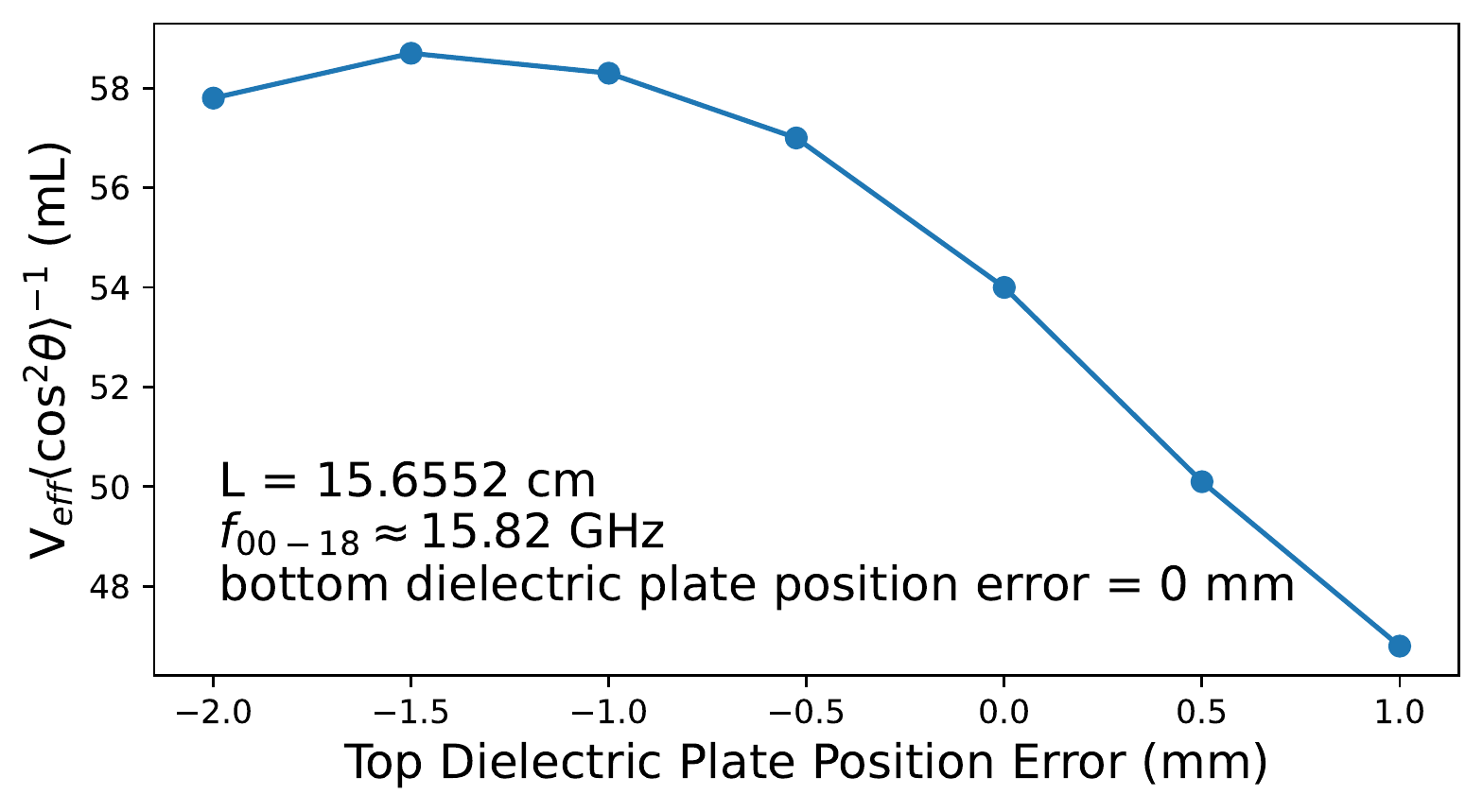}}
  \caption{The effect of position error on the effective volume. This study suggests that it would actually be more optimal to deviate from the evenly-spaced configuration.}
  \label{fig:veff_position_err}
\end{figure}

\begin{figure}[htp]
  \centering
  \subfloat[]{\includegraphics[height=48mm]{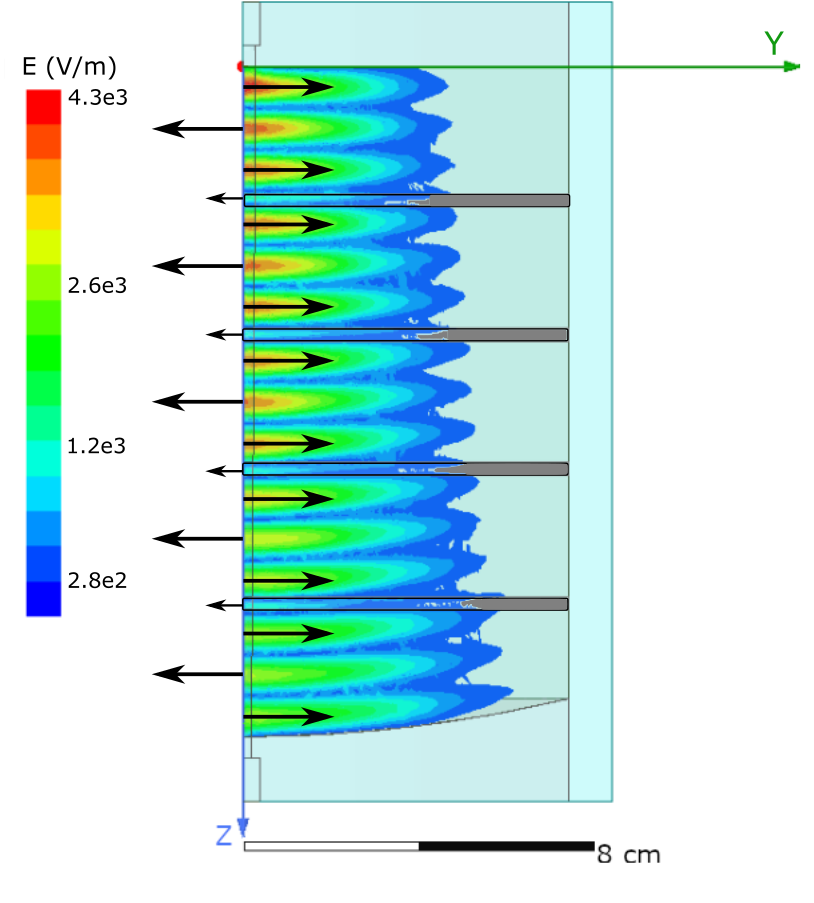}}\hfil
  \subfloat[]{\includegraphics[height=48mm]{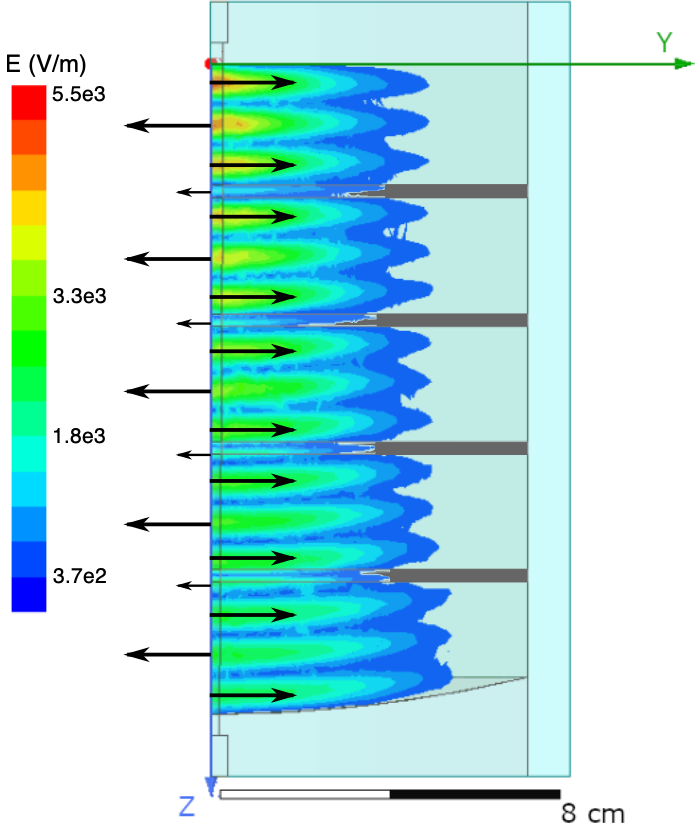}}
  \caption{The effect of dielectric plate position error on the electric field, simulated at about \SI{15.82}{GHz}. (a) No position error. (b) $\delta_{top} = \SI{-0.52}{mm}$, $\delta_{bottom} = \SI{-2}{mm}$. The electric field magnitude scales linearly from blue to red in accordance with the color bar. }
  \label{fig:field_error_position}
\end{figure}

These perturbation studies also reveal that $\veff$ increases with negative position errors $\delta_{top}$ and $\delta_{bottom}$, as shown in Fig.~\ref{fig:veff_position_err}. Comparing the unperturbed and perturbed cases in Fig.~\ref{fig:field_error_position} provides intuition for why that's the case. For the unperturbed case, the portion of the wavefront radially farther away from the beam axis gets pushed towards the dielectrics, reducing $\veff$. Pulling the dielectrics away from the curved mirror also pulls it away from the curved wavefront. 
For positive position errors, the curved wavefront gets pushed more into the dielectrics, and $\veff$ decreases. It appears that a position error of $\sim \SI{-1}{mm}$ optimizes $V_{eff}$. 

\subsection{Estimating expected sensitivity}\label{sec:expected_sensitivity}
\begin{table}[htp]
\begin{tabular}{l|l}
parameter   & value               \\ \hline 
$\beta$     & 1                       \\\hline
$V_{eff,max}$   & \SI{60}{cm^3}      \\\hline
$\rho_{\Ap}$ & \SI{0.45}{GeV/cm^3} \\\hline
$m_{\Ap}$    & \SI{66.2e-15}{GeV}    \\\hline
$\Delta t$           & \SI{1.5e26}{GeV^{-1}} (\SI{100}{s})\\\hline
Q           & 10000 \\\hline
$T_n$       & \SI{8.617e-13}{GeV} (\SI{10}{K}) \\\hline
SNR         & 3                   \\\hline
b           & \SI{6.6e-20}{GeV} (\SI{16}{kHz}) 
\end{tabular}
\caption{Operating parameters for Orpheus used to estimate Orpheus science reach.}
\label{tab:operating_parameters}
\end{table}

Before assembling the combined spectrum to derive an exclusion limit for $\chi$, the detector sensitivity is estimated from the operating parameters shown in Table~\ref{tab:operating_parameters}. The equation for sensitivity in $\chi$ is 

\begin{align}
  \chi &= \sqrt{\frac{1}{\cost}\frac{\beta+1}{\beta}\frac{\snr \times b T_n}{m_{\Ap}\rho_{\Ap}V_{eff}Q_L}}\left ( \frac{1}{b \Delta t} \right )^{1/4}\label{eqn:dp_power_estimate}.
\end{align}
This is derived from rearranging the SNR equation. The bandwidth $b$ is chosen to be comparable to dark matter signal bandwidth $b\sim \SI{16}{GHz}/10^6$. So for randomly polarized dark photons, ${\cost = 1/3}$ and $\chi = \num{1.36e-13}$.

\subsection{Processing Individual Raw Spectrum}
\subsubsection{Removing Spectral Baseline}\label{sec:baseline_removal}
\begin{figure}[htp]
  \centering
  \includegraphics[width=\linewidth]{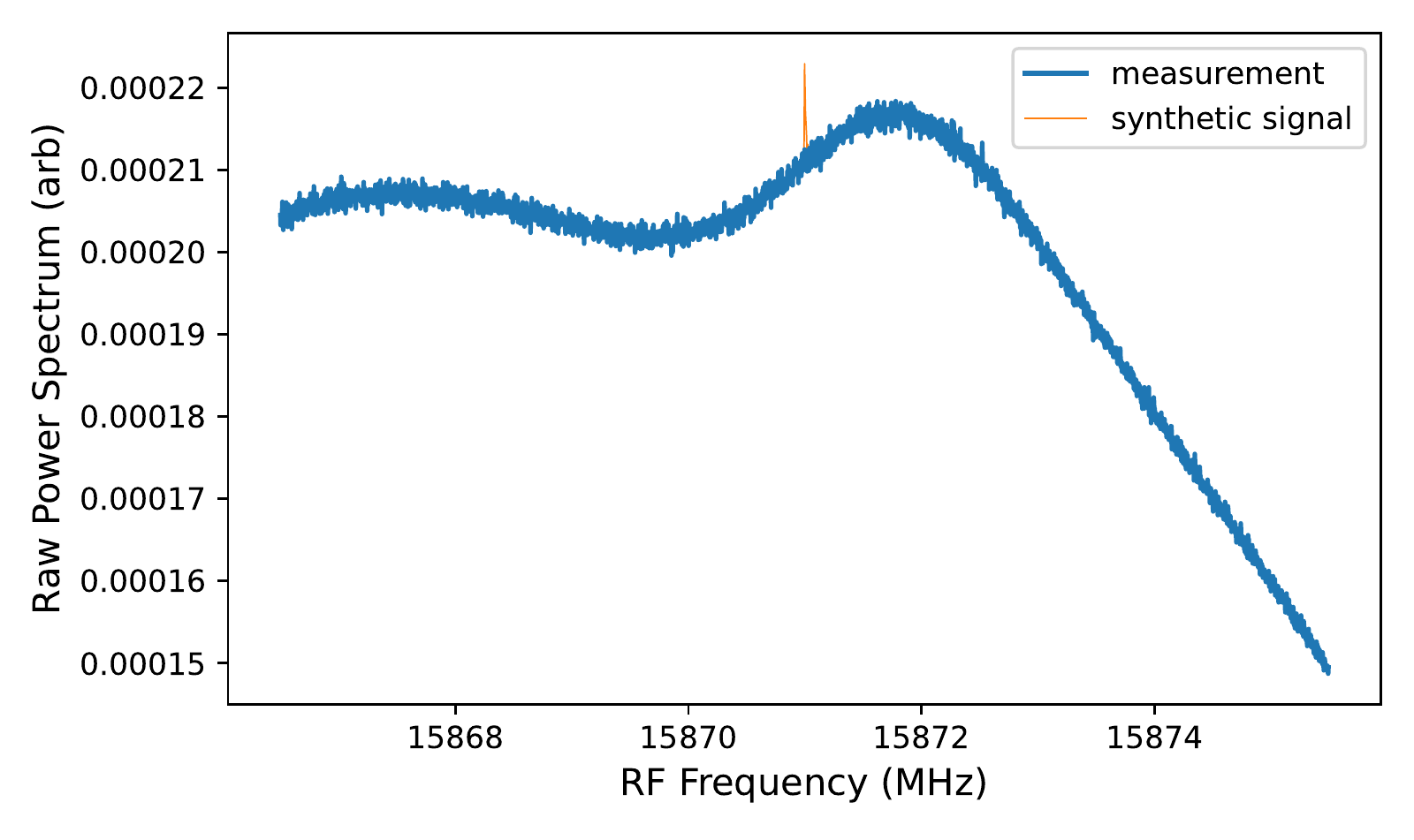}
  \caption{Raw spectrum for a single cavity tuning as measured by the digitizer. Assuming the spectrum is just noise, the raw spectrum is the noise power of the system. A hypothetical dark photon signal of arbitrary power is superimposed on the raw spectrum to show that a signal would be much narrower than the large-scale structure of the noise background.}
  \label{fig:raw_spectrum}
\end{figure}

The raw spectrum consists (Fig.~\ref{fig:raw_spectrum}) of the system noise power $P_n = G b k_b T_n$, plus fluctuations about this noise power. $T_n$ affects the SNR of the dark photon data. The system gain, i.e., transfer function, does not affect the SNR of the dark photon data but causes the large-scale gain variation seen in Fig.~\ref{fig:raw_spectrum}. The first step of processing the raw spectra is to remove this gain variation. The gain variation is caused by both the IF and RF electronics. The IF gain and RF gain are thought of as separate because the IF gain applies equally to all spectra, whereas the RF gain depends on the frequency of the cavity. The gain variation is removed by first dividing the spectrum by $G_{IF}$ and then by $G_{RF}$. 

Figure~\ref{fig:raw_spectrum} shows a hypothetical dark photon signal superimposed on the raw spectrum. 
The dark photon signal is much narrower (seven bins) than the large-scale gain variation caused by the electronics. Thus it is safe to apply filters to remove this gain variation. 
In more detail, the spectral shape of the dark photon signal is proportional to the dark photon kinetic energy distribution. The most conservative energy distribution assumes a virialized, isothermal halo that obeys a Maxwell-Boltzmann distribution,

\begin{align}
  \mathcal{F}(f) = \frac{2}{\sqrt{\pi}}\sqrt{f - f_a} \left ( \frac{3}{f_a \langle \beta^2 \rangle }\right )^{3/2} \exp \frac{-3(f-f_a)}{f_a\langle \beta^2 \rangle}
  \label{eqn:lineshape}
\end{align}
where $f$ is the measured frequency, $f_a$ is the frequency of the associated SM photon, $\langle \beta^2 \rangle = \frac{\langle v^2 \rangle}{c^2}$, and $\langle v^2 \rangle$ is the root mean square (RMS) velocity of dark matter halo. The ``quality factor'' of the dark photon lineshape is $\mathcal{O}(10^6)$.
In the lab frame, the lineshape takes a more complicated form, but is well approximated by Equation~\ref{eqn:lineshape} if $\langle \beta^2 \rangle \rightarrow 1.7\langle \beta^2 \rangle$~\cite{PhysRevD.96.123008}.

\begin{figure}[htp]
  \centering
  \includegraphics[width=\linewidth]{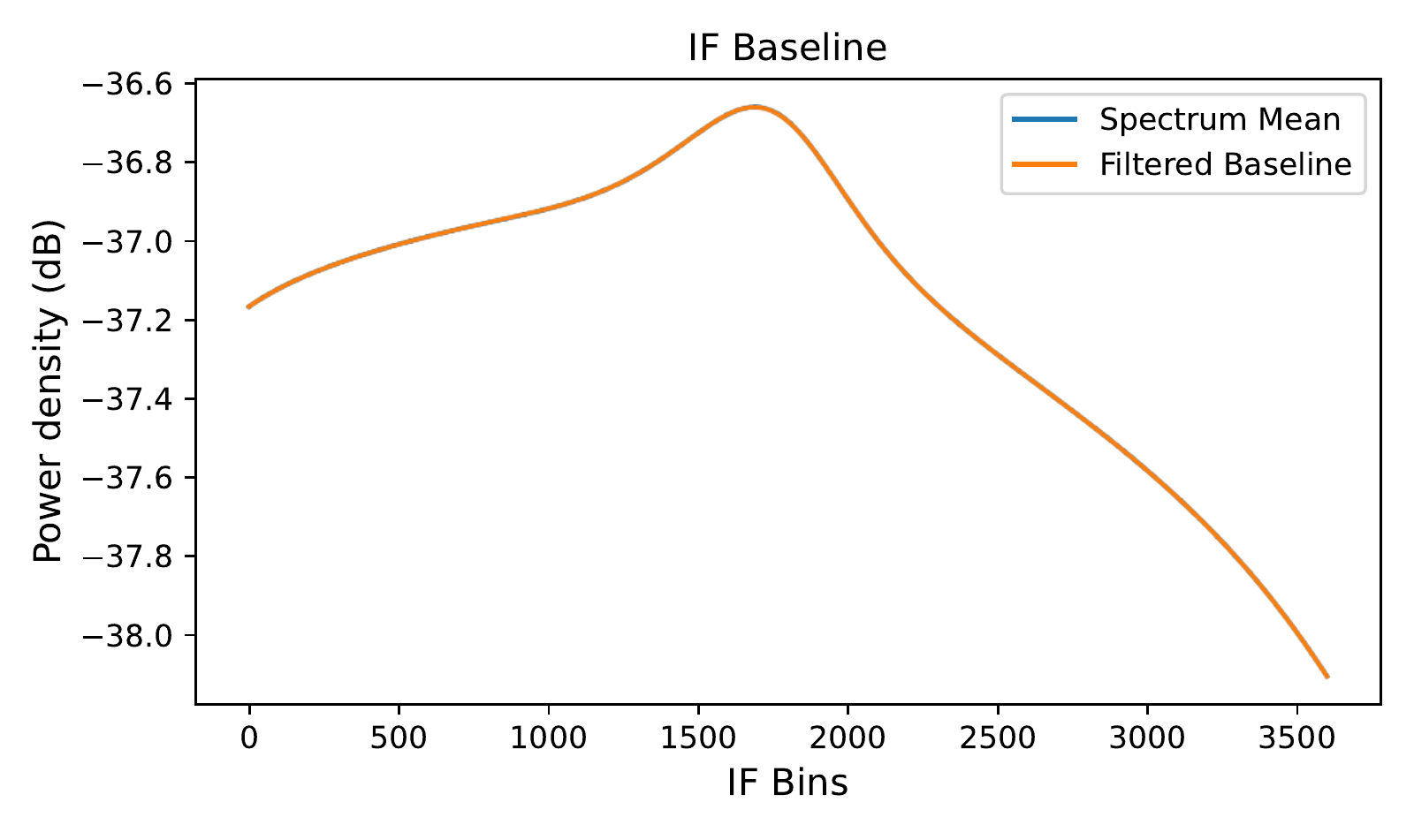}
  \caption{The average of all the recorded spectrum for all cavity tunings (the reference level is arbitrary). This is thought of as a good estimation of the gain variation caused by the IF electronics.}
  \label{fig:if_baseline}
\end{figure}

\begin{figure}[htp]
  \centering
  \includegraphics[width=\linewidth]{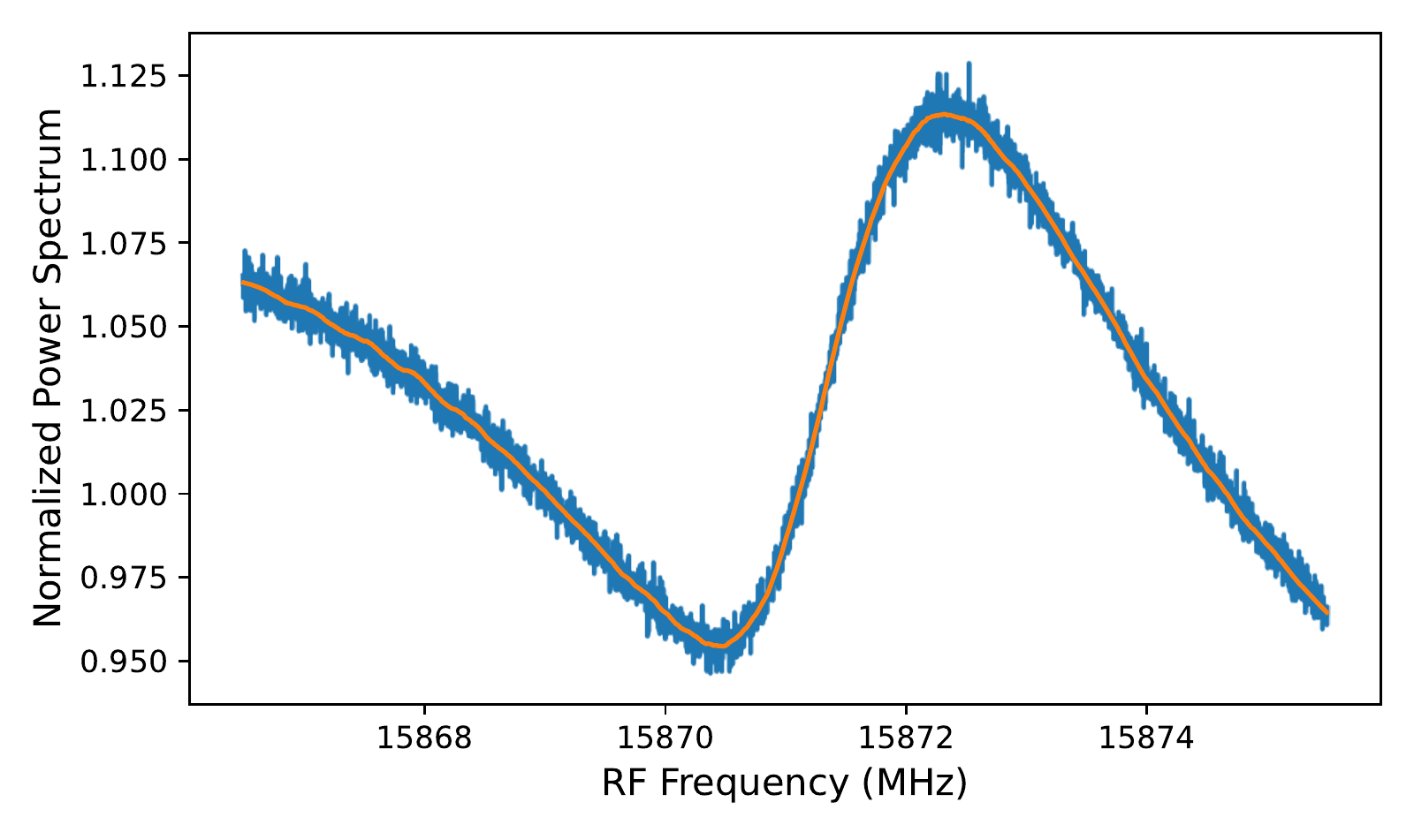}
  \caption{The raw spectra for a single cavity tuning divided by the IF baseline (Fig.~\ref{fig:if_baseline}) to form a normalized spectrum. A Savitzky-Golay filter is then applied (orange) to obtain gain variation from the RF electronics.}
  \label{fig:spectra_if_baseline}
\end{figure}

The IF baseline is estimated by averaging all the power spectrum while the cavity was actively tuning. The IF baseline is then smoothed using a fourth-order Savitzky-Golay software filter with a window length of 501 bins. Note that the Savitzky-Golay window is much larger than the hypothetical dark photon signal in Fig.~\ref{fig:raw_spectrum}, which is about seven bins wide. The resulting IF baseline is shown in Fig.~\ref{fig:if_baseline}. Every spectrum is divided by this IF baseline, and the result is a unitless power spectrum that is still affected by the RF gain variation, as shown in Fig.~\ref{fig:spectra_if_baseline}. 

\begin{figure}[htp]
  \centering
  \includegraphics[width=\linewidth]{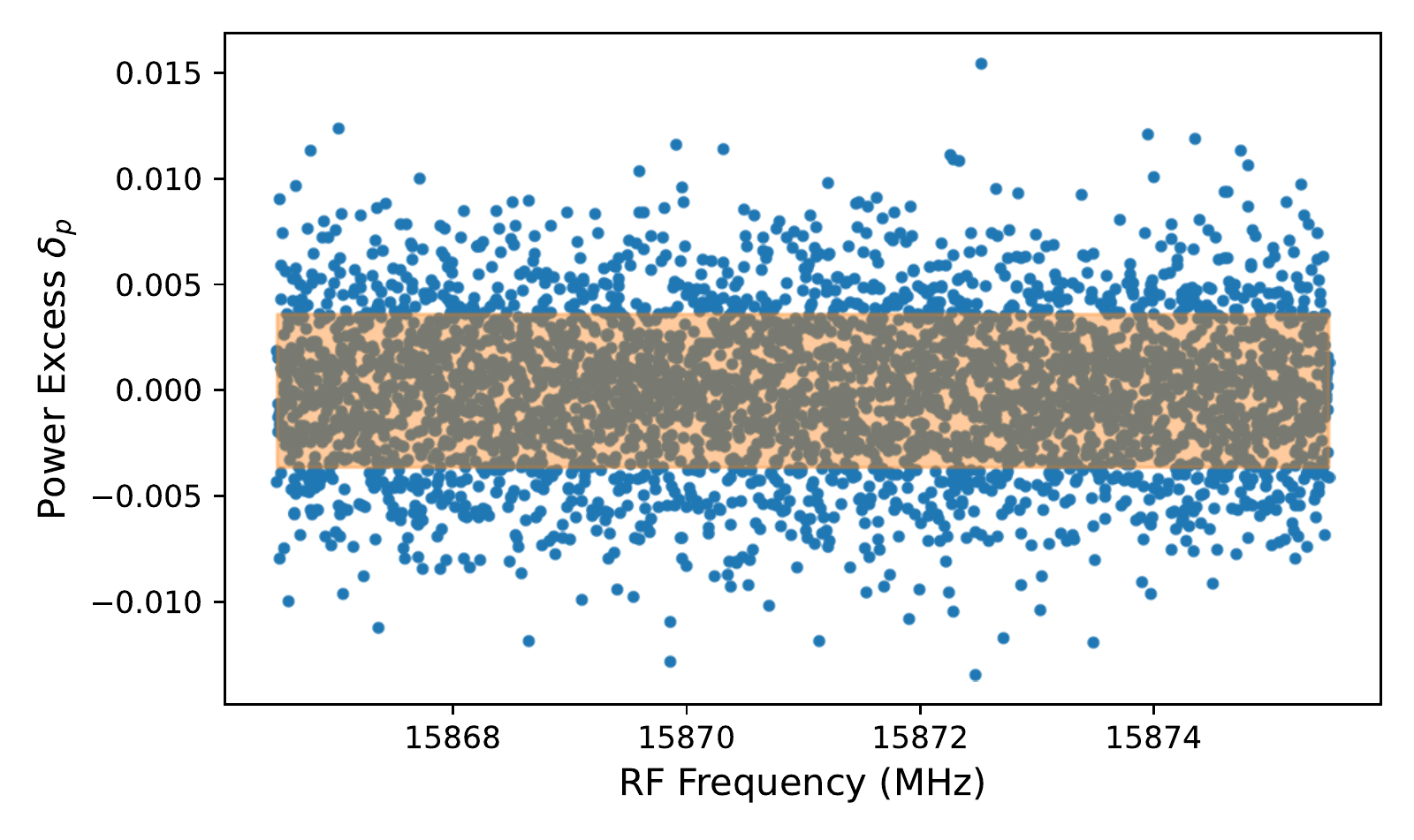}
  \caption{The power excess for a single cavity tuning in units of system noise power. The orange band represents the standard deviation.}
  \label{fig:delta_p}
\end{figure}

The RF gain variation is removed by applying a second-order Savitzky-Golay software filter with a window length of 151 bins to the individual spectrum~\cite{PhysRevD.96.123008}, as shown in the orange curves in Fig.~\ref{fig:spectra_if_baseline}. The spectra are then divided by the filtered curve to remove the RF gain variation. The processed spectra are now normalized such that the mean of the bins is one. The dark photon signal would show up as a power excess, so the process spectra are subtracted by one to yield the flat spectra fluctuating about zero, shown in Fig.~\ref{fig:delta_p}. Each bin is either an average of \num{75000} spectra or \num{250000} spectra, and so the fluctuations about zero are Gaussian by the Central Limit Theorem. The power excess for each bin in the processed spectrum is denoted as $\delta_p$. All bins in a spectrum are pulled from the same Gaussian distribution, and so the uncertainty of each bin $\sigma_{p}$ is taken as the standard deviation of all the bins in a spectrum. $\delta_p$ is the power excess normalized to units of system noise power $P_n$.

\begin{figure}[htp]
  \centering
  \includegraphics[width=\linewidth]{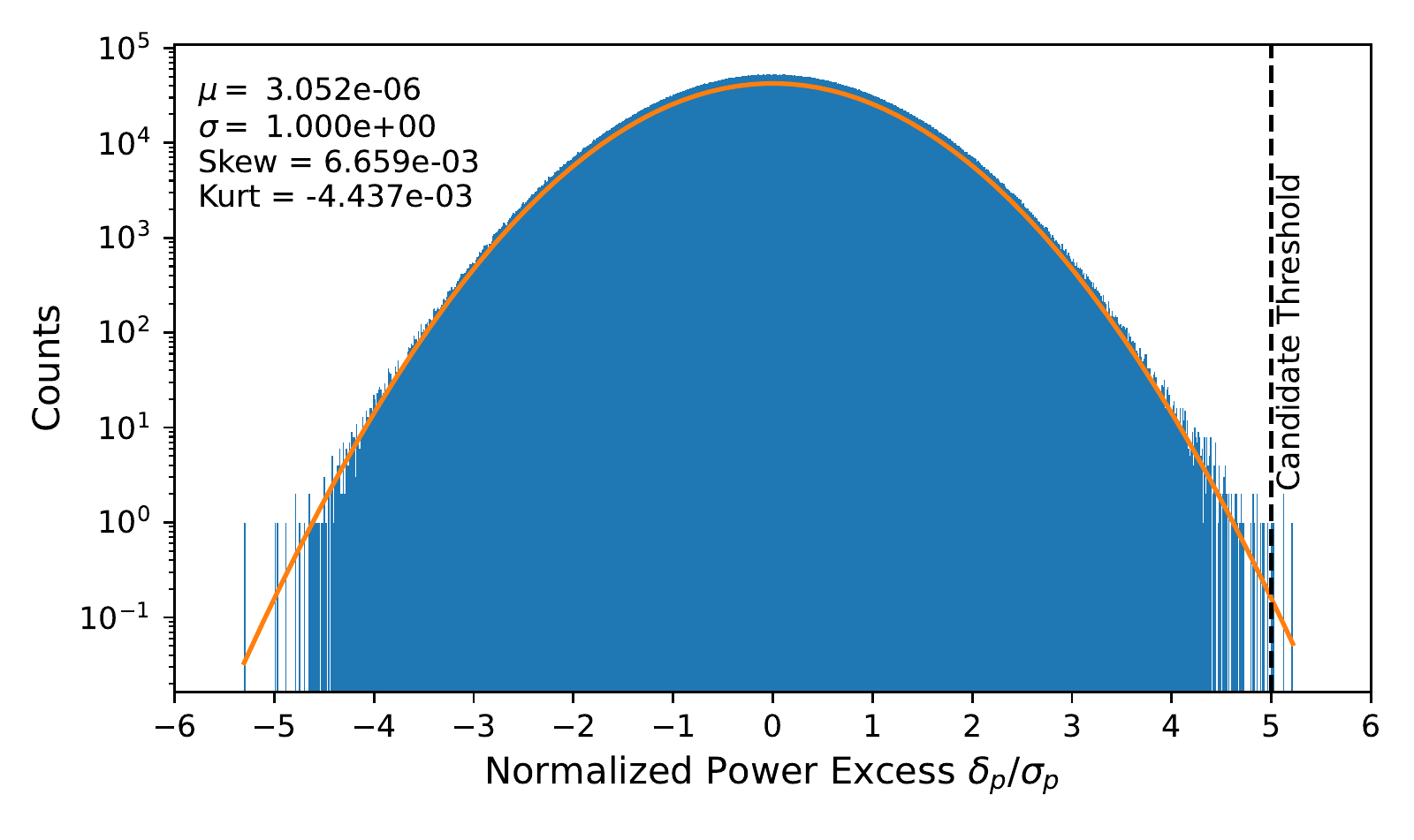}
  \caption{The power excess $\delta_p$ of every processed spectrum. This histogram contains \num{10825200} entries. For comparison, the orange line is a standard normal distribution with the amplitude rescaled to the maximum number of entries in one RF bin. The distribution has Gaussian statistics and follows the orange line well. Six $\delta_p$ bins surpass the $\snr = 5$ candidate threshold. These candidates are ruled out as dark matter signals through further analysis, as explained in Section~\ref{sec:limits}.}
  \label{fig:spectra_histogram}
\end{figure}

Once all spectra are processed, one can obtain a histogram of the power fluctuations normalized to the standard deviation $\delta_p/\sigma_p$, as shown in Fig.~\ref{fig:spectra_histogram}. The histogram follows a normal distribution with zero mean and unit standard deviation. This demonstrates that power measured from the cavity is consistent with Gaussian noise.

\subsubsection{Filter-Induced Attenuation $\eta$}\label{sec:eta}
\begin{figure}[htp]
  \centering
  \includegraphics[width=0.8\linewidth]{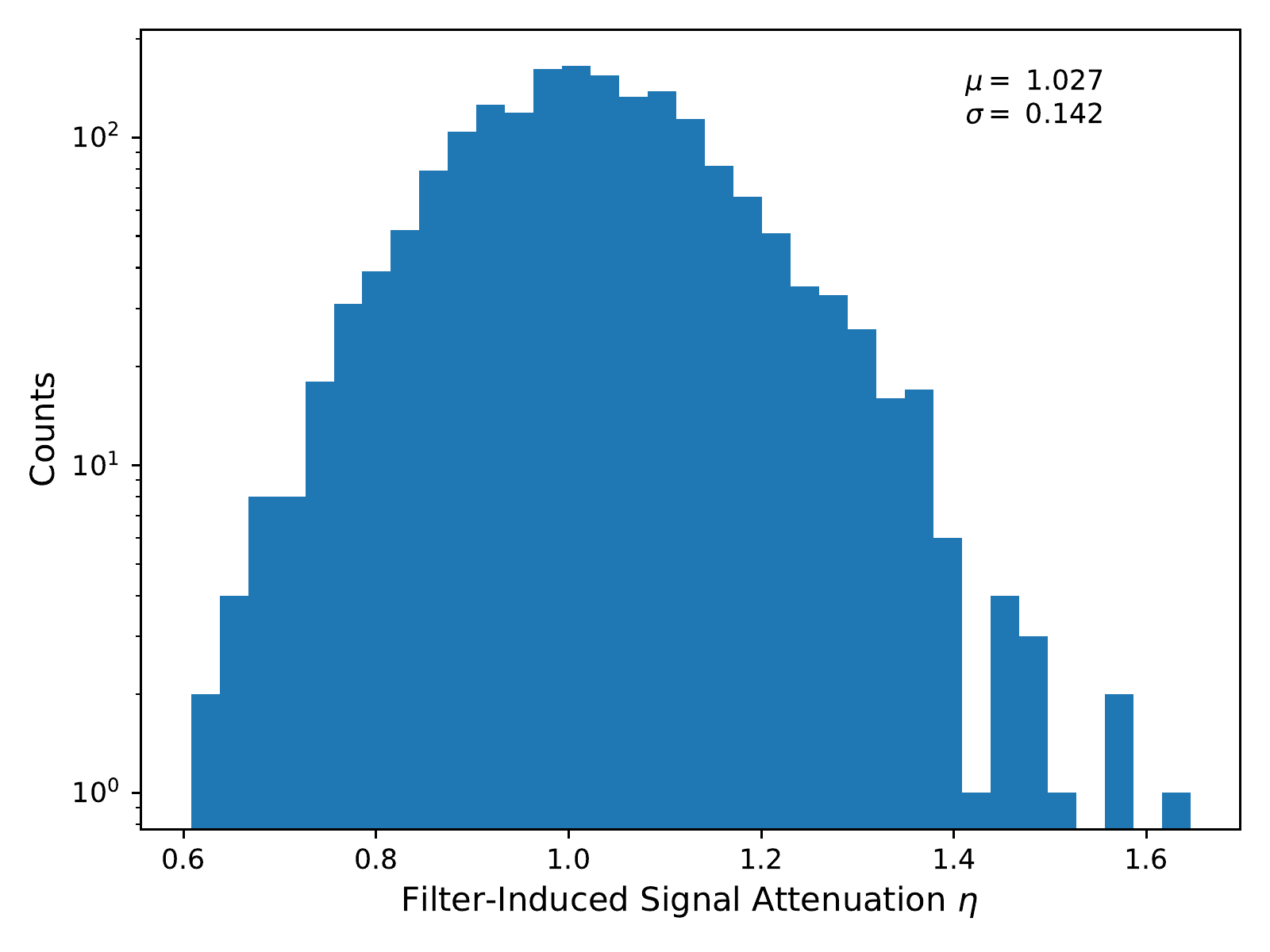}
  \caption{The simulated signal attenuation factor $\eta$.}
  \label{fig:synthetic_snr}
\end{figure}

Applying the two Savitzky-Golay filters affects the SNR of a dark matter signal. This effect is captured by the filter-induced signal attenuation $\eta$ in Equation~\ref{eqn:dp_power}. $\eta$ is determined using simulated dark photon signals with $\snr=5$ that are superimposed on the raw spectrum. The simulated dark photon signals are calculated as

\begin{align}
P(f)_{synthetic} = \snr \times \frac{\ff}{\max{(\ff)}} \sigma_P P_n. \label{eqn:simulated_signal}
\end{align}
$\sigma_P P_n$ is the uncertainty of the raw spectrum. $P_n$ is obtained from multiplying the IF and RF baseline. There are 50 simulated dark photon signals evenly spaced between \SI{15.8}{GHz} and \SI{16.7}{GHz}. A simulated signal is superimposed on the raw spectrum if its frequency falls within a Q-width of the center of the spectrum, and Equation~\ref{eqn:simulated_signal} is used to calculate the signal power vs. RF frequency. After the signal is superimposed, the baseline removal procedure is repeated, and the signal-to-noise ratio of the simulated signal in the processed spectrum ($\textrm{SNR}^{\prime}$) is obtained. Let $\delta_{ps}$ be the height of the simulated signal after baseline removal. Then $\textrm{SNR}^{\prime}= \delta_{ps}/\sigma_p$, and $\eta = \textrm{SNR}^{\prime}/\snr$. Figure~\ref{fig:synthetic_snr} shows $\eta$ for all the simulated signals. The baseline removal procedure should not increase the SNR, so the signal attenuation factor is determined to be $\eta = 1.00 \pm 0.14$.

The rest of the analysis proceeds without the injected simulated signals.

\subsubsection{Rescaling Spectra to be in units of dark photon power}
Next, the data is rescaled to be in units of expected dark photon power. Different operating conditions, such as different temperatures or $Q_L$, will change the noise temperature or expected signal power. The data is rescaled so that a single-bin dark photon signal would have one unit of excess power, regardless of the operating conditions. This rescaling makes the SNR (the signal being that from the dark photon) the true figure of merit for which potential dark photon candidates are sought. The processed spectra are rescaled by multiplying each bin by the noise power and dividing by the hypothetical single-bin dark photon expected power,

\begin{align}
  \delta_s = \delta_p \frac{k_b b T_n}{P_{s}(\chi = 1)}.
\end{align}

$\chi=1$ is arbitrarily chosen because, unlike the QCD axion, there is no benchmark model for the dark photon signal power. The analysis procedure will self-correct for the arbitrarily chosen $\chi$. Also, within the uncertainty, $T_{cav} \sim T_{amp,input} \sim T_{rec}$. So for a given spectrum measurement, $T_n$ can be thought of as constant and independent of the detuning factor $\Delta$. Thus, for simplicity, each processed spectrum will be rescaled using the noise temperature on resonance $T_n \approx T_n(f_0)$.

Another way to understand the rescaling of the power excess for each processed spectrum $\delta_p$ has a Gaussian distribution with $\mu = 0$ and $\sigma = 1/\sqrt{N_p}$, where $N_p$ is the number of bins in the processed spectrum. That is, $\delta_p\sim \text{Gaus}(0, 1/\sqrt{N_p})$. Multiplying $\delta_p$ by $P_n$ scales the power excess to a calibrated power, and dividing by $P_s(\chi = 1)$ normalizes the power excess to a standard power.

The uncertainty for each rescaled spectra bin is similarly rescaled as 
\begin{align}
  \sigma_s = \sigma_p \frac{k_b b T_n}{P_{s}(\chi = 1)}\label{eqn:rescaled_uncertainty}.
\end{align}
There are systematic uncertainties in the signal power and noise temperature. The relative uncertainty in $Q_L$, $T_n$, and $\beta$ are recorded for each spectrum. The relative uncertainty in $f_0$ is subdominant to the other parameters and is not tracked. The relative uncertainty for $\veff$ is $0.076\%$ (Section~\ref{sec:position_error}), and the relative in $\eta$ is $0.14\%$ (Section~\ref{sec:eta}). But the systematic uncertainties are subdominant to the statistical uncertainties, and introducing them in Equation~\ref{eqn:rescaled_uncertainty} is complicated by the bin-to-bin correlations. These complications are avoided by introducing the systematic uncertainties after all the processed spectra are combined (Section~\ref{sec:combined_spectrum}). 

\begin{figure}[htp]
  \centering
  \includegraphics[width=\linewidth]{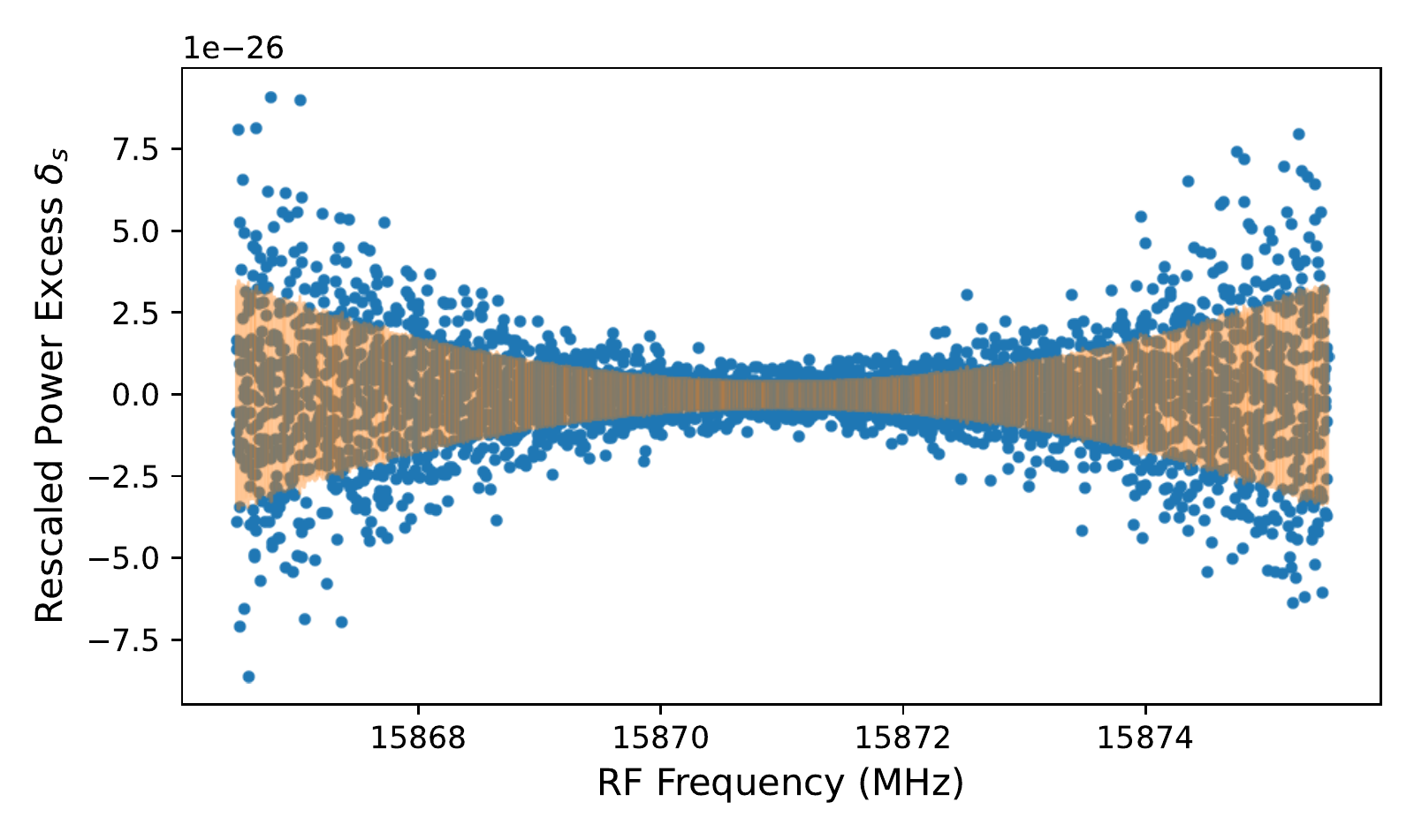}
  \caption{The power excess for a single cavity tuning rescaled to units of single-bin dark photon power. The orange band is the standard deviation.}
  \label{fig:rescaled_spectrum}
\end{figure}

The resulting rescaled spectrum is shown in Fig.~\ref{fig:rescaled_spectrum}.

The spectra are rescaled so that a dark photon signal would have unit height if its power is confined to a single bin. However, the expected dark photon lineshape is spread over about seven bin widths, and this reduces the SNR of each bin. 

A matched filter is used to increase the SNR of a potential dark photon signal. Each spectrum is convolved with the lineshape as the kernel (Equation~\ref{eqn:lineshape}). For this analysis, the kernel consists of 31 bins, with the peak of the dark photon lineshape at the center. However, there is a subtlety because neighboring bins have different uncertainties. Thus, the elements of the convolution are weighted to minimize the $\chi^2$. 

\begin{figure}[htp]
  \centering
  \includegraphics[width=\linewidth]{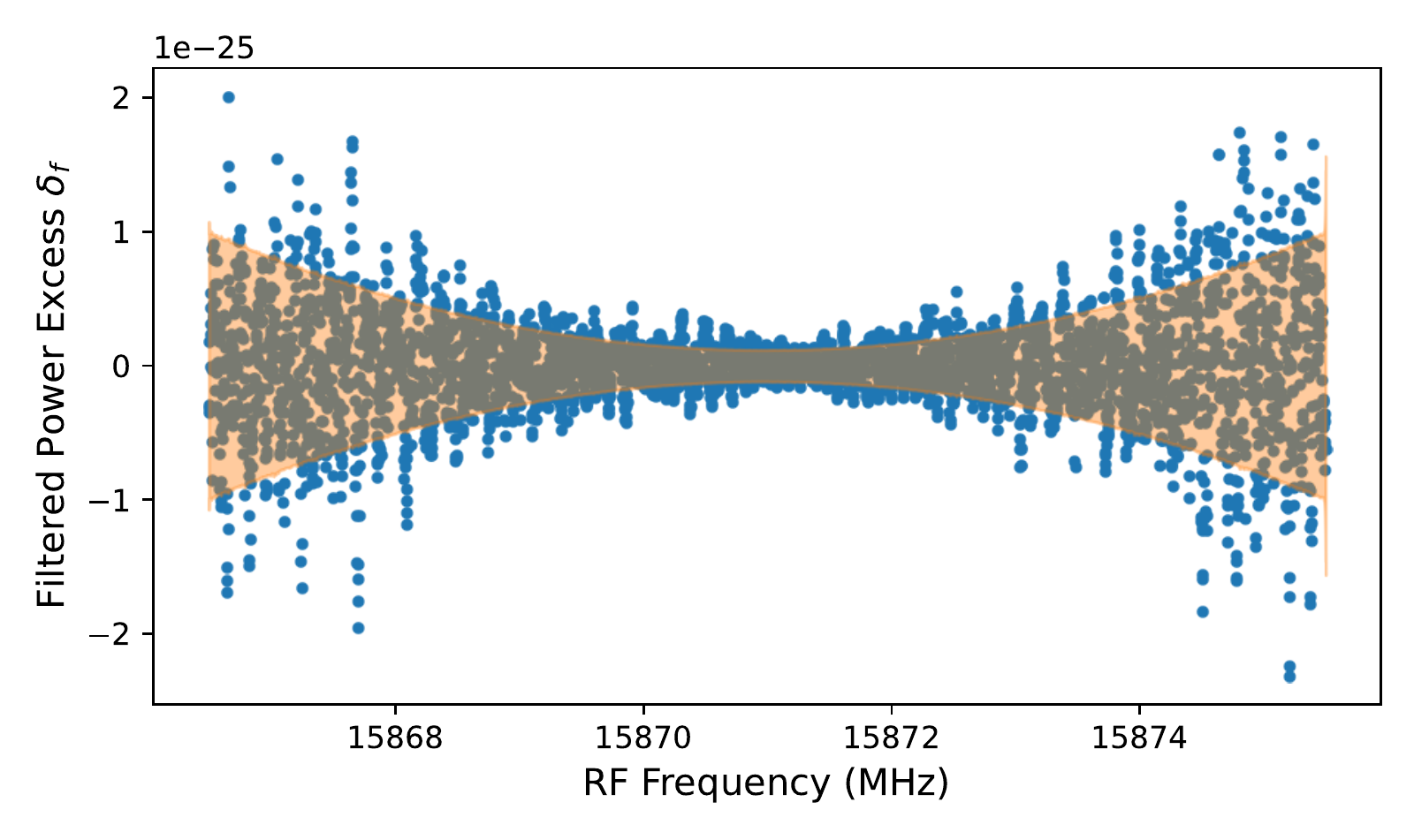}
  \caption{The filtered spectrum with power excess $\delta_f$ for a signle cavity tuning. $\delta_f$ results from applying  matched filter to the rescaled spectrum with power excess $\delta_s$. The orange band is the standard deviation.}
  \label{fig:filtered_spectra}
\end{figure}

Applying the matched filter with the lineshape kernel leads to the filtered spectra with power excesses $\delta_f$ in Fig.~\ref{fig:filtered_spectra}. 

\subsection{Combined Spectrum}\label{sec:combined_spectrum}
The individual filtered spectra single are combined into a single spectrum. The Maximum-Likelihood estimate of the mean $\delta_c$ and uncertainty $\sigma_c$ of the combined spectrum is obtained by a weighted average of all IF bins corresponding to a particular RF bin. The weights of each contributing bin are the inverse variance, and $\delta_c$ and $\sigma_c$ are calculated as

\begin{align}
  \delta_c = \frac{\sum_i \frac{\delta_{fi}}{\sigma^{2}_{fi}}}{\sum_i \frac{1}{\sigma^{2}_{fi}}};\quad \quad \sigma_c = \sqrt{\frac{1}{\sum_i \frac{1}{\sigma_{fi}^2}}}
\end{align}

This method of combining spectrum is well-established in haloscope data analysis~\cite{PhysRevD.96.123008, PhysRevD.103.032002, PhysRevD.64.092003}, and finding the Maximum-Likelihood estimations of means and standard deviations is a general and ubiquitous problem in experimental data analysis.

\begin{table}[htp]
\begin{tabular}{l|c}
parameter   & mean relative uncertainty               \\ \hline 
$\eta$     & 0.14                       \\\hline
$V_{eff}$   & \num{0.076}      \\\hline
$T_n$       & \num{0.052} \\\hline
Q           & \num{0.008} \\\hline
$\beta$     & \num{0.012}                       \\\hline
\end{tabular}
\label{tab:systematics}
\caption{Systematic uncertainties.}
\end{table}

The systematic uncertainties are now introduced into the analysis. Table~\ref{tab:systematics} shows the mean of the relative uncertainties recorded for each processed spectra. The most dominant systematic uncertainties are $\eta$, $\veff$, and $T_n$. However, these are subdominant to the statistical uncertainties. The mean of $\sigma_c/|\delta_c|$ for all bins in the combined spectrum is $\sim 14$. The systematic uncertainties add in quadrature to the statistical uncertainties. The new uncertainties for the combined spectrum are calculated as $\sigma_c^{\prime} = \sqrt{\sigma_c^2 + \delta_c^2\left( 0.14^2 + 0.076^2 + 0.052^2 \right)}$

\begin{figure}[htp]
  \centering
  \includegraphics[width=\linewidth]{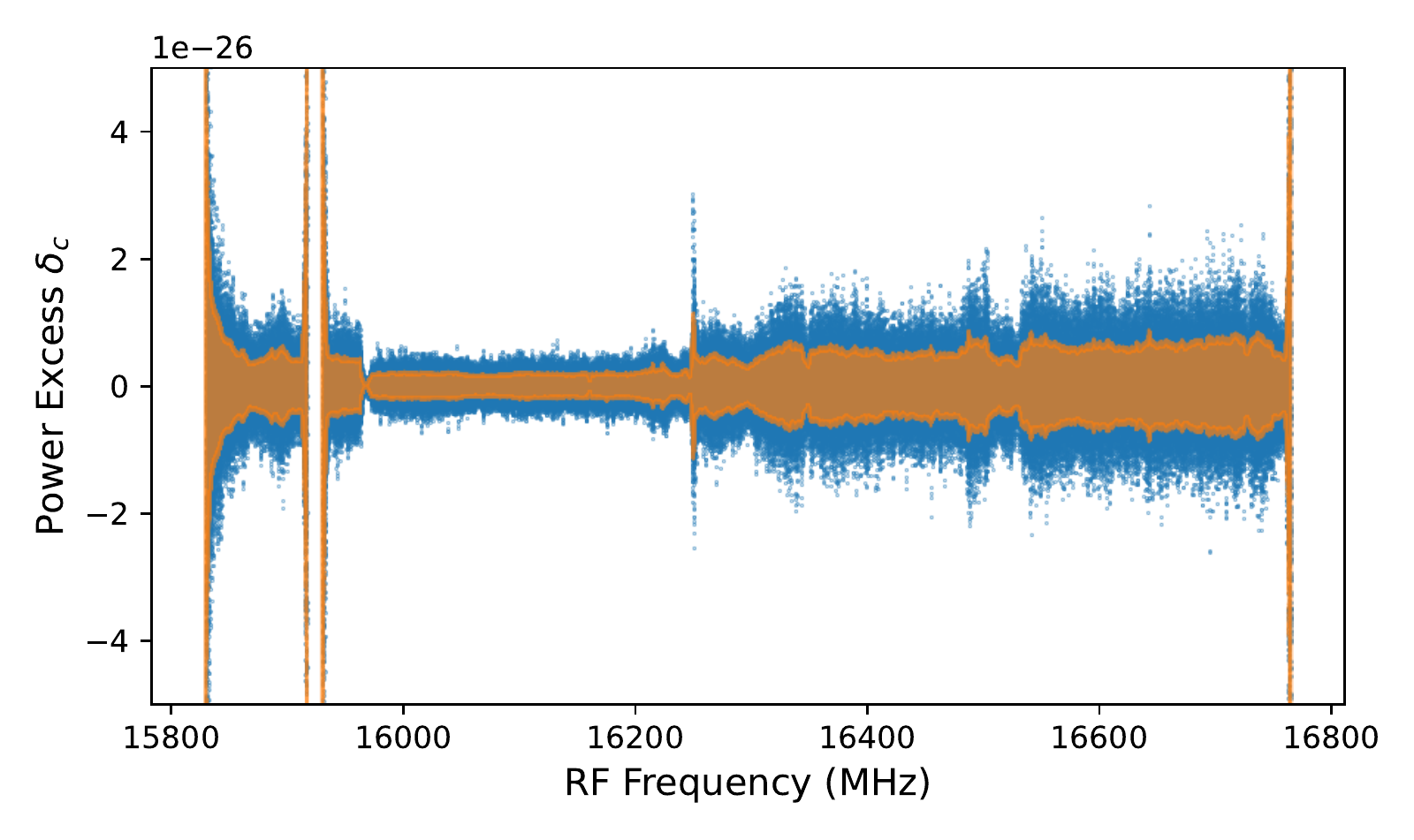}
  \caption{The combined spectrum. The orange band is the standard deviation and includes both statistical and systematic uncertainties.}
  \label{fig:combined_spectrum}
\end{figure}

The combined power excess is shown in Fig.~\ref{fig:combined_spectrum}. Note that the gap near \SI{15.95}{GHz} is a result of manually tuning the top dielectric plate.  

\subsection{Placing 90\% Exclusion Limit}\label{sec:limits}
A statistically significant event for this experiment is defined to be a power excess with an $\snr > 5$. A detection is defined to be a statistically significant event that is persistent after the frequency range is rescanned and that passes additional laboratory tests. These tests include determining that the signal is not radio frequency interference (RFI) and confirming that the signal power decreases appropriately when the cavity is detuned from the signal (references that implement these laboratory tests include~\cite{PhysRevD.64.092003, PhysRevD.103.032002}). One can also investigate the potential dark matter signal with different modes to determine if the signal power scales with the varying $\veff$~\cite{PhysRevLett.127.261803}. 

The histogram in Figure~\ref{fig:spectra_histogram} shows 6 bins from the normalized spectrum above the candidate threshold. However, these bins are inconsistent with a dark matter signal upon closer inspection. These candidates do not persist across spectra with overlapping RF frequencies (the tuning steps are approximately a fourth of the cavity bandwidth). The power excess of these candidates also do not maximize when the signal frequency matches closer to the cavity resonance. The power excess of the neighboring seven bins are also not statistically significant, which is inconsistent with the Standard Halo Model. In the absence of any apparent dark photon candidates, the statistics of the combined spectrum are used to set an upper limit on what dark photon signals could exist\footnote{This SNR threshold is arbitrary. The chances of discovery are increased by lowering the SNR threshold at the cost of the operational complexity from having to perform more rescans.}.

The probability distribution $p$ of measuring a power excess $\delta_c$ given a haloscope signal power $P_S$ is a normal distribution centered around the dark photon signal power.

\begin{align}
  p(\delta_c|P_S) = \frac{1}{\sqrt{2\pi\sigma_{c}}}\exp\left ( -\frac{(\delta_{c} - P_S)^2}{2\sigma_{c}^2} \right )
\end{align}
Note that $P_S=0$ if no dark photon exists in the RF bin. 

However, to place a limit on any possible dark photon signal power, it is more useful to have the probability distribution of $P_S$ given a measured power excess $\delta_c$, $p(P_S|\delta_c)$. Using Bayes Theorem

\begin{align}
  p(P_S|\delta_c) = p(\delta_c|P_S) \frac{p(P_S)}{p(\delta_c)}
  \label{eqn:bayes}
\end{align}
where $p(P_S)$ is the probability of measuring dark photon signal power S, and $p(\delta_c)$ is the probability of measuring power excess $\delta_c$. $\delta_c$ in Equation~\ref{eqn:bayes} is a parameter and not a continuous variable, so $P(\delta_c)$ is a constant. The prior used for this analysis is that $P_S \geq 0$. In the absence of any other information, any value for $P_S$ above zero is equally likely\footnote{$P_S$ also has an upper limit based on previous exclusions, but this limit has a negligible effect on the subsequent analysis.}. Consequently, 

\begin{align}
  \frac{p(P_S)}{p(\delta_c)} \propto H(P_S)
\end{align}
where $H(P_S)$ is the unit step function. That means $p(\delta_s|P_S)$ is just a normal distribution truncated at zero,

\begin{align}
  p(P_S|\delta_c) \propto \frac{1}{\sqrt{2\pi\sigma_c}}\exp\left ( -\frac{(\delta_c - P_S)^2}{2\sigma_c^2} \right ) H(P_S)\label{eqn:dp_gaussian_distribution}.
\end{align}
Readers may refer to~\cite{PhysRevD.101.123011, PhysRevD.57.3873} for additional discussion of Bayesian analysis.

The 90\% confidence limit for the dark photon signal power is the value of $\delta_c$ for which 90\% of the probability distribution lies below. The inverted, but equivalent, statement is that signal powers above this $\delta_c$ are excluded with 90\% confidence. The 90\% confidence limits for dark photon signal power are determined by applying the percent point function to the truncated normal distribution in Equation~\ref{eqn:dp_gaussian_distribution}. 

\begin{figure}[htp]
  \centering
  \includegraphics[width=\linewidth]{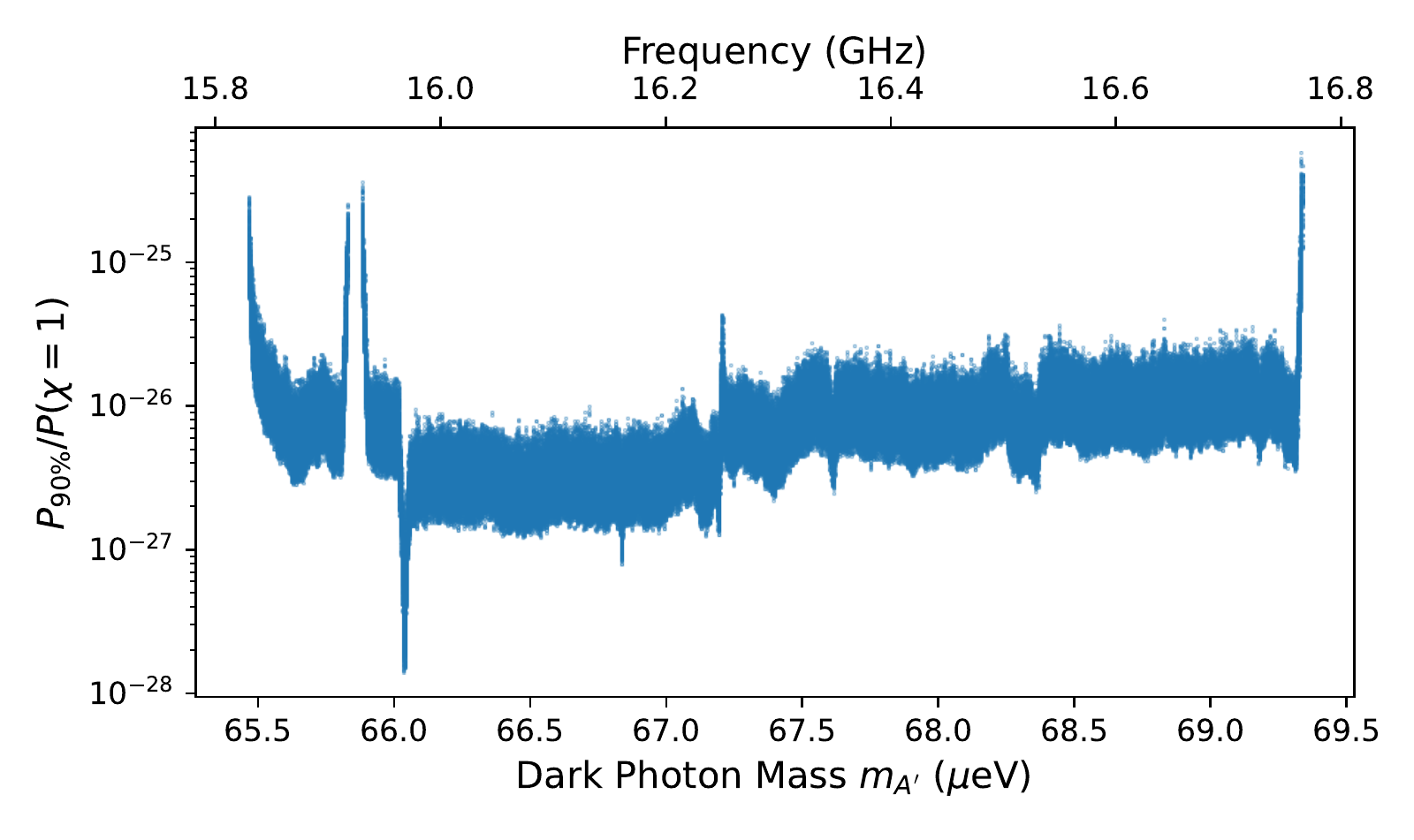}
  \caption{The 90\% confidence limit on the dark photon power, normalized to dark photons signal power of mixing angle $\chi=1$.}
  \label{fig:dp_power_exclusion}
\end{figure}

\begin{figure}
  \centering
  \subfloat{\includegraphics[width=\linewidth]{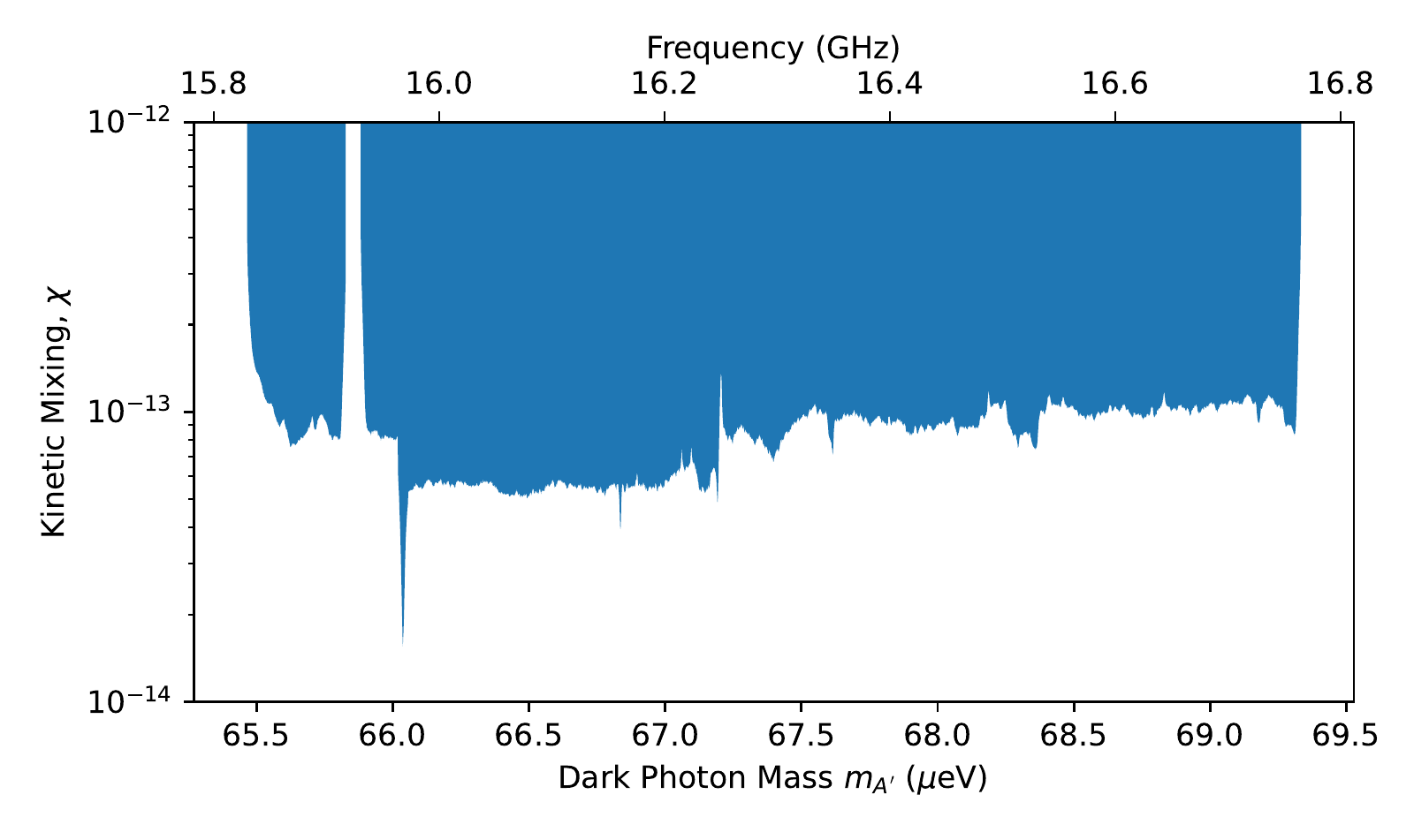}}\\
  \subfloat{\includegraphics[width=\linewidth]{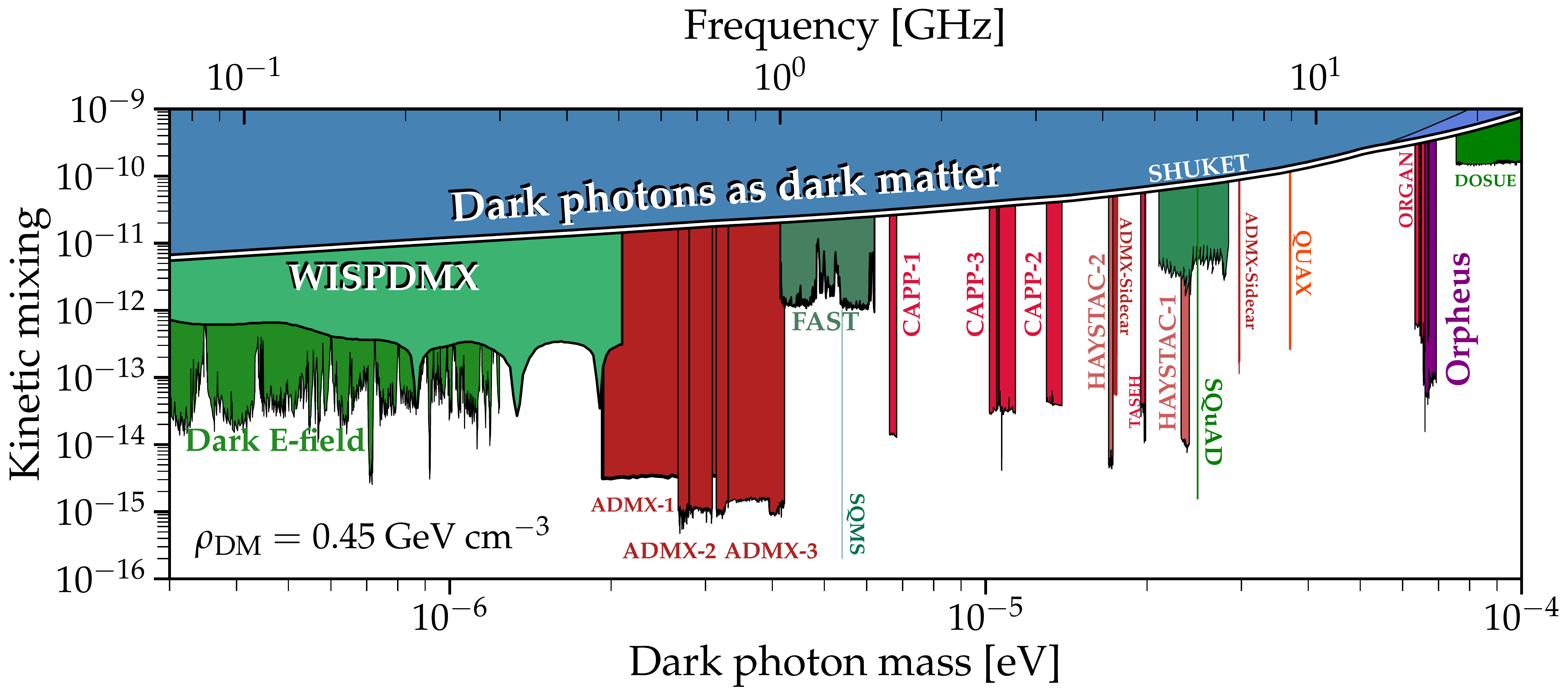}}
  \caption{Top: A 90\% exclusion on the mixing angle parameter space. Bottom: Orpheus limits in the context of other microwave cavity haloscopes. Figure adapted from~\cite{ciaran_o_hare_2020_3932430}.}
  \label{fig:dp_limits}
\end{figure}

By examining Equation~\ref{eqn:dp_power}, one sees that 

\begin{align}
  \frac{\chi_{90\%}}{\chi=1} = \sqrt{\frac{P_{90\%}}{P(\chi=1)}}
\end{align}
Finding $\chi_{90\%}$ amounts to taking the square root of Fig.~\ref{fig:dp_power_exclusion}. Between $\SI{65.5}{\mu eV}$ and $\SI{69.3}{\mu eV}$, the excluded dark photon mixing angle is $\chi_{90\%} \sim \SI{1e-13}{}$ for the unpolarized dark photon case ($\cost = 1/3$). The limits are shown in Fig.~\ref{fig:dp_limits} and are better with expectations laid out in Section~\ref{sec:expected_sensitivity} (perhaps because of the overlapping spectra). Orpheus is the highest-frequency tunable microwave haloscope search to date and achieves sensitivities comparable to other haloscope experiments. 

The true result for the excluded $\chi$ is as jagged as the power exclusion in Fig.~\ref{fig:dp_power_exclusion}. To produce a smoother plot, the $\chi$ exclusion is downsampled by a factor of 100 using the Fourier method (Python's scipy.signal.resample method). The resampling of limits for smoother plots is common practice~\cite{PhysRevD.103.032002}.

If the dark photon polarization is fixed or varying on timescales much longer than the measurement, the scenario implies $\langle \cos^2\theta \rangle \geq 0.076$ for a 90\% confidence limit~\cite{PhysRevD.104.092016}, and the results can be appropriately rescaled. In the fixed polarization scenario, the integration time and timing information, e.g., repeated measurements at the same cavity frequency taken hours apart, can be used to improve the lower bound of $\langle \cos^2\theta\rangle$ and thus improve the DPDM exclusion limit~\cite{PhysRevD.104.095029, PhysRevD.104.092016}. In the case of Orpheus, the deepest exclusion near $\SI{15.95}{GHz}$ in Fig.~\ref{fig:dp_limits} resulted from about one day of integration.

The Orpheus limits also demonstrates the cavity's potential advantages over a cylindrical haloscope operating at similar frequencies such as ORGAN~\cite{doi:10.1126/sciadv.abq3765}. Orpheus has three times $\veff$ and $Q_L$ compared to ORGAN, and achieved as much as almost an order of magnitude better in sensitivity with less experimental run time (Fig.~\ref{fig:dp_limits}).

\section{Future Direction}\label{sec:future}

\subsection{Upcoming Axion Run}
\begin{figure}[htp]
  \centering
  \includegraphics[width=0.95\linewidth]{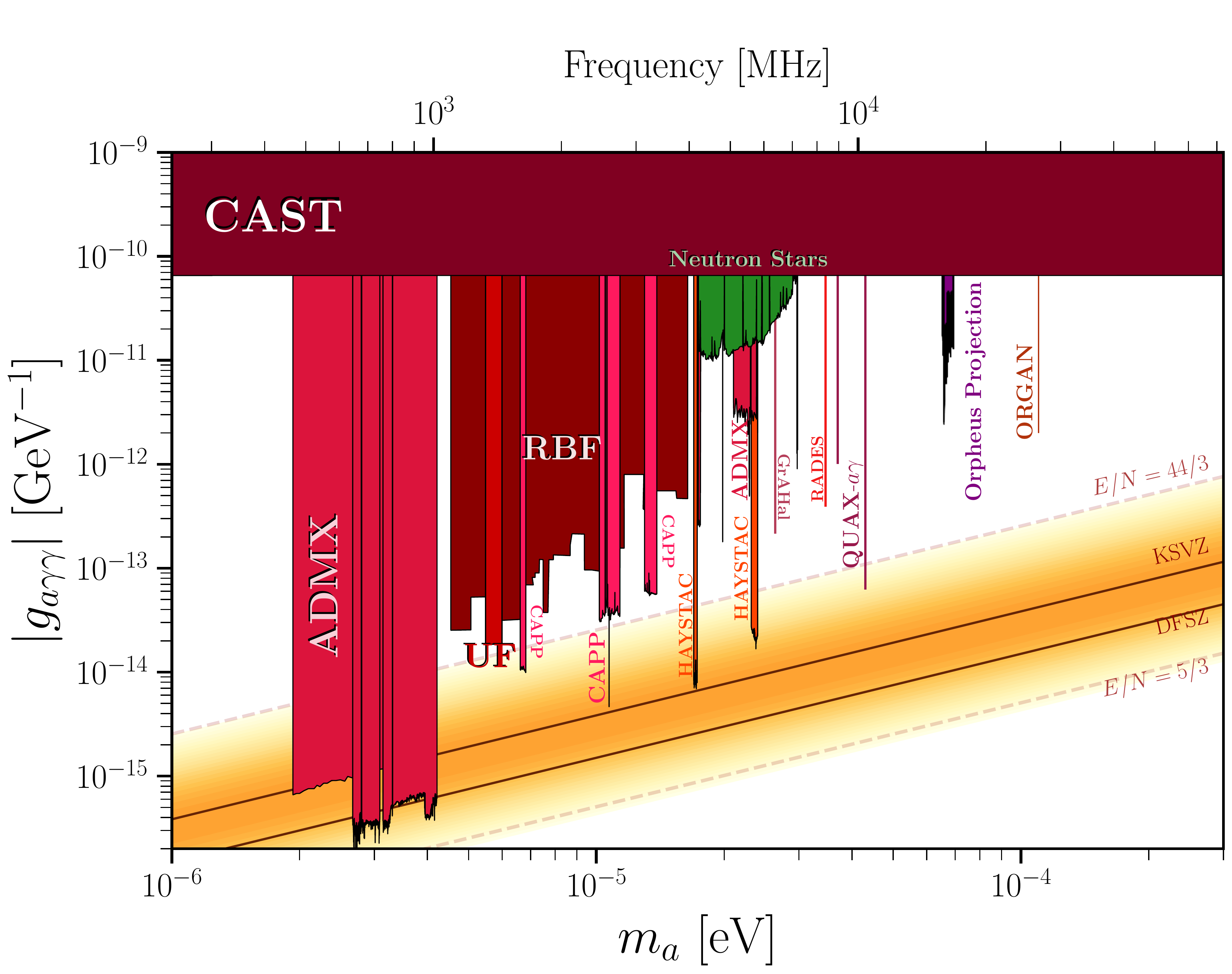}
  \caption{The projected axion limits assuming a \SI{1.5}{T} dipole magnet. Figure adapted from~\cite{ciaran_o_hare_2020_3932430}.}
  \label{fig:big_axion_projection}
\end{figure}

Commissioning a magnet would allow Orpheus to search for axions in addition to dark photons. If the inaugural search had implemented a \SI{1.5}{T} magnet, Orpheus would have been sensitive to axions with $\gagg\sim\SI{3e-12}{GeV^{-1}}$ from \SI{15.8}{GHz} to \SI{16.8}{GHz} (Fig.~\ref{fig:big_axion_projection}), over an order of magnitude more sensitive than CAST. A \SI{1.5}{T} superconducting dipole magnet is currently being fabricated to prepare for an axion-data taking run (magnet design in~\cite{cervantes2021search}). The insert hardware is also being upgraded to improve tuning reliability and thermalization.

\subsection{Electrodynamic Optimizations}
Orpheus has many design parameters that can be adjusted to optimize $Q_0$ and $\veff$, including the mirror radius of curvature, dielectric plate thickness, and dielectric plate positions. As currently configured, the Orpheus $V_{eff}$ is about $2\%$ of the physical volume. However, Fig.~\ref{fig:veff_position_err} demonstrates that a more optimal placement of the dielectric plates increases $\veff$ by more than 7\%. $\veff$ can also be increased by placing dielectrics at every other antinode instead of every fourth of an antinode.

This iteration of Orpheus achieved $Q_L\approx 10000$, which is comparable to the $Q_L$ expected of a cylindrical copper cavity operating at the TM$_{010}$ mode operating in a similar frequency range\footnote{This is calculated from the ADMX Run1c cavity~\cite{PhysRevLett.127.261803}, which has $Q_L\sim\num{80000}$, and using the anomalous skin effect frequency scaling relation $\num{80000}\times \left(\SI{16}{GHz}/\SI{1}{GHz}\right )^{2/3}=\num{13000}$.}. But the literature suggests that quality factors of \num{5e4}~\cite{Dunseith_2015} and even \num{2e5}~\cite{Clarke_1982} are achievable for GHz range Fabry-Perot cavities. Both simulations and analytical estimations suggest that diffraction is a dominant source of loss~\cite{cervantes2021search} and will only become worse as the cavity length increases relative to the mirror size. Diffraction losses can be mitigated by optimizing the mirror radius of curvature (likely by decreasing it). One can also reduce diffraction by increasing the mirror size, but that increases the physical volume of your cavity, which would necessitate a larger, more expensive magnet. It may be possible to mitigate diffraction losses by curving the surface of the dielectric plate so that they act as lenses that collimate the field. However, given that the dielectrics are only a few mm thick, its radius of curvature would have to be about $\sim \SI{200}{cm}$ and may be difficult to machine. One workaround is to have dielectrics with a smaller diameter than the mirrors, but it may add more mechanical complications. It is also possible that curving the dielectric plate surface will reduce the cavity tuning range. 

The dielectric losses can be reduced by using sapphire instead of alumina. Sapphire has a dielectric loss tangent $\tan_d\sim0.00002$ compared to alumina's $\tan_d \sim 0.0001$. But sapphire's birefringence may complicate the cavity mode structure.

From the scan rate equation (Equation~\ref{eqn:scan_rate}) and the relationship $Q_L = Q_0/(1+\beta)$, one can determine that $\beta = 2$ optimizes the scan rate. Unfortunately, with the current coupling mechanism, the aperture size would have to be comparable to or greater than the waveguide cross section, and increasing the aperture size decreases $Q_0$. This is especially unfortunate for designing larger Orpheus detectors because, for a fixed coupling hole size, the cavity coupling coefficient reduces for longer cavities. Luckily, any cavity can be impedance-matched with the appropriate network~\cite{pozar}. But this impedance matching network must be engineered and implemented for Orpheus if this design is to be scaled to reach Dine-Fischler-Srednicki-Zhitnitsk (DFSZ) sensitivity.

Only the \tem mode was used to search for DM signals. It is possible that other modes have substantial coupling to the axion or dark photon. The $V_{eff}$ of these modes should be simulated. If multiple modes couple to dark matter, then collecting data around these modes would be an easy way to increase the scanned mass range of a data-taking run. 

\subsection{Scanning More Dark Matter Parameter Space and Reaching DFSZ Sensivity}
The Orpheus detector can be modified to be sensitive to axions and dark photons at different frequencies. The cavity tuning range can be adjusted by changing the dielectric thicknesses and mirror curvature appropriately.

Orpheus can also become sensitive to the QCD axion by making it larger and colder. From Equation~\ref{eqn:axion_power}, the axion-photon coupling constant can be estimated as

\begin{align*}
  \gagg &= \frac{1}{B_0}\sqrt{\frac{\beta+1}{\beta}\frac{\snr \times \Delta f T_n m_{a}}{\rho_a V_{eff}Q_L}}\left ( \frac{1}{b \Delta t} \right )^{1/4}.
\end{align*}

Orpheus can achieve Kim-Shifman-Vainshtein-Zakharov (KSVZ) sensitivity if ${V_{eff}\sim \SI{120}{mL}}$ and $Q_L\sim \num{2e4}$, $T_n\sim\SI{1}{K}$, and $B_0 = \SI{10}{T}$. That would require the optimizations mentioned earlier, cooling the cavity with a dilution refrigerator, quantum noise limited amplifiers, and technological advances in winding superconducting dipole magnets. Except for the dipole magnet, this is all achievable with current technology. DFSZ may be reached by increasing the cavity size even further, such that ${V_{eff}\sim \SI{600}{mL}}$. Detection mechanisms that subvert the Standard Quantum Limit, such as vacuum squeezing~\cite{Backes2021} and superconducting qubit photon counters~\cite{PhysRevLett.126.141302}, may also be developed in this frequency range to increase sensitivity further.

\section{Conclusion}
Orpheus has been able to exclude an impressive amount of parameter space half an order of magnitude higher in frequency than other haloscope experiments while also having a larger tuning range. 
Between $\SI{65.5}{\mu eV}$ and $\SI{69.3}{\mu eV}$, the excluded dark photon kinetic mixing strength is ${\chi_{90\%} \sim \num{1e-13}}$ for the unpolarized dark photon case. With modest alterations and several experimental iterations, the same apparatus may be used to exclude larger parameter space from \SI{45}{\mu eV} to \SI{80}{\mu eV}  with similar sensitivities. A \SI{1.5}{T} magnet is currently being wound to allow Orpheus to scan for axions in addition to dark photons.

Ultimately, Orpheus is a pathfinder experiment with limited scope. May the hard-earned lessons prove useful to other dielectric haloscope experiments.

\section{Acknowledgements}
This work was supported by the U.S. Department of Energy through Grants No. DE-SC0011665 and by the Heising-Simons Foundation. Pacific Northwest National Laboratory (PNNL) is operated by Battelle Memorial Institute for the DOE under Contract No. DE-AC05-76RL01830. Prepared by LLNL under Contract DE-AC52-07NA27344 with release \#: LLNL-JRNL-834496. Many of the parts were fabricated by the University of Washington Physics Machine Shop and CENPA machine shop. CENPA administration and engineers helped with developing the infrastructure to commission the Orpheus test stand. We thank M. Baryakhtar for helpful discussions and clarification on dark photon cosmology.

\appendix
\section{More about the Simulated and Measured Scattering Parameters S$_{21}$ and S$_{11}$}
\begin{figure}[htp]
  \centering
  \includegraphics[width=0.95\linewidth]{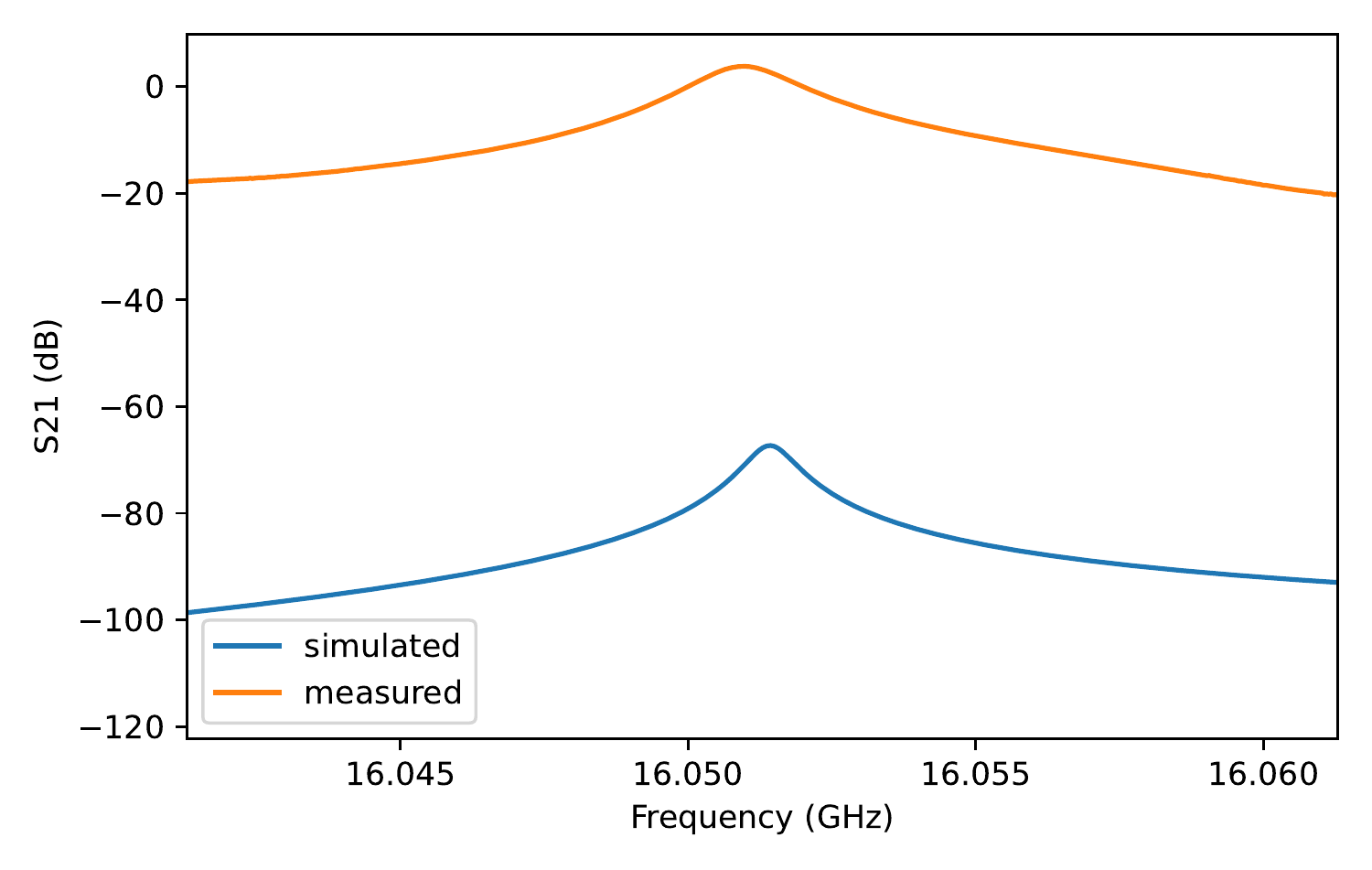}
  \caption{The measured and simulated S$_{21}$.}
  \label{fig:s21}
\end{figure}

\begin{figure}[htp]
  \centering
  \includegraphics[width=0.95\linewidth]{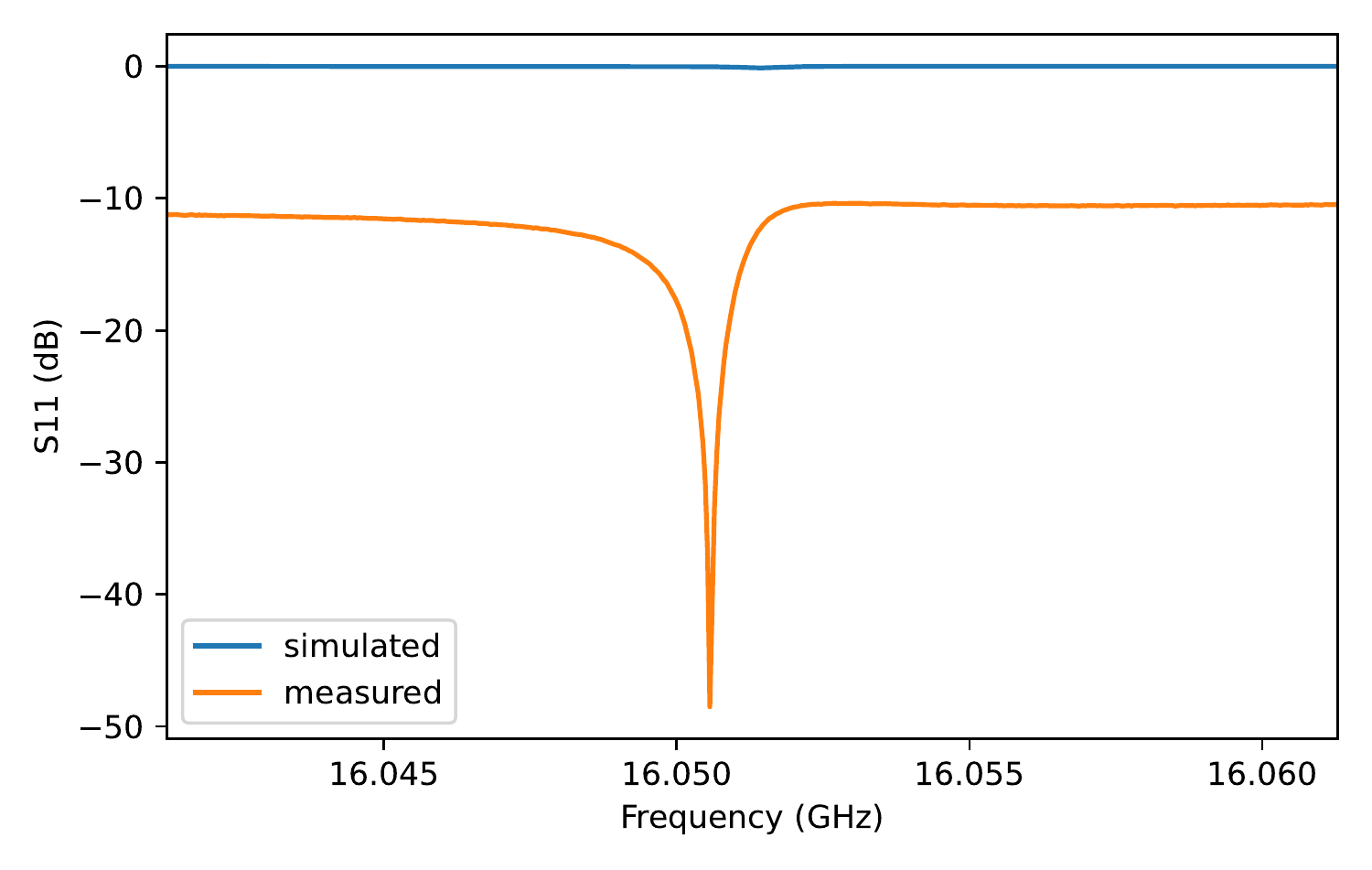}
  \caption{The measured and simulated S$_{11}$.}
  \label{fig:s11}
\end{figure}

Fig.s~\ref{fig:s21} and~\ref{fig:s11} show both the measured and simulated scattering parameters S$_{21}$ and S$_{11}$. 

The scattering parameters were simulated using HFSS's driven modal solution. The eigenmode solver is untenable due to the large density of modes, particularly due to the subresonances of the dielectrics. The majority of these modes are irrelevant. The driven modal solution filters out modes that do not couple well to the aperture.

The number of mesh elements required for simulation is reduced by simulating thicker mirror apertures than what was implemented in the experiment. The thicker apertures reduced the throughput of the energy from the simulated wave port, and so the simulated cavity is unsurprisingly very undercoupled ($\beta \approx 0$). However, this undercoupling does not affect the field distribution, only the amplitude of the field. In other words, the cavity coupling coefficient does not affect the simulated $\veff$ or $f_0$. Ultimately, the simulations are only used to extract $\veff$ and to identify the \tem mode. Other parameters for determining the dark matter signal strength, such as $Q_L$, $\beta$, and $f_0$, are measured directly.

The width of the measured resonance is roughly double the simulated resonance width because the measurement corresponds to a critically-coupled port ($\beta = 1$). The width corresponds to $f_0/Q_L$, and $Q_L = Q_0/(1+\beta)$.

The absolute values of the scattering parameters differ between the measurement and simulation. One reason is because of the different cavity coupling coefficients. The other reason is that the measured scattering parameters include the amplifier gains and insertion losses along the transmission line. However, the effects of the transmission line do not impact the science as they affect the potential signal power and the noise floor in the same manner. Because of the effects of the transmission line, the absolute values of the scattering parameters are not indicative of the losses of the cavity. Rather, it is the width of the resonance that is a direct measurement of the losses in the cavity (Equations~\ref{eqn:lorentzian_transmission} and~\ref{eqn:lorentzian_reflection}). 

\FloatBarrier
\bibliography{orpheus_thesis}

\end{document}